\documentclass[a4,useAMS,usenatbib,usegraphicx]{mn2e}
\usepackage{subfig}
\usepackage{lscape}
\usepackage{rotating}
\usepackage{natbib}
\usepackage{amssymb,amsmath}
\usepackage{fixltx2e}
\usepackage{longtable}
\usepackage[T1]{fontenc}
\usepackage{hyperref}
\hypersetup{
    colorlinks=true,
    linkcolor=blue,
    filecolor=magenta,      
    urlcolor=blue,
}

\title[Isolated HI clouds or dark galaxies ?]{Kinematic clues to the origins of starless HI clouds : dark galaxies or tidal debris ?}

\author[R. Taylor, J. I. Davies, P. J\'{a}chym, O. Keenan, R. F. Minchin, J. Palou\v{s}, R. Smith, R. W\"{u}nsch]{R. Taylor$^{1}$\thanks{Email: rhysyt@gmail.com}, J. I. Davies$^2$, P. J\'{a}chym$^1$, O. Keenan$^2$, R. F. Minchin$^3$, J. Palou\v{s}$^1$, \newauthor R. Smith$^4$, R. W\"{u}nsch$^1$\\
$^1$Astronomical Institute of the Czech Academy of Sciences, Bocni II 1401/1a, 141 31 Praha 4\\
$^2$School of Physics \& Astronomy, Cardiff University, Queens Buildings, The Parade, Cardiff CF24 3AA, U.K.\\
$^3$Arecibo Observatory, HC03 Box 53995, Arecibo, Puerto Rico 00612\\
$^4$Yonsei University, Graduate School of Earth System Sciences-Astronomy-Atmospheric Sciences, Yonsei-ro 50, Seoul 120-749, Korea;\\}

\begin{document}

\newcommand{\HI}{H\textsc{i}}
\newcommand{\Msolar}{M$_{\odot}$}
\newcommand{\kms}{km\,s$^{-1}$}

\date{2016}

\pagerange{\pageref{firstpage}--\pageref{lastpage}} \pubyear{2016}

\maketitle

\label{firstpage}

\begin{abstract}
Isolated \HI{} clouds with no optical counterparts are often taken as evidence for galaxy-galaxy interactions, though an alternative hypothesis is that these are primordial `dark galaxies' which have not formed stars. Similarly, certain kinematic features in \HI{} streams are also controversial, sometimes taken as evidence of dark galaxies but also perhaps explicable as the result of harassment. We numerically model the passage of a galaxy through the gravitational field of cluster. The galaxy consists of SPH particles for the gas and n-bodies for the stars and dark matter, while the cluster includes the gravitational effects of substructure using 400 subhalos (the effects of the intracluster medium are ignored). We find that harassment can indeed produce long \HI{} streams and these streams can include kinematic features resembling dark galaxy candidates such as VIRGOHI21. We also show that apparent clouds with diameter $<$20 kpc and velocity widths $<$50 \kms{} are almost invariably produced in these simulations, making tidal debris a highly probable explanation. In contrast, we show that the frequency of isolated clouds of the same size but velocity width $>$100 \kms{} is negligible - making this a very unlikely explanation for the observed clouds in the Virgo cluster with these properties.

\end{abstract}

\begin{keywords}
galaxies: evolution - surveys: galaxies.
\end{keywords}

\section{Introduction}
\label{sec:intro}
As we described in \cite{me16}, hereafter T16, starless clouds of neutral atomic hydrogen (\HI{}) appear to be very rare in the local Universe, but they are not entirely absent. The leading hypothesis is that the majority of those clouds which do exist are tidal debris produced during galaxy-galaxy encounters, with simulations by \cite{bekki} (B05) and \cite{duc} (DB08) supporting this. However we also showed that the clouds in the Virgo cluster described in \cite{me12} (T12) and \cite{me13} (T13) have properties which make them extremely difficult to explain in this way (a similar cloud was recently reported in \citealt{sorgho}). Specifically they are isolated ($>$100 kpc from the nearest galaxy), compact ($<$17 kpc in diameter) and of high velocity width ($>$150 \kms{}). Additionally, there are no indications of any more extended \HI{} features in their vicinity. We demonstrated that producing clouds with those particular properties is extremely difficult and the only similar features produced in our simulations are transient, seldom lasting longer than 50 Myr.

We also tested an alternative hypothesis in T16 that the clouds could be primordial `dark galaxies' : rotating discs of gas embedded in dark matter halos. These have been proposed as a possible solution to the well-known `dwarf galaxy problem', e.g. \cite{moore}, \cite{d04}. We demonstrated that such objects can easily explain the small size and high velocity widths of the observed clouds, and would be stable against the effects of harassment both in terms of disruption and triggering of star formation. This supports the speculation that recently discovered `ultra diffuse' galaxies (which have similar baryonic masses) require a high dark matter content in order to survive in cluster environments (\citealt{vandok}, \citealt{koda}). 

The nature of the clouds could have important consequences for cosmological models. The idea that some dark halos do not accrete enough baryons to form stars has recently been revived in new simulations (e.g. \citealt{eagle}) and extremely dark matter dominated galaxies appear to be common in clusters, e.g. \citealt{vandok}, \citealt{koda}, \cite{van16}, \cite{d16}, \cite{munoz}. If, on the other hand, the clouds are stripped gas from galaxies, they might explain why so few \HI{} streams are currently observed in Virgo despite the large numbers of gas-depleted galaxies (see T16 for a full discussion). Thus we continue our goal of T16 to establish whether it is more likely that the clouds are primordial or non-primordial objects.

The nature of our simulations in T16 and those of B05 and DB08 are slightly different. The previous works demonstrated that interactions between two galaxies could produce starless \HI{} features by gas stripping directly out of the galaxies. Our simulations examined the evolution of a stripped gas stream in a cluster environment. We dropped a stream of gas through the simulated gravitational field of a Virgo-mass cluster (extracted from a n-body cold dark matter (CDM) simulation) on 27 different trajectories, the aim being to assess the probability (as opposed to the possibility) of producing isolated clouds, especially those with properties similar to the T13 clouds. B05, DB08 and T16 all neglect the role of any external medium.

The results of T16 support the conclusion that some isolated clouds can indeed be explained by purely gravitational interactions. Compact clouds with velocity widths $<$ 50 \kms{} were almost ubiquitous in our simulations - but those with widths $>$ 100 \kms{} were only found 0.2\% of the time. Whilst the simulations of B05 and DB08 produced features with high velocity widths, as discussed in T16 this was only true for relatively large features $>$ 100 kpc in extent. Thus although it is common to cite B05 and/or DB08 as a possible explanation for starless clouds (e.g. \citealt{hcg44}, \citealt{cannon}), one should be very careful - the velocity gradient of the clouds is not a trivial detail but a parameter which very strongly determines whether tidal debris is a likely explanation or not. 

Our simulations in T16 produced results consistent with those of B05 and DB08, often producing large features with high velocity widths. However we also emphasised that small features with such widths are very rare, an important detail which has hitherto been neglected. For example, \cite{wong} state that B05 and DB08 `demonstrated that the majority of gas clouds identified with no known optical counterpart can be easily reproduced by galaxy-galaxy interactions'. Similarly \cite{pisano}, citing B05, state that, ``all claims of intergalactic \HI{} clouds without associated stars, or `dark galaxies'... have turned out to be, on closer inspection, either low surface brightness galaxies... or tidal debris connected with a bright galaxy.'' \cite{jan15} state that, ``many `dark' galaxy candidates turn out to be tidal features'', citing DB08. Our models in T16 agreed with this interpretation regarding large features but showed that the tidal debris hypothesis has significant problems explaining smaller structures.

In T16 we modelled the stripped gas as a simple cylinder. This is clearly unrealistic, so it is preferable to model the gas removal process as well as its subsequent evolution. Here we replace the infalling gas stream with a model galaxy, with stars, gas and dark matter all represented by particle distributions. This enables us to quantifiably estimate whether the parent galaxies would appear disturbed - an important parameter since the AGES clouds are all near undisturbed galaxies. It also allows us to test if the clouds are optically dark or if significant stellar material is also removed, examine the possibility that compact clouds may be removed without the formation of a longer stream, and directly measure their separation from their parent galaxy.

Our basic procedure is the same as in T16, except instead of dropping toy model gas cylinders into a simulated cluster, we now use galaxies on multiple trajectories. Our goal is not to determine if there is some carefully fine-tuned encounter which can reproduce the observed clouds, but whether or not such clouds form naturally (i.e. on random trajectories), how frequently, and for how long they would match the parameters of the observed clouds. 

We also investigate whether our simulations produce features similar to the object known as VIRGOHI21, which is quite similar to the T13 clouds except that it is found in the middle of a 250 kpc stream (\citealt{m07}, \citealt{haynesV21}). The VIRGOHI21 object is a sharp `kink' in the velocity gradient of the stream, prompting speculation that it was a kinematically distinct dark galaxy. The DB08 model explains this object as the result of a high velocity encounter disturbing the gas of NGC 4254. However, their model of NGC 4254 was rather unusual, so it is worthwhile to demonstrate if the same results can be obtained with more typical spiral galaxies. Thus we investigate two kinds of potentially fake dark galaxies : those embedded in long \HI{} streams, and those which appear to be isolated.

It is important to emphasise that as in T16, our simulations are still very limited, particularly with regards to the very important intracluster medium (ICM). This is partly due to the technical challenge of including the harassing galaxies and ICM in a self-consistent simulation, and partly because the previous analyses have been interpreted to claim that the observed features can be explained solely by tidal interactions. Ultimately we also need to incorporate the ICM, heating and cooling, and perhaps gas in the harassing galaxies, but it is sensible to increase the complexity of the simulations gradually to try and understand which effect is the most significant.

The three questions we will try and answer in this work can be expressed as follows :\\
1) Can the \HI{} clouds described in T13 be produced by harassment of a spiral galaxy, and if so, how probable is this scenario ?\\
2) How probable is the production of a system similar to VIRGOHI21 (i.e. with a sharp velocity kink) by harassment\,?\\
3) How do these results depend on the properties of the harassed spiral galaxy ?

The remainder of this paper is organized as follows. In section \ref{sec:sims} we describe the simulation setup and analysis techniques, in section \ref{sec:results} we describe the results, and in section \ref{sec:conc} we discuss our conclusions. Throughout this paper we assume a distance to the Virgo cluster of 17 Mpc and a Hubble constant of 71 \kms{} Mpc$^{-1}$.

\section{Numerical Simulations}
\label{sec:sims}
Both our simulation code and analysis methods are described in detail in T16. Therefore we only briefly summarise them here, except for the description of the spiral galaxy which is obviously very different from the streams and dark galaxies we used in T16.

\subsection{Simulation setup}
\subsubsection{The cluster}
\begin{figure}
\centering
\includegraphics[width=85mm]{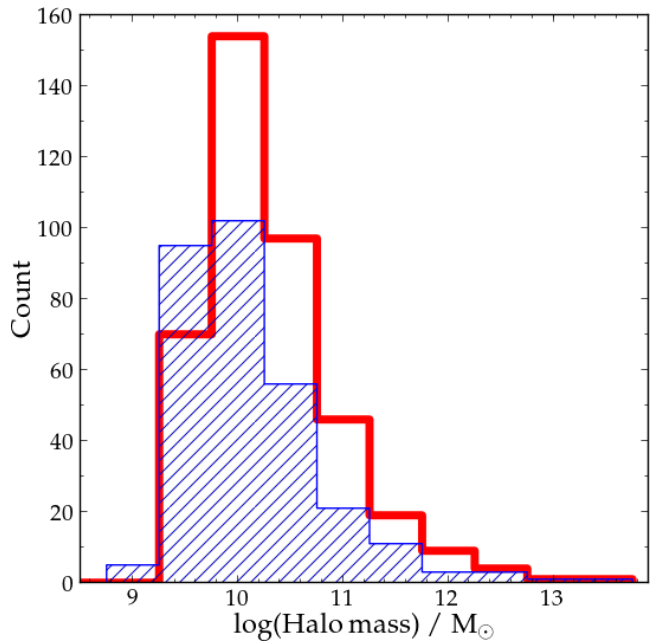}
\caption[hmass]{Distribution of the masses of the NFW subhalos in the simulation. The open red histogram shows the distribution at the beginning of the simulation and the blue hatched histogram shows the distribution after 5 Gyr of evolution. Subhalos tend to merge with the main cluster potential (not shown) over time, hence the mass in the substructure is less by the the end of the simulation.}
\label{fig:halomass}
\end{figure}

\begin{figure}
\centering
\includegraphics[width=85mm]{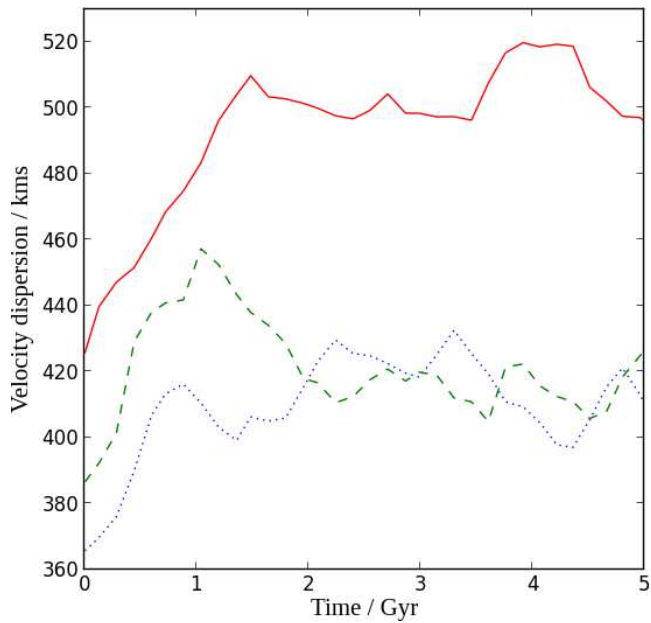}
\caption[vevo]{Evolution of the \textit{rms} velocity dispersion of the cluster along three different (arbitrary) axes : $x$ in solid red, $y$ in dotted blue, and $z$ in dashed green. Time is measured from the beginning of the simulation. The mean velocity dispersion after the first 1 Gyr (when the velocity dispersion is still rising) is 490 \kms{} along the $x$ axis, 420 \kms{} along the $y$ axis, and 410 \kms{} along the $z$ axis.} 
\label{fig:vdispevo}
\end{figure}

Our model of the cluster is identical to that already described in T16 and in detail in \cite{rory15} and \cite{warnick}. Briefly, the original simulation was a pure n-body CDM simulation using 512$^{3}$ particles. The simulation domain was 64 $h^{-1}$ Mpc on a side, and the effects of cosmological expansion were included. Re-simulating with this many particles is prohibitively computationally expensive, so this was reduced to 400 subhalos using the structure-finding algorithm of \cite{gill}. In the simulations described here, each subhalo was approximated by a Navarro-Frenk-White potential (\citealt{nfw}) based on their measured mass and concentration. The properties of each subhalo are allowed to vary with time based on the original simulation data. In the interests of computation time we only use 5 Gyr from the original 6.8 Gyr of simulation time. The overall cluster mass and virial radius at $z$ = 0 are comparable to the real Virgo cluster (1.1$\times$10$^{14}$\Msolar{} and 973 kpc; see \citealt{rory15} and references therein); the distributions of the subhalo masses are shown in figure \ref{fig:halomass}. The velocity dispersion depends on the choice of the direction of measurement but along the (arbitrary) $x$ axis averages 485 \kms{}, which compares favourably (though slightly smaller than) the observed Virgo value of 530 \kms (\citealt{mei}). The time evolution of the velocity dispersion is shown in figure \ref{fig:vdispevo}. 

We then construct a spiral galaxy using the smooth particle hydrodynamics code `gf' (\citealt{gf}) with properties described in the next section. Full details of the code are given in \cite{williams}, see also \cite{gf}. As with the streams in T16, we position the galaxy at 27 initial positions (at the corners and mid-points of a cube; see T16 figure 6). The size of this cube is set so that the galaxy is initially $\approx$500 kpc from the cluster centre\footnote{We found in T16 that infall from 1 Mpc does not significantly affect the evolution of a gas stream - the same changes occur, with the only difference being that this takes longer from 1 Mpc than 500 kpc. However this might not be the case for a more massive and/or more extended cluster.}. The galaxy then undergoes radial infall through the cluster as it evolves. Thus the galaxy experiences a gravitational field which is a reasonable approximation to the overall properties (though of course not the precise details) of the Virgo cluster. The essentially random trajectories allow us to evaluate what should actually happen to an infalling galaxy; rather than fine-tuning the initial conditions to prove a particular result is possible, we want to estimate the probability of a given result occurring naturally. 

\subsubsection{The infalling spiral galaxy}
Our model spiral galaxy consists of SPH particles for the gas and n-bodies for the stars and dark matter. For all models we use 50,000 dark matter particles, initially 20,000 stars and 20,000 gas particles. The numbers are chosen to give high enough resolution to study features with the mass of the dark clouds we are interested in, which would consist of about 100 particles given the total gas mass (see below), but low enough to be acceptably computationally cheap (though a higher resolution would be preferable for reasons discussed in section \ref{sec:simanalysis} - this is one advantage of the DB08 model which used 1,000,000 particles for each component of the galaxy). As in T16 we use a gravitational softening length of 100 pc and the gas is set to be isothermal. As discussed in T16, this resolution should be sufficient for modelling features similar to the T13 clouds, which are likely to be at least 2 kpc in diameter in order to remain optically dark (with an upper limit of 17 kpc diameter) as below this their column density would exceed the threshold for star formation\footnote{Also, as in T16, to be self-bound by their \HI{} gas alone, the clouds would have to be so small their column densities would be several orders of magnitude higher than the \HI{} in most galaxies so they should be star-forming. This means that we are only interested in unbound objects, so higher resolution simulations would not change the conclusions.}. The dark matter distribution in the simulations is an isothermal sphere truncated at a radius of 50 kpc.

Star formation is employed using a Schmidt law with an index of 1.5. Star formation in \textit{gf} is not very sophisticated (see \citealt{rory10} for details) but is useful to indicate whether gas would remain optically dark or not. For star formation to occur the density of gas must exceed a threshold of 2.5$\times$10$^{-3}$\Msolar{}\,pc$^{-3}$. Above this threshold star particles are created at a rate given by the Schmidt law while the mass of the associated gas particles is decreased (total mass is conserved). We set the mass of the new star particles to equal the original star particles. Our models typically have a star formation rate of $\sim$0.6 \Msolar{}\,yr$^{-1}$ in isolation.

To ensure the disc is stable, we allow all models to evolve in isolation for 2 Gyr. After some slight redistribution of mass in the first few hundred megayears, the disc quickly settles down and the final radial profile remains very close to the initial parameters. Examples of the evolution of the radial profiles are shown in figures \ref{fig:gasprofileM2} and \ref{fig:starprofileM2} at 0, 1 and 2 Gyr. Since the disc has achieved stability by 1 Gyr, we use this as the initial conditions for the harassment runs. 

As discussed, our goals are partially motivated by the DB08 study, which attempted to reproduce the VIRGOHI21 feature associated with the spiral galaxy NGC 4254. Since we wish to see if such a feature can arise in our harassment simulations, we choose NGC 4254 to provide the basic parameters of our galaxy, i.e. the gas and stellar mass, circular velocity and radial density profile. However, many of the parameters of NGC 4254 are not well determined observationally and it is worthwhile to examine how changing the parameters influences the end result - in particular, how the results change using parameters similar to those adopted in DB08 compared to those motivated directly from the observations (see below). 
 
\begin{figure}
\centering
\includegraphics[width=85mm]{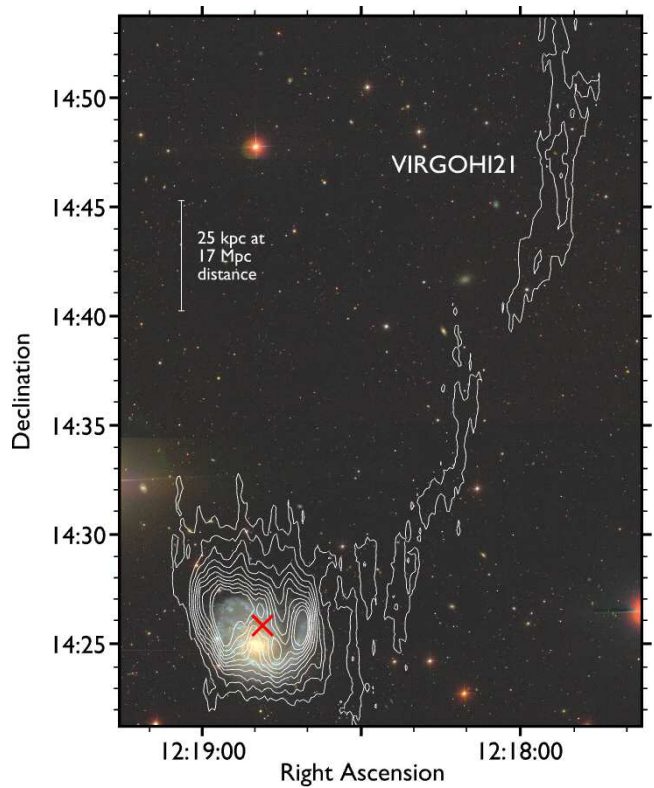}
\caption[vmap]{NGC 4254 \HI{} contours in white overlaid on an RGB image from the SDSS. The contours are from a moment 0 map integrated over the velocity range 2,200 - 2,600 \kms{}, ranging from 0.1 to 1.0 Jy \kms{} in steps of 0.1 Jy \kms. The red cross indicates the \HI{} centre of the galaxy as determined with \textit{mbspect}. A position-velocity diagram of this system is shown in figure \ref{fig:ducvsrealvhi21}.}
\label{fig:4254Map}
\end{figure}

\begin{figure}
\centering
\includegraphics[width=84mm]{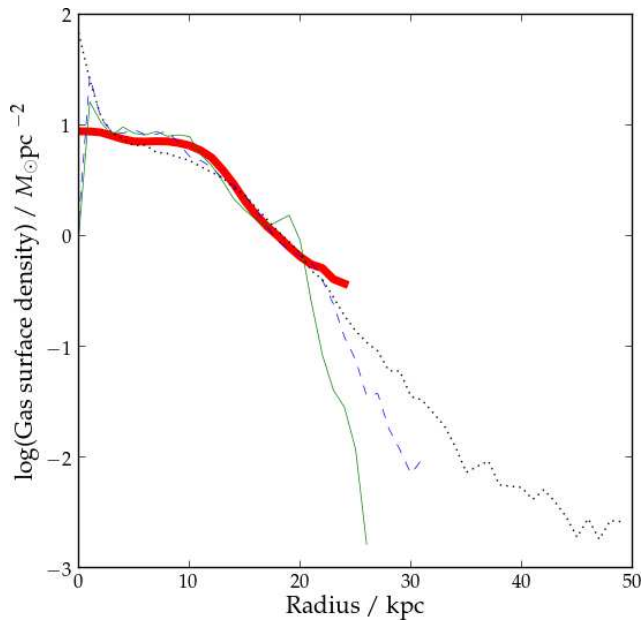}
\caption[vmap]{Evolution of the surface density profile of the gas component in our M2 simulation. The thick red line shows the initial conditions, which are taken from an azimuthally-averaged profile of NGC 4254 from the Westerbork data cube (\citealt{m07}). The thin green solid line shows the profile extracted from the simulation with the galaxy in isolation after 1 Gyr and the dashed blue line shows the profile at 2 Gyr. The black dotted line shows the median profile of all 27 simulations after 5 Gyr in the cluster.}
\label{fig:gasprofileM2}
\end{figure}

\begin{figure}
\centering
\includegraphics[width=84mm]{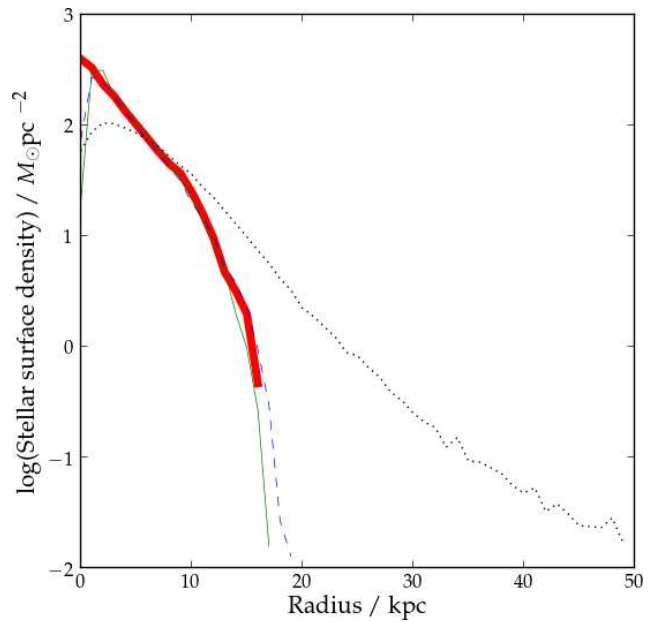}
\caption[vmap]{Evolution of the surface density profile of the stellar component in our M2 simulation in isolation, using the same colour scheme as figure \ref{fig:gasprofileM2}.}
\label{fig:starprofileM2}
\end{figure}

While its long \HI{} stream and single prominent spiral arm clearly indicate that the galaxy has experienced an unusual event, the nature of that event is controversial. \cite{sofue} show that the stellar arm (see figure \ref{fig:4254Map}) could be the result of ram pressure stripping, while the model of \cite{d08} shows that it could be induced by a low-speed tidal encounter. Signatures of ram pressure stripping have been found in the radio continuum by \cite{kanth} as well as in the displacement of the \HI{} and optical discs (figure \ref{fig:4254Map}), despite the rather high (1 Mpc) projected distance from the cluster centre. This makes it especially difficult to determine the characteristics of NGC 4254 prior to its interaction. Therefore we use a range of values compatible with the constraints from the observations.

\paragraph*{\textnormal{\textit{Gas in NGC 4254}}}\mbox{}\\
Estimates of the gas mass of NGC 4254 vary. After correcting for our assumed distance of 17 Mpc, \cite{dandl}'s flux measurement gives an \HI{} mass of 1.0$\times$10$^{10}$\Msolar{}. More recent values have tended to be lower : \cite{hucht} gives 7.0$\times$10$^{9}$\Msolar{}, while \cite{phook} and subsequent others give values around 5.0$\times$10$^{9}$\Msolar{} (e.g. \citealt{m07}, \citealt{chung09}, \citealt{alfalfavirgo}, \citealt{chemin}). Since the Westerbork data given in \cite{m07} is well-resolved and we use this to determine the gas distribution, we use their value of 5.5$\times$10$^{9}$\Msolar{}, which is very close to the other recent values. 

The molecular gas in the galaxy is substantial, estimated to be between 1.0$\times$10$^{9}$\Msolar{} (\citealt{ober}) and 6.5$\times$10$^{9}$\Msolar{} (\citealt{wilson}); \cite{chemin} estimates 5.0$\times$10$^{9}$\Msolar{}. Including the molecular gas is not possible in $gf$ and would require extremely high resolution, and in any case would be unlikely to affect the amount of stripped gas significantly (since the molecular gas is much more centrally concentrated than the \HI{} but much less massive than the stellar component). Since we wish to create synthetic observations to quantify how much \HI{} we would detect in any resulting features (see section \ref{sec:simanalysis}), we choose to ignore the molecular gas.

The observed \HI{} shows a roughly flat profile (figure \ref{fig:gasprofileM2}) within the stellar disc (a radius of approximately 13 kpc) then an exponential decrease out to around 25 kpc (the limit at which the \HI{} can be measured from the Westerbork data due to sensitivity). We use this measured profile to set our initial profile for the gas, scaling the surface density according to the total gas mass used in each of our models.

\paragraph*{\textnormal{\textit{Stars in NGC 4254}}}\mbox{}\\
As with the gas, estimates for the stellar mass vary. \cite{kranz} estimate 2.0-3.5$\times$10$^{10}$\Msolar{}, \cite{chemin} says 4.2$\times$10$^{10}$\Msolar{}. We estimate a slightly higher value of 6.1$\times$10$^{10}$\Msolar{} based on (somewhat crude) SDSS $g$ and $i$ band photometry (see T12). We use the profile shown in figure \ref{fig:starprofileM2}, which was extracted from an SDSS $g$ band image, with a stellar mass of 3.5$\times$10$^{10}$\Msolar{} (the upper limit from \citealt{kranz}, who appear to have done the most detailed investigation of this parameter).

\paragraph*{\textnormal{\textit{Dynamical mass of NGC 4254}}}\mbox{}\\
Dynamical mass is determined by the velocity width of the galaxy, which must be corrected for inclination angle. Estimates vary widely, to some extent depending on the component used for the measurement. The lowest estimate is 20$^{\circ}$ in \cite{chemin} and \cite{makarov}, based on the stellar component; \cite{goldmine} optical data gives 28$^{\circ}$; \cite{hucht}, \cite{chung09}, \cite{sofue} and \cite{wilson} say 28-30$^{\circ}$ using optical and CO data; \cite{kranz}, \cite{phook} and \cite{sofue} all give 41 or 42$^{\circ}$ using NIR or \HI{} data. These are significant variations, causing the estimated deprojected circular velocity to vary by a factor of two, and therefore the dynamical mass estimate to vary by a factor of four. 

The difficulty is that the galaxy is experiencing some kind of interaction : not only is there a single prominent spiral arm but the \HI{} and optical centres are offset (by approximately 1$'$ or 5 kpc), and the \HI{} extends well beyond the optical disc but only towards the north (see figure \ref{fig:4254Map}). The nature of the interaction is unclear (e.g. ram pressure stripping according to \citealt{kanth} and \citealt{sofue}; various sorts of tidal encounter according to \citealt{vol09}, \citealt{d08} and DB08), making it impossible to estimate how it may have affected the \HI{} inclination angle and line width. For our standard models we measure a W20 of 270 \kms{} from the Westerbork data (using the \textit{mbspect} task from the \textsc{miriad} data analysis package. This would give a rotation velocity range from 210 \kms{} at 40$^{\circ}$ inclination to 395 \kms{} at 20$^{\circ}$.

\paragraph*{\textnormal{\textit{Our chosen parameters for NGC 4254}}}\mbox{}\\
The major uncertainty in the observations is the inclination angle (and hence circular velocity and dynamical mass), and (arguably) the initial gas distribution. There is not much margin for error in the gas or stellar mass. Therefore we use galaxies corresponding to three different inclination angles for NGC 4254 : 20, 30 and 40$^{\circ}$ for our M1, M2 and M3 models respectively.

Our M2 model represents a typical spiral galaxy, with an observationally-derived gas distribution (albeit being rather gas rich) and typical circular velocity - we take this as our standard model. M3 is similar to M2 except for the higher circular velocity and correspondingly greater dynamical mass. 

Our M1 model has the lowest inclination angle, giving a circular velocity similar to that used in DB08. The DB08 approach is different to ours, adopting parameters for NGC 4254 they believe were consistent with the galaxy prior to its interaction. They used a slightly higher gas mass of 8.5$\times$10$^{9}$\Msolar{} with a more extended density profile - flat to a radius of 25 kpc then linearly decreasing to zero at 30 kpc. We argue that this combination of a more extended gas disc with a low dynamical mass may make the galaxy more susceptible to gas stripping and thus more readily allows the formation of a relatively massive pure gas feature such as VIRGOHI21. Therefore we investigate this possibility by using the DB08 gas mass and profile for the M1 model.

The full parameters are given in table \ref{tab:stab}. In this way we compare the effects of harassment on normal (M2), gas-rich (M1), and very massive (M3) spiral galaxies. We note, however, that the M1 model is far from a typical spiral galaxy : its gas content would be exceptionally high for a galaxy of this size, and the flat distribution of its gas is atypical (as shown in \citealt{bandb}, which show that gas profiles are normally exponential rather than flat). Although the \HI{} in the real NGC 4254 is, as discussed, slightly offset from the stellar disc and more extended towards the north, overall its surface density profile (see figure \ref{fig:gasprofileM2}) is not at all unusual - flat within the stellar disc then decreasing exponentially. 

%In our view this argues against the gas in the stream originating from NGC 4254, especially considering that NGC 4254 is quite gas rich and would have been exceptionally so had the gas in the stream originated within its disc\footnote{One alternative origin of the stream is discussed in \cite{d08}. In this model VIRGOHI21 is a massive, gas-rich but optically dark galaxy. Its interaction with NGC 4254 creates the single prominent spiral arm and the gas in the stream originates from VIRGOHI21 itself rather than the spiral galaxy. Numerical modelling showed this could reproduce the \HI{} stream joining NGC 4254 and VIRGOHI21 but not the \HI{} seen to the north of VIRGOHI21.}

\begin{table}
\centering
\caption[models]{Parameters of our model galaxies, as follows : (1) Name of model; (2) \HI{} mass in solar masses; (3) Shape of the \HI{} profile; (4) Stellar mass in solar masses; (4) Deprojected circular velocity in \kms{}, assuming a W20 of 270 \kms{}; (3) Dynamical mass in solar masses at 25 kpc radius. For all galaxies we use a total stellar mass of 3.5$\times$10$^{10}$\Msolar{} with a radial profile matching the observations of NGC 4254.}
\label{tab:stab}
\begin{tabular}{c c c c c}
\hline
  \multicolumn{1}{c}{Model} &
  \multicolumn{1}{c}{\HI{} Mass} &
  \multicolumn{1}{c}{\HI{} Profile} &
  \multicolumn{1}{c}{V$_{circ}$} &
  \multicolumn{1}{c}{M$_{dyn}$} \\
\hline
  M1 & 8.5E9 & Flat+linear & 210 & 2.6E11\\
  M2 & 5.5E9 & Observed & 270 & 4.2E11\\
  M3 & 5.5E9 & Observed & 395 & 9.1E11\\
\hline
\end{tabular}
\end{table}

\subsection{Simulation analysis}
\label{sec:simanalysis}
We use the same techniques as those fully described in T16 section 4.2. In brief, while we inspect the raw particle data, our main technique is to grid the data to create virtual position-velocity data cubes. Since these can be of arbitrary resolution and sensitivity we can not only directly compare the simulations with existing data, but also (in principle) predict what future, higher resolution observations should detect. This is a straightforward procedure since we obviously know the gas particle mass, and the \HI{} line intensity depends only on the \HI{} mass. 

The only limitation of this method is the particle number. For an AGES-class survey, with a spatial resolution of 3.5$'$, a velocity resolution of 10 \kms{} and a sensitivity of 0.6 mJy, a S/N of 4.0 (which we set as our threshold for detectability) in a single cell would require just 6 particles. Thus, unfortunately, the detectable\footnote{As discussed in T12 and T13, this is a subjective value owing to the typical search methods used - a single-channel 4$\sigma$ peak might be detected by an automatic program, but a human observer would probably reject this from its catalogue.} low-mass features will be strongly influenced by individual particles.

We visualise both the particle and gridded data using \textsc{frelled}, as described in \cite{me15}. We produce gridded P-V cubes of each simulation for each timestep, and although these are used in the automatic parameter measurements it is obviously impractical to visually inspect the resulting 5,400 data cubes. Instead, each cube is clipped to remove anything below a S/N of 4.0 in each velocity channel, and integrated along the velocity axis to create a detectability map for each timestep.

While the realtime view in \textsc{frelled} displays particles as simple points, for the figures they are shown as diffuse circular Gaussians. A minor change to the \textsc{frelled} code was to make the size of the Gaussians dependent on the SPH kernel size. This greatly enhances the visibility of the low-density material while preserving the smaller structures in dense regions, though it can have the adverse affect of making the discs appear more disturbed than they really are.

\section{Results}
\label{sec:results}
As the simulations use 90,000 particles in multiple components, they are considerably more computationally expensive than the simulations of T16. Therefore we only run the full set of 27 simulations using our `M2' model, which has the parameters of a typical spiral galaxy. We visually inspect the simulations and select 9 initial positions which we then use for our more extreme examples of the M1 and M2 models. These initial positions are selected on the basis of what happens to the galaxy during its infall : we choose 3 in which there is negligible disturbance to the galaxy, 3 in which the galaxy is slightly disturbed (i.e. a grand design structure is evident in the disc, and the gas becomes distributed into tails or other structures of comparable size to the initial disc or smaller), and 3 in which the galaxy shows a major disturbance (i.e. with tails longer than 100 kpc). The idea is to test whether the more extreme cases show behaviour which is significantly different from the standard galaxy.

\subsection{The M2 model}
\label{sec:M2results}
\subsubsection{Global properties}
\begin{figure*}
\centering
\includegraphics[width=175mm]{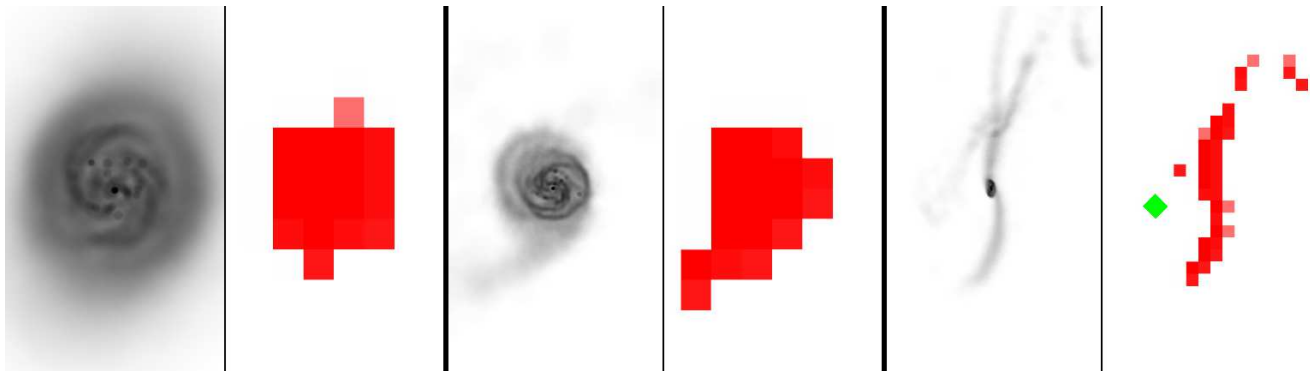}
\caption[vmap]{Snapshots of three representative cases of the M2 simulations, each after 4.2 Gyr in the cluster. Each pair of images shows the particle distribution (greyscale) on the left with the gridded data (each pixel is equivalent to 3.5' across) in red on the right, set to show anything with a S/N $>$ 4.0 to an AGES-class survey. The green diamond on the far right panel shows an \HI{} cloud detectable to AGES at least 100 kpc and 500 \kms{} in projected space/velocity from the nearest other \HI{} detection. The left pair of images shows an example of an undisturbed galaxy and the middle panel a moderately disturbed case, with a field of view 120$\times$200 kpc. The right panel shows a heavily disturbed galaxy, with a field of view 300$\times$500 kpc. The complete set of simulations, with movies for the complete 5 Gyr of evolution, are shown in figures \ref{fig:M2run1}-\ref{fig:M2run2}.}
\label{fig:M2run}
\end{figure*}

The results of our standard galaxy can be seen in figure \ref{fig:M2run}, which shows the particle distribution and the gridded data. A wide variety of effects can be easily seen from a visual inspection. Quite unlike the much less massive discs we examined in T16, some of these galaxies show very little damage from being in the cluster even for the full 5 Gyr. Others show only minor changes, for example temporary evidence of grand design structures or short tails which quickly dissipate or fall back into the disc. A few show rather stronger disturbances, with tails produced that are $>$ 100 kpc in extent which are harassed into an assortment of very different shapes. There are occasionally stellar streams devoid of gas and gas streams with low stellar content, as well as one-sided tails and isolated clouds. 

Galaxies can move between all three of these phases - long tails eventually disperse and become undetectable, while grand design structures fade in a few rotations of the galaxy after the perturbation (for a detailed discussion on the formation of spiral arms in galaxy clusters see \citealt{schmuk}). As a crude estimate, we made a visual inspection of all the simulations to determine how much time each galaxy spends in each phase. On average, galaxies spend about 60\% of their time in the quiescent state, 25\% in a mildly disturbed phase, and 15\% being strongly disturbed. Such extended \HI{} features are certainly not seen in anything like 15\% of galaxies in the real Virgo cluster. As noted in T16, there are around 350 late-type galaxies in Virgo but only four long \HI{} streams. This supports our earlier assertion that the process of gas removal must be more complicated than simply displacing the gas from its parent galaxy.

\begin{figure*}
\centering 
  \subfloat[]{\includegraphics[height=40mm]{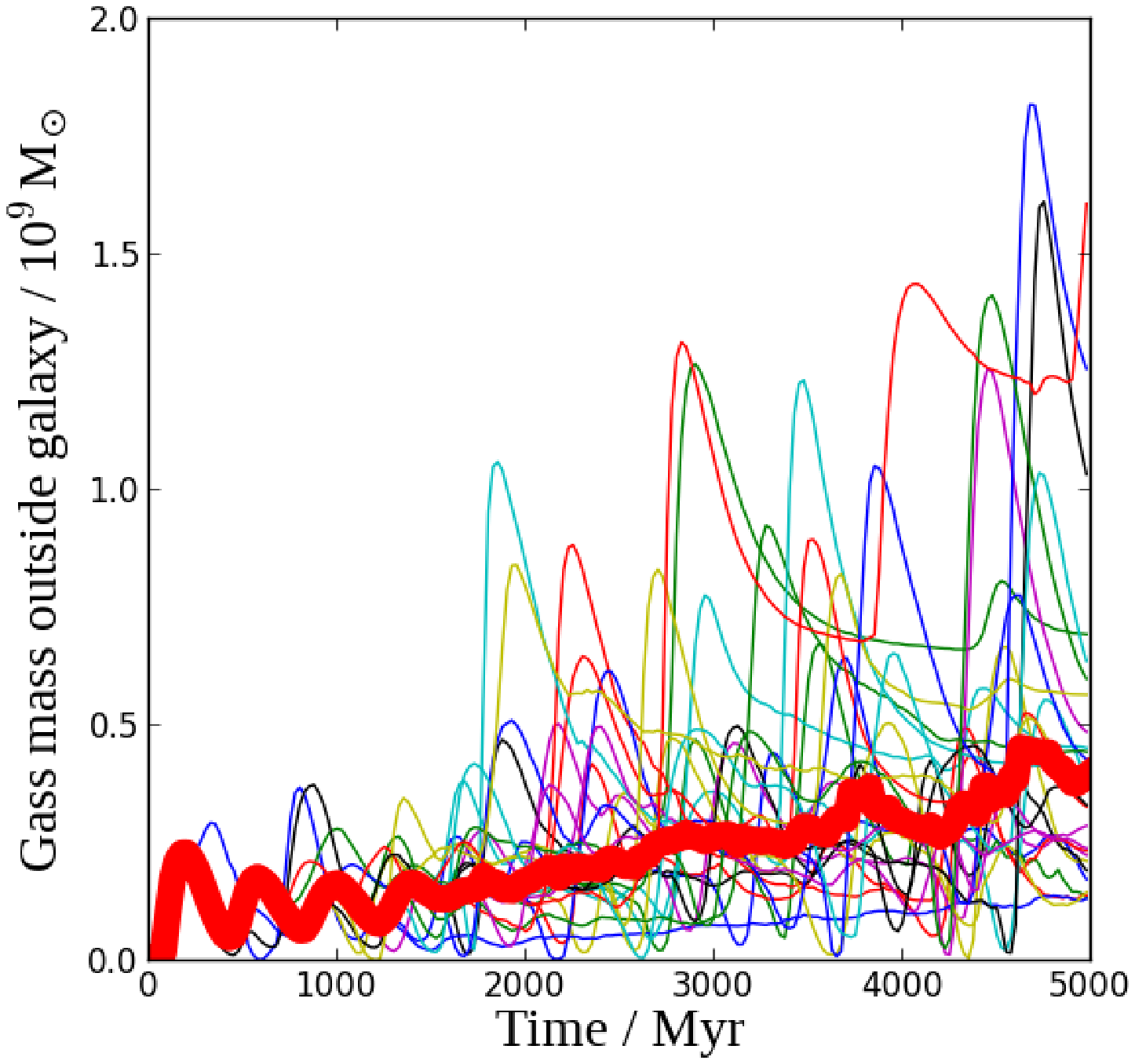}}
  \subfloat[]{\includegraphics[height=40mm]{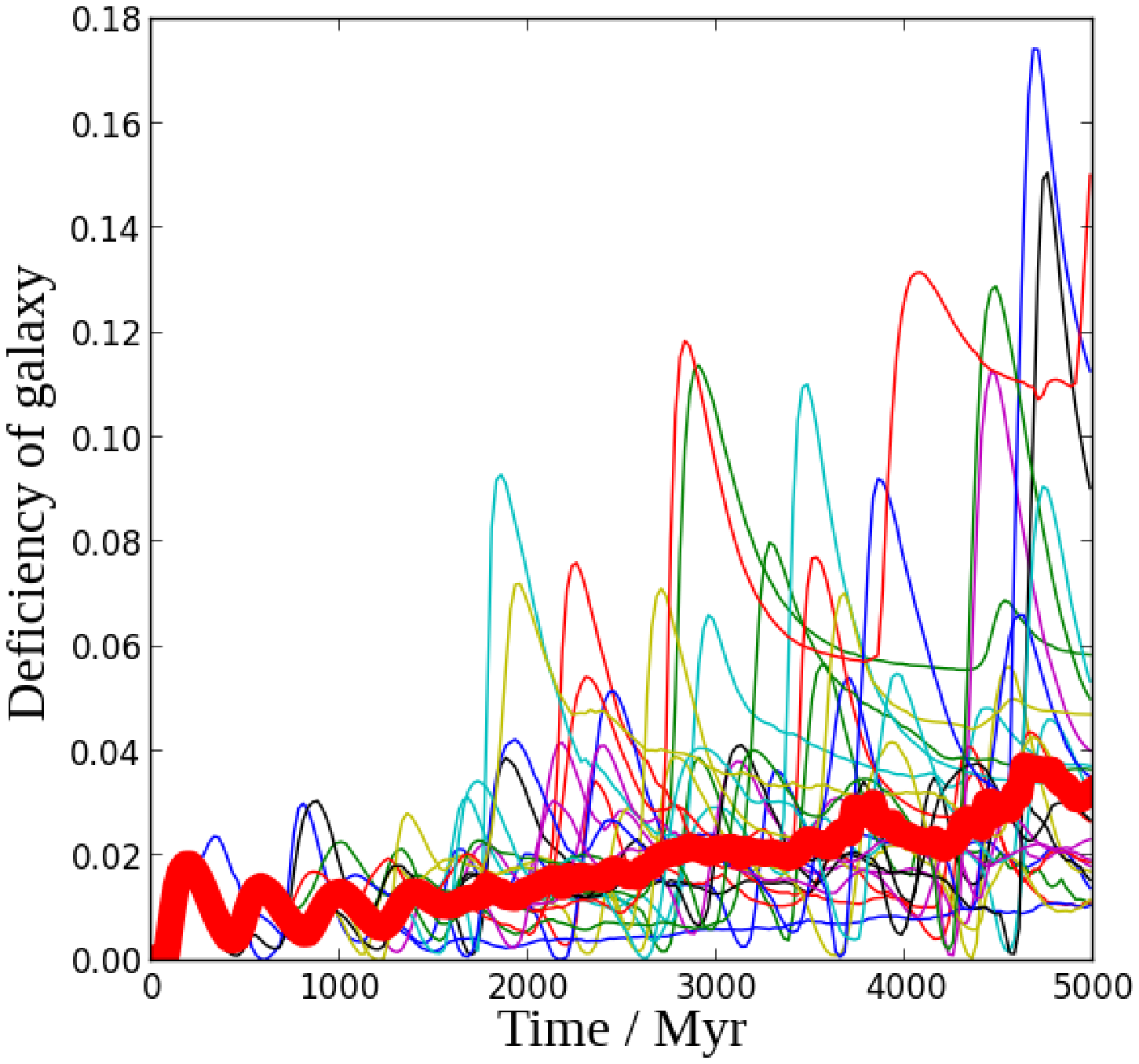}} 
  \subfloat[]{\includegraphics[height=40mm]{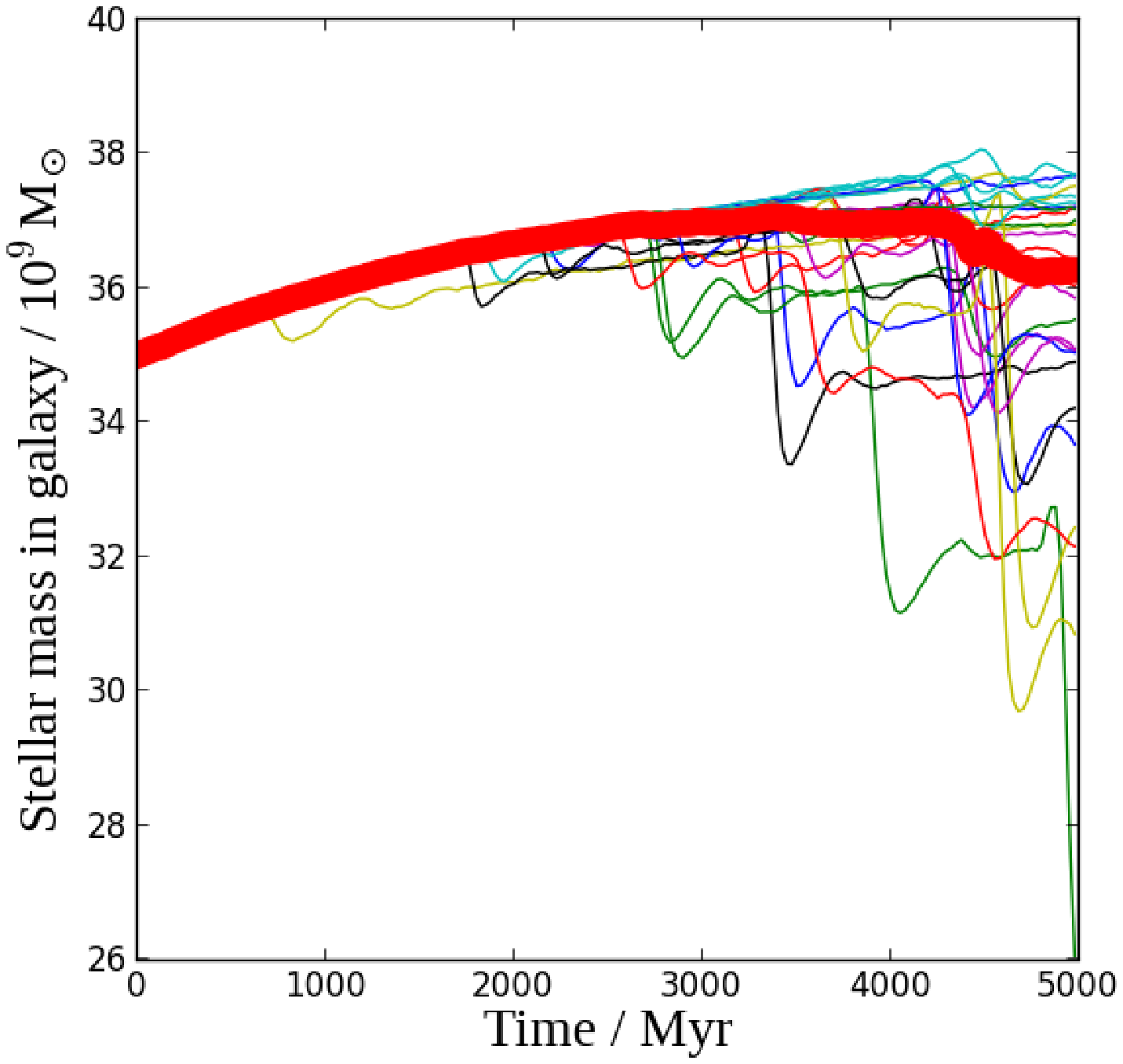}}
  \subfloat[]{\includegraphics[height=40mm]{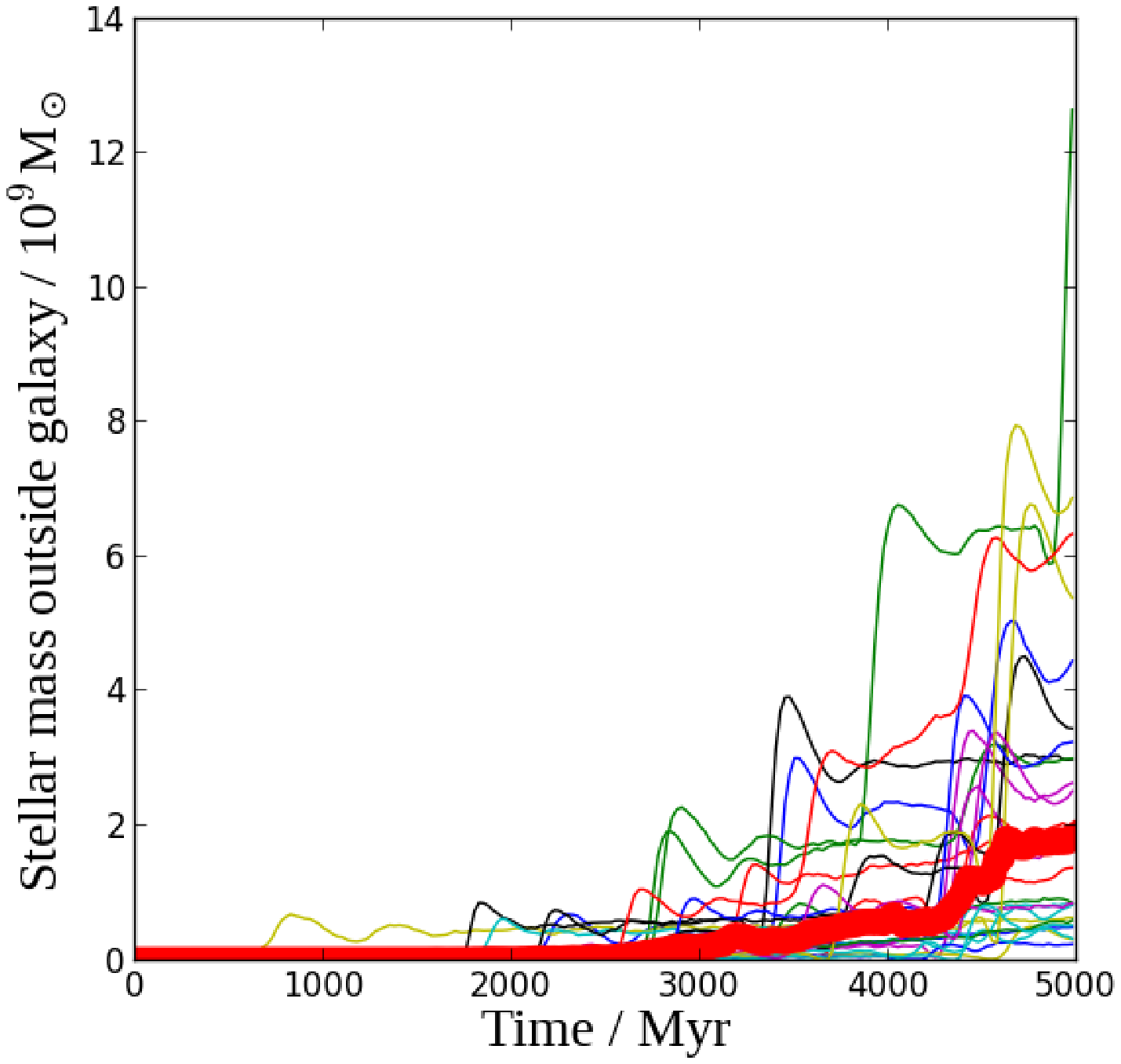}}  
\caption[Streams]{Evolution of the major properties of the stars and gas in the M2 model. In all cases we the define, `the galaxy' to be a sphere of radius 25 kpc centred on the median particle position. Panel (a) shows the gas mass outside the galaxy; (b) shows the measured \HI{} deficiency of the galaxy assuming its initial deficiency is zero; (c) shows the total stellar mass within the galaxy; (d) shows the total stellar mass outside the galaxy. The thick red line indicates the median value while the thin coloured lines show individual simulations.}
\label{fig:v290props}
\end{figure*}

The evolution of the overall properties of the simulated galaxy can be seen in figure \ref{fig:v290props}. The stripped gas mass typically reaches 4.0$\times$10$^{8}$\Msolar{} (increasingly linearly with time, on average, but with a very strong scatter). This is somewhat higher than in the DB08 simulation (2.0$\times$10$^{8}$\Msolar{}) and only slightly less than in the real tail containing VIRGOHI21 (5.0$\times$10$^{8}$\Msolar{} according to \citealt{haynesV21}), the other known long streams in Virgo (see T16 table 1) or the low-mass streams used in our previous simulations in T16. However it typically takes several gigayears of being in the cluster to strip this much material, so it seems very unlikely that a single encounter could be responsible.

All galaxies in these simulations experience some gas loss, but typically they retain 90-95\% of their original gas content. The observed deficiency of these galaxies never exceeds +0.2, even with the (very generous) assumption that their original gas content corresponds to a deficiency of zero. In fact, as discussed, these initial conditions represent a particularly gas-rich galaxy with a deficiency of -0.5. The effects of harassment are nowhere near strong enough to explain the strong deficiencies of many galaxies in the real Virgo cluster ($>$+0.5 in T12), which likely requires ram-pressure stripping (e.g. \citealt{vol01}, \citealt{roe}). Rather, these results are consistent with the observation that most long streams in Virgo are not associated with highly deficient galaxies (e.g. \citealt{koop}, \citealt{m07}), and those which are have masses insufficient to explain the presumed parent galaxy's strong deficiency (\citealt{oo05}).

Perhaps surprisingly, there is not so much difference in the typical fraction of gas and stars which are stripped. By the end of the simulation the median fractions remaining in the discs are 92\% for the gas but 95\% for the stars. However, there is considerably more variation in the gas, as is evident by comparing figures \ref{fig:v290props}(b) and (c) - more extreme stripping events occur more frequently for the gas than for the stars. The star formation rate of the galaxy also does not vary dramatically, with the median total number of star particles being around 21,500 after 5 Gyr while in the most extreme case this only rises to 22,500 (star formation tends to dominate over stellar removal, at least for the first 2-3 Gyr, so the stellar mass in the disc increases). These numbers correspond to average star formation rates of 0.5 - 0.9\Msolar{}\,yr$^{-1}$. While there are a few instances where the star formation rate briefly exceeds this, overall the amount of gas affected by harassment is simply not enough to cause any major alterations to the star formation activity of the galaxy. Harassment neither induces a significant density increase nor a decrease, thus the star formation rate (and the amount of gas converted into stars) is largely unaltered. 

The radial profile evolution of the gas and stars can be seen in figures \ref{fig:gasprofileM2} and \ref{fig:starprofileM2} respectively. Interestingly, while the gas profile remains broadly simlar to the initial conditions even after 5 Gyr in the cluster, there is significant evolution in the stellar profile. While initially there is a clear change in the gradient at $r\!\approx$ 9 kpc, by 5 Gyr the profile is well-described by an exponential with a scale length of 4 kpc even to $r\!>$ 40 kpc. This change in the stellar profile is slow and steady over the 5 Gyr. Despite this, according to this model the real NGC 4254 could have been in the cluster for 1-2 Gyr with no discernible effects on its stellar or gaseous discs.

As noted above the morphology of the gas which is stripped is another story, with a wide variety of structures produced which are reminiscent of the results of T16. As in our previous study, isolated clouds are not uncommon - however, those which match the specific criteria of the AGES clouds ($<$ 17 kpc diameter, $>$ 100 kpc from the nearest \HI{} detection, W50 $>$ 100 \kms{}) are again extremely rare. The stripped gas mass is comparable to the low-mass streams in T16, and consequently the evolution of the streams here is very similar. 

Those isolated clouds which do form have velocity widths due to streaming motions along the line of sight (just as in the B05 and DB08 models, and the low-mass streams in T16), rather than the self-gravitating features seen in the more massive streams of T16. They are generally not truly isolated - they are slight overdensities in streams which are just below the AGES sensitivity limit. They are basically similar in nature to the clouds in the low-mass streams in the T16 simulations. This is not surprising since the same argument we have mentioned against the observed clouds being self-bound also applies to simulated clouds : at the high velocity widths we are interested in, their size would have to be so small their column densities would be so high that they would rapidly form stars.

\begin{figure*}
\centering 
  \subfloat[]{\includegraphics[height=55mm]{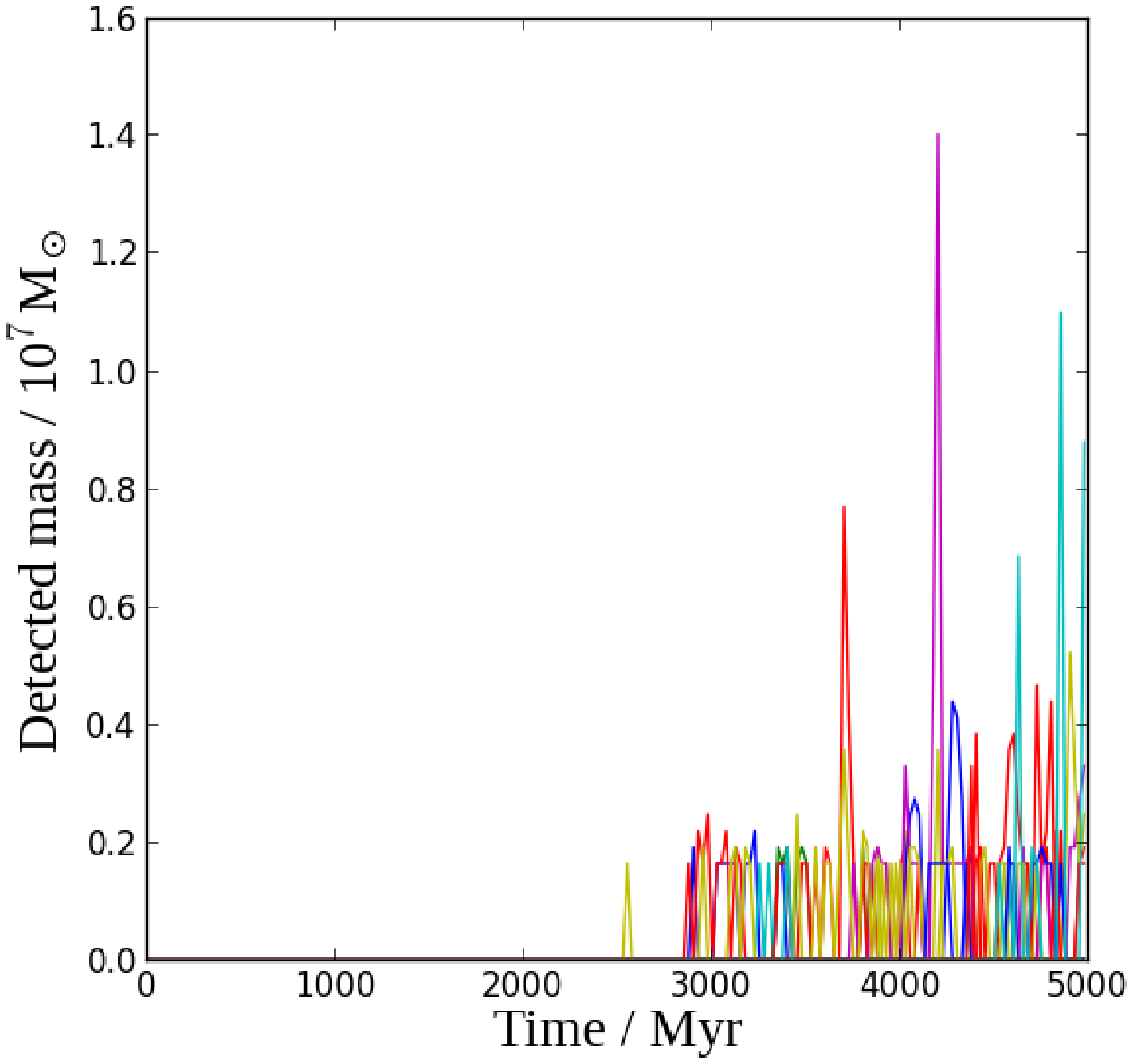}}
  \subfloat[]{\includegraphics[height=55mm]{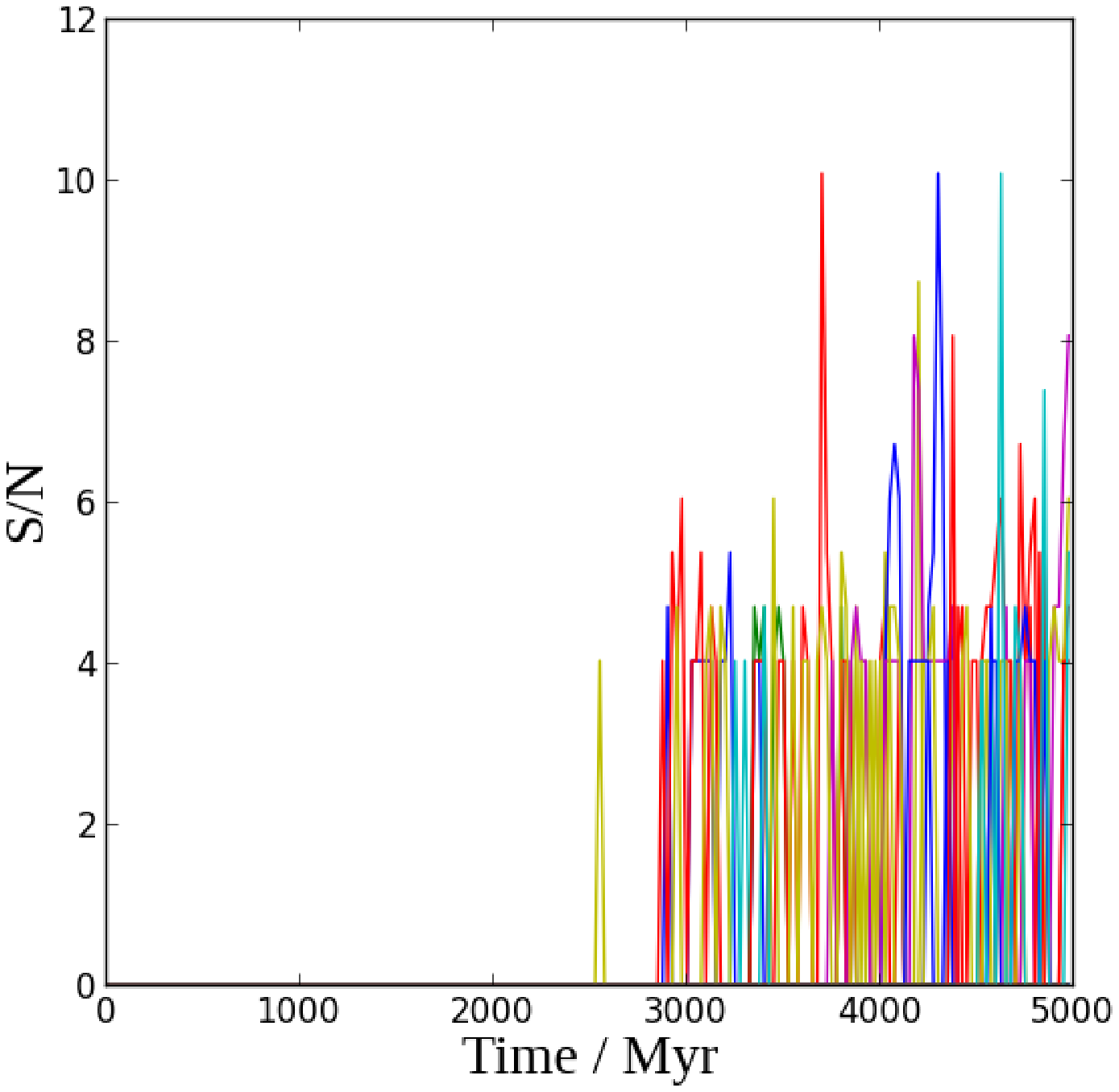}} 
  \subfloat[]{\includegraphics[height=55mm]{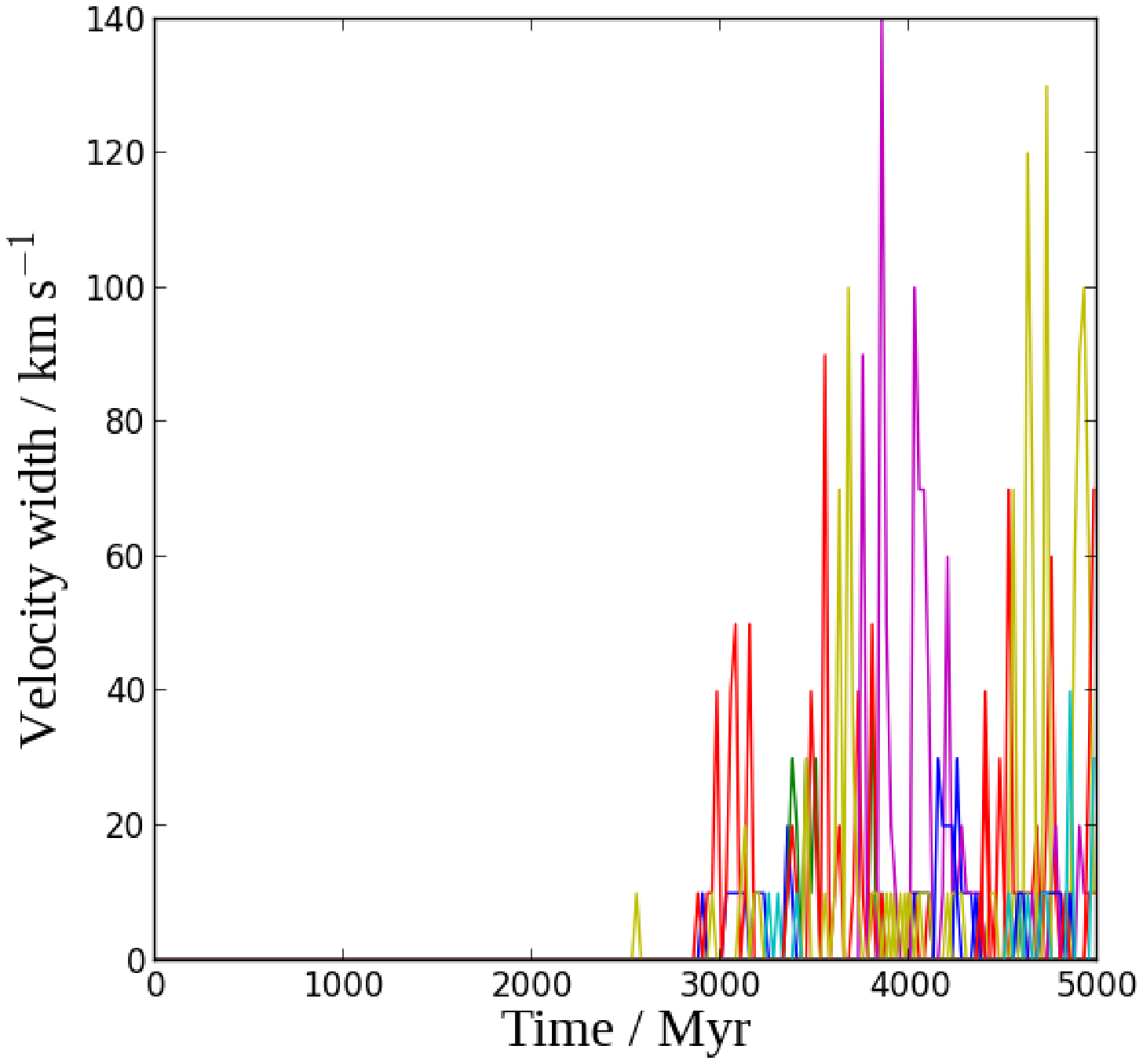}}
\caption[Streams]{Evolution of the unresolved, isolated \HI{} clouds as seen with an AGES-class survey for the M2 model. As in T16 we plot only the properties of the cloud with the highest velocity width for each simulation (each plotted as a different coloured line), since the high velocity widths appear to be the limiting factor in reproducing clouds similar to those described in T12 and T13. Panel (a) shows the detected mass in the cloud; (b) shows its signal to noise ratio; (c) shows its velocity width.}
\label{fig:v290clouds}
\end{figure*}

As shown in figure \ref{fig:v290clouds}, only for six timesteps for the whole run of 27 simulations did any clouds ever equal or exceed a velocity width of 100 \kms{} (0.1\% of the total simulated time). Each simulation timestep is 2.5 Myr though for reasons of disk space we only output every tenth interval, so each output timestep (which we measure) contains 25 Myr of evolution. Thus the six timesteps for which the clouds are produced is equivalent to a \textit{maximum} of 150 Myr out of the total 135 Gyr (27 simulations each of 5 Gyr). Or, to put it another way, 25 of our 27 simulations never produced such clouds at all. Of the two that did, one had high width clouds for just 1\% of the time while the other did so for 2\% of its duration. There was never more than a single detectable high width cloud in any simulation at any time, and then only very rarely.

Once again altering the initial conditions has made little difference to main result : it is almost impossible to produce clouds like this purely by tidal encounters. Technically these new simulations have made the conclusion from T16 stronger, decreasing the fraction of timesteps featuring AGES-like dark clouds from 0.2\% (in T16) to 0.1\%. The reasons are that in T16 every simulation contained an initial stream and only a stream, whereas here they begin with a galaxy - not every simulation forms a stream, and when they do form, the galaxy is always detectable so it is more difficult to form an isolated cloud. In any case the conclusion from T16 was already decisive and these results do not alter that.

It should be emphasised that as in T16, it is only clouds with properties similar to the AGES clouds which are rare. Clouds with velocity widths $<$50 \kms{} are much more common, appearing for 9\% of the total simulated time. In individual simulations, isolated clouds (detectable to AGES) with velocity widths $<$50 \kms{} were present for as much as 36\% of the time. Clouds with low velocity gradients (i.e. total velocity width per unit physical size) can be easily explained by the tidal debris hypothesis - it's the clouds with high velocity gradients that this model has difficulty with.   

Unlike in T16, we are now able to examine the condition of the parent galaxy. We inspected the simulations for those six timesteps when a cloud has W50 $\geq$ 50 \kms{}. In three cases there were also extended \HI{} features visible in the synthetic observations (spanning three or more Arecibo beams at S/N $>$ 4.0) that clearly indicated the disturbed nature of the galaxy, while in the other three cases the \HI{} of the galaxy was not visibly perturbed. The situation is similar for the low velocity width clouds (which are much more frequent so we only examined a small random subset), with the galaxy being clearly disturbed in some cases but not in others. We also found that the clouds of any velocity width would remain optically dark, with only one cloud having two star particles and all the rest being pure gas. 

\subsubsection{Kinky curves as fake dark galaxies}
We now turn our attention to the `kinkiness' of the streams. As in T16, high velocity width features are common in the detectable streams but rare in isolation. One might think that perhaps this means there are potentially far more fake dark galaxies lurking in the streams, but this is not so. In isolation, a single high velocity width feature could be mistaken for rotation. This is not necessarily true within a stream : if the adjacent pixels are of similar velocity width, no part of the stream will appear as kinematically distinct, nor would there be any reason to assume the stream could potentially fragment into separate high-width features. Rather, it is the change in velocity gradient of the stream that determines whether an observer might mistake a feature for a dark galaxy.

\begin{figure}
\centering 
  \subfloat[]{\includegraphics[height=75mm]{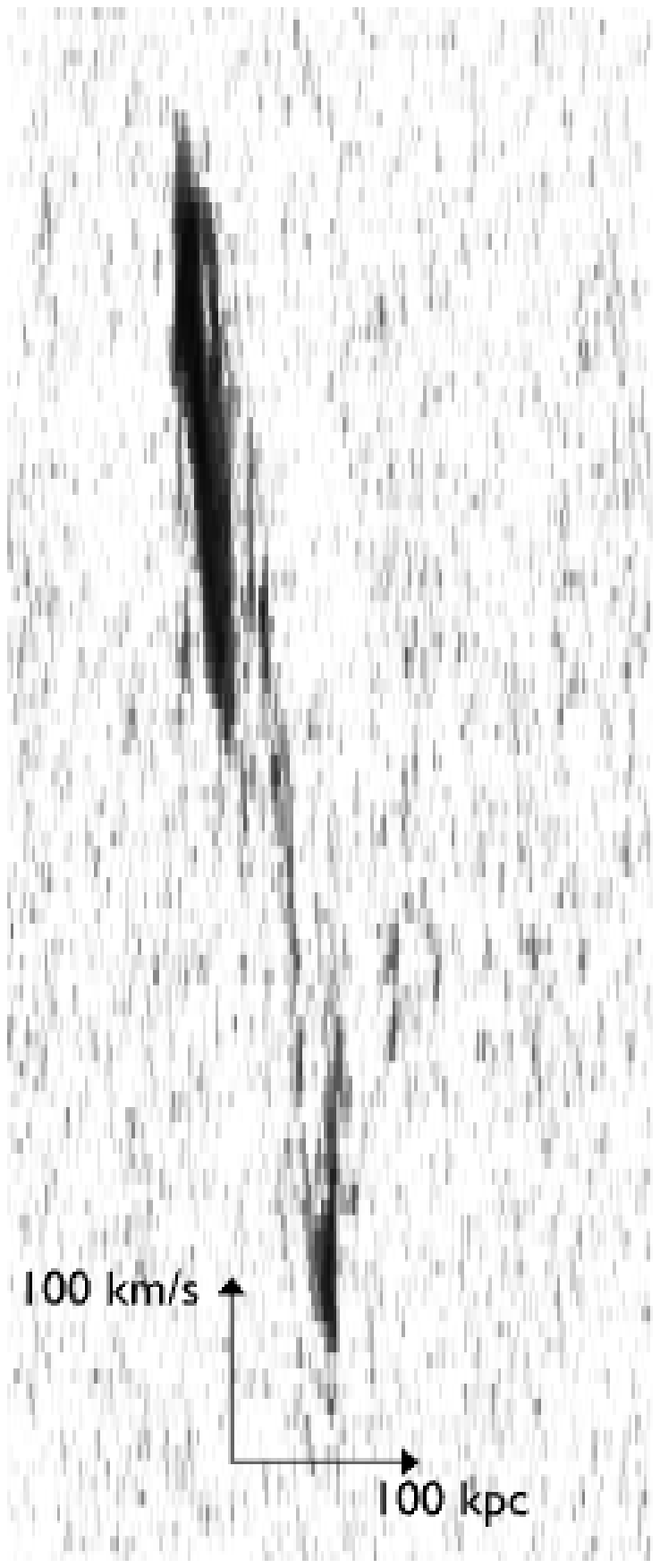}}
  \subfloat[]{\includegraphics[height=75mm]{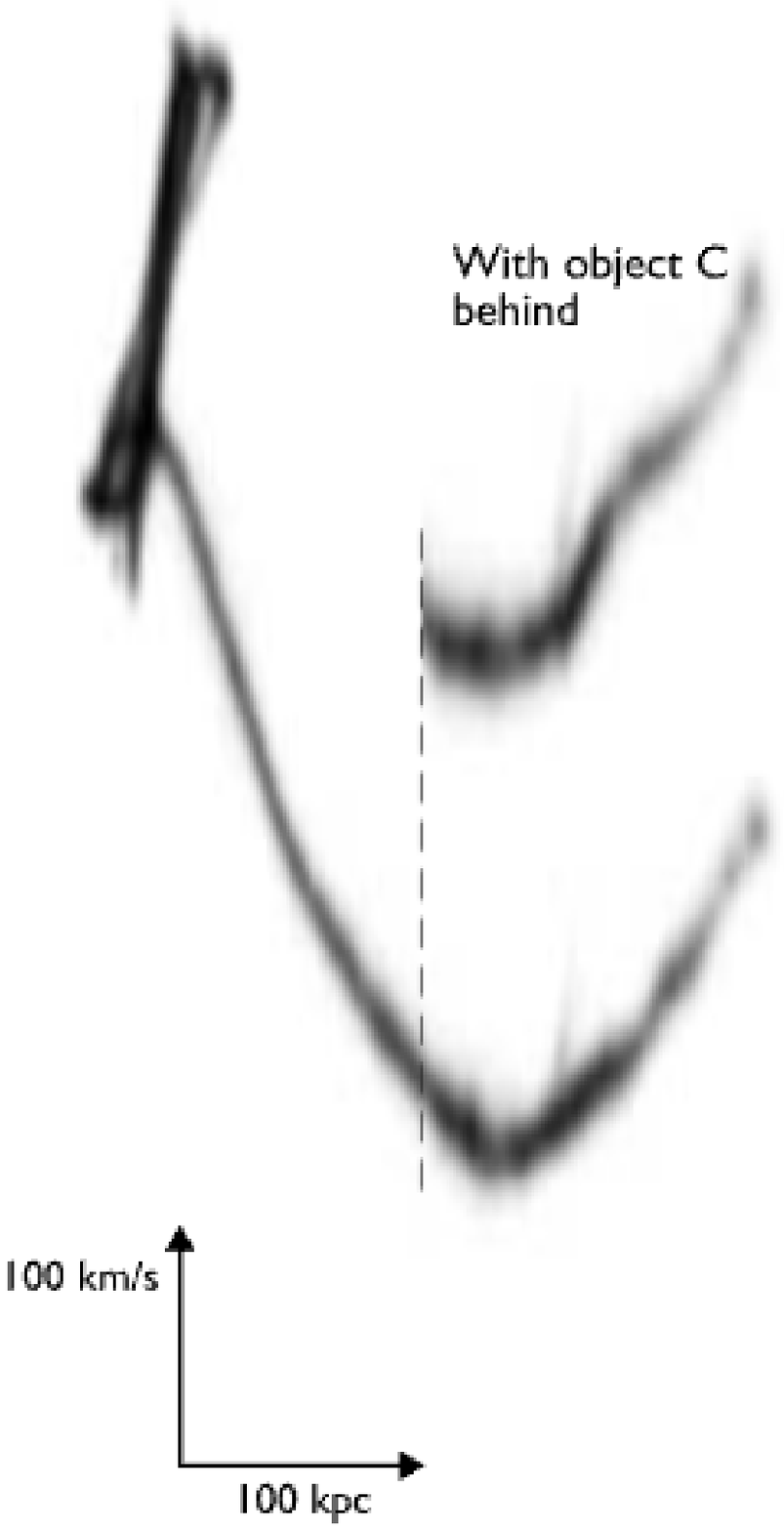}} 
\caption[Streams]{Comparison of the P-V diagrams of the real VIRGOHI21 system (panel (a), taken from the publicly available Westerbork data cube described in \citealt{m07}) and the simulation of DB08 (panel (b), stretching their figure 6 to have the same aspect ratio as panel (a)). The DB08 model does not reproduce the very sharp change in velocity seen in the real system or our simulations (figure \ref{fig:vhi21analogues}), although their use of a third object did suggest this was possible.}
\label{fig:ducvsrealvhi21}
\end{figure}

We argued in T16 that the results of DB08 do not actually reproduce the steep velocity gradient (or sharp change in gradient) of VIRGOHI21, which can be seen in figure \ref{fig:ducvsrealvhi21}. It therefore seems worthwhile to examine the results of our own simulations to see if such steep gradients are produced. Parametrising the sharpness of the change in gradient is non-trivial, fortunately it is also unnecessary for this investigation. A cursory glance in position-velocity space at one simulation showing a long tail revealed that sharp `kinks' in the velocity profile were common. We therefore visually inspected all the simulations in P-V space alongside the P-V diagram of VIRGOHI21 (set to the same scale as the simulations) for reference. Here we examined the particle data, not the synthetic observations, as the goal was only to establish if features of this type were produced at all rather than assess whether they are detectable. We searched only for sharp velocity gradient changes, not precise analogues to the whole NGC 4254/VIRGOHI21 system - as we shall show, it is not necessary to produce the exact P-V shape of the VIRGOHI21 stream to create a fake dark galaxy.

\begin{figure}
\centering 
  \subfloat[]{\includegraphics[height=55mm]{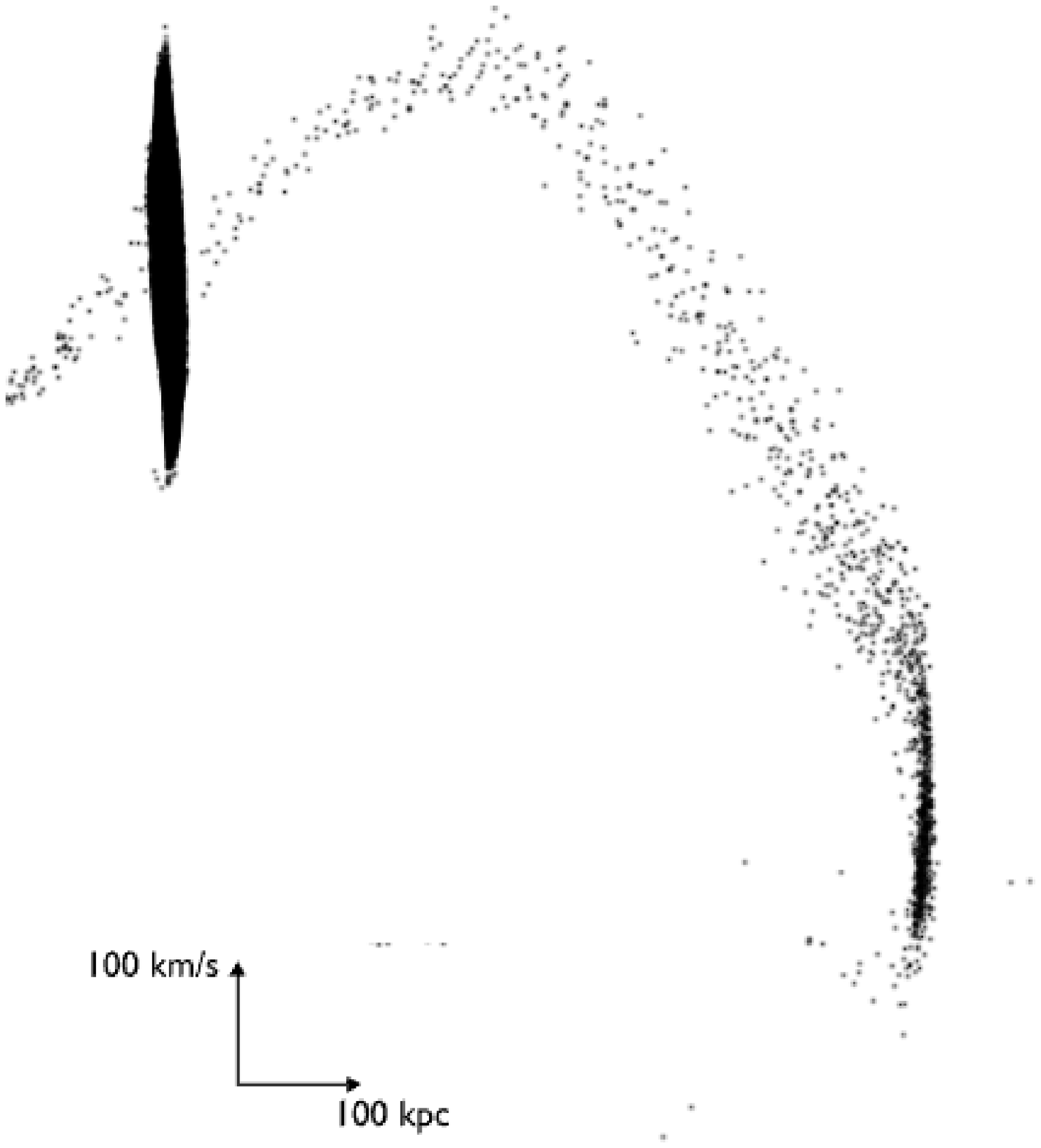}}
  \subfloat[]{\includegraphics[height=55mm]{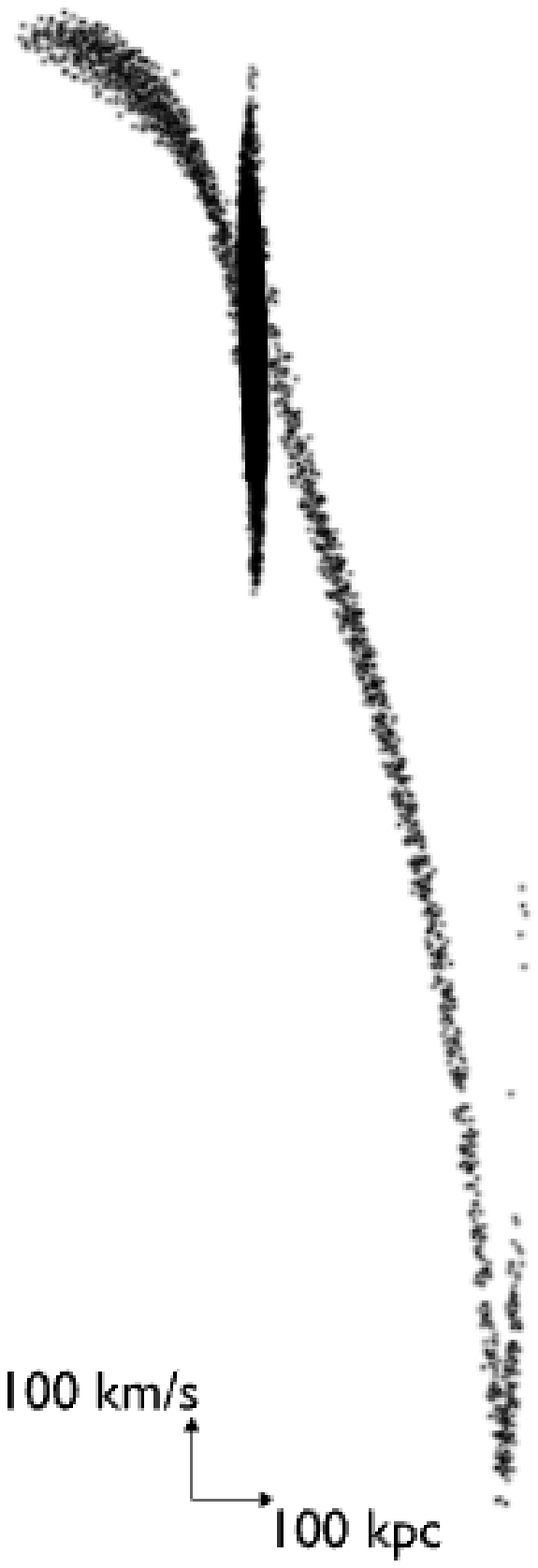}} 
\caption[Streams]{Examples of VIRGOHI21 analogues seen in our simulations of harassed discs, with similar velocity widths and sharp changes of velocity gradients. Panel (a) shows the simulation which began at the initial position -500.0, 0.0, -500.0 kpc from the cluster centre. The tail on the left is truncated by the field of view, it has a similar length to the visible tail but does not show any sharp velocity changes. (b) Shows a system of the initial position 0.0, -500.0, -500.0 kpc from the cluster centre. Unlike panel (a) the field of view is not truncated - the system has a long one-sided tail similar to VIRGOHI21 as well as the sharp velocity kink.}
\label{fig:vhi21analogues}
\end{figure}

We found that such sharp velocity kinks are common (verging on ubiquitous) wherever long streams are produced - we show a couple of examples in figure \ref{fig:vhi21analogues}. Thus the notion that the steep velocity gradient of VIRGOHI21 \textit{could} be a tidal structure is vindicated, though we strongly caution that this does not necessarily mean this is the true explanation (see also section \ref{sec:conc}). These VIROGHI21-mimics are transient but can last for appreciable timescales, $\sim$250 Myr in the case of the example of figure \ref{fig:vhi21analogues}(a). It is also worth noting that we saw features with even higher velocity widths ($>$ 1,000 \kms{}) and sudden gradient changes, but an observer would probably not regard these as being dark galaxy candidates. They appear in systems where there are many other \HI{} streams present and are not isolated features, and their sheer width would warn an observer (if they were even detectable) that they do not resemble galaxies.

The origin of these sharp velocity kinks varies and it not always easily attributable to a single interaction. For instance in figure \ref{fig:vhi21analogues}(b), the kink occurs as the stream closely approaches ($<$ 50 kpc) the cluster centre and is strongly accelerated. The high velocity width of the kink quickly expands, reaching 1,000 \kms{} in approximately 125 Myr. In contrast, the feature in figure \ref{fig:vhi21analogues}(a) has a quite different origin. The overdensity arises as the stream climbs out of the cluster potential and begins to fall back in, but there is no single clear interactor that can be identified as the cause of the high velocity width. This feature is rather more persistent than that in panel (b), having a similar velocity width (and small physical size, never being much more than $\sim$20 kpc across in any dimension) for around 250 Myr.

Neither of the two features discussed is a perfect analogue of the real VIRGOHI21, though both reproduce some aspects of the system quite closely. Feature (a) is an overdensity at the end of a long stream with a sharp change in the velocity structure. It is also found at a considerable distance from the cluster centre (400 kpc), though not as high as the real VIRGOHI21 (1 Mpc). The major difference compared to the real VIRGOHI21 is that the velocity gradient of the stream does not change sign, however, it could still appear as a convincing dark galaxy candidate. Feature (b) is a change in the sign of the velocity gradient of the stream, as in the real VIRGOHI21, and the stream is also much more asymmetrical than feature (a). The main difference from the real system is that the density of the stream is roughly uniform, so the S/N of the kink would be lower than the rest of the stream because of its high velocity width - which is contrary to the observations. It is also very much closer to the cluster centre than the real VIRGOHI21. Additionally, both features are at the end of a stream rather than in the middle as in the real system. We leave modelling a more exact recreation of VIRGOHI21 to a future work.

\subsubsection{Tidal history of the galaxy}
\begin{figure*}
\centering 
  \subfloat[]{\includegraphics[height=55mm]{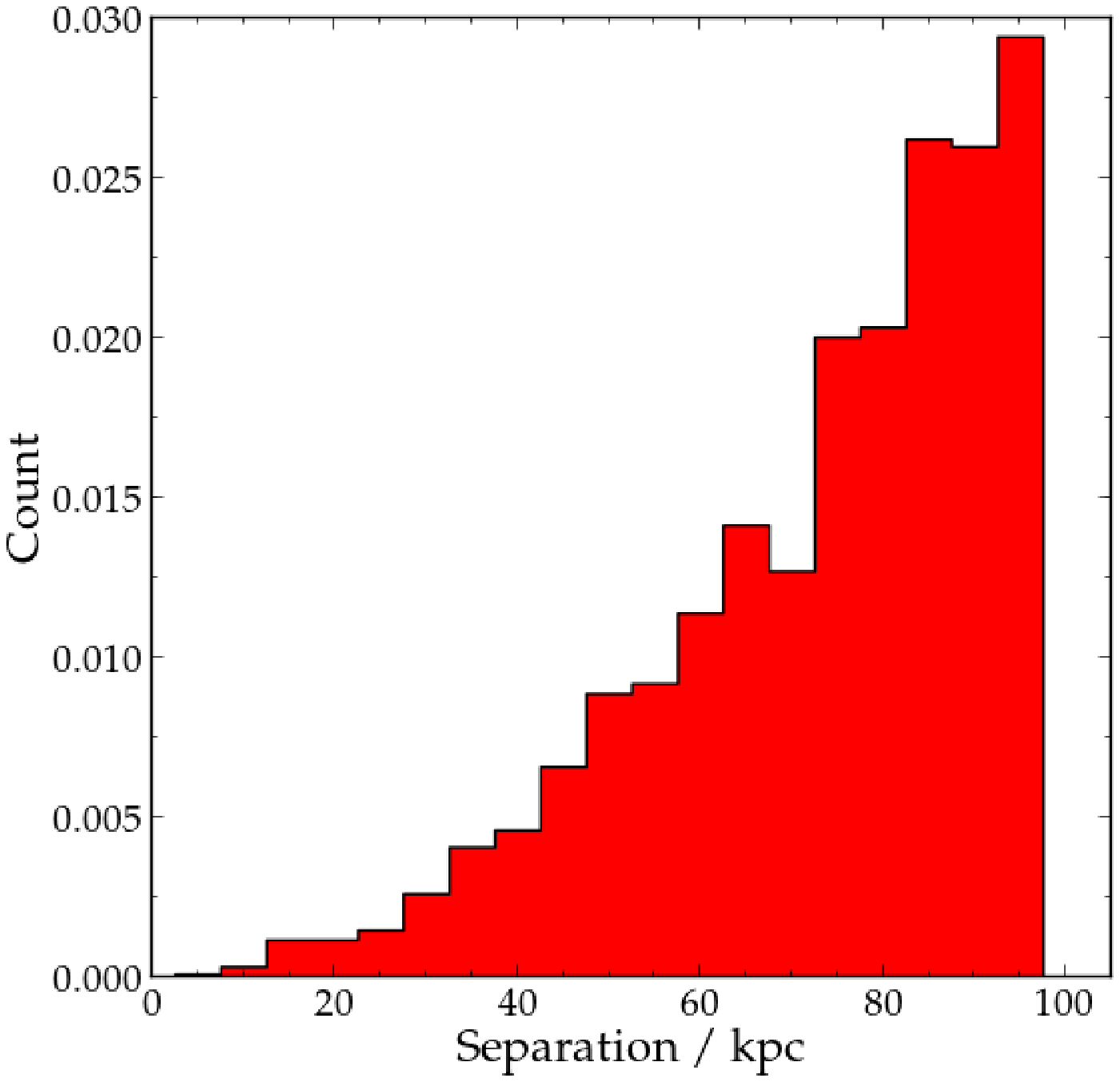}}
  \subfloat[]{\includegraphics[height=55mm]{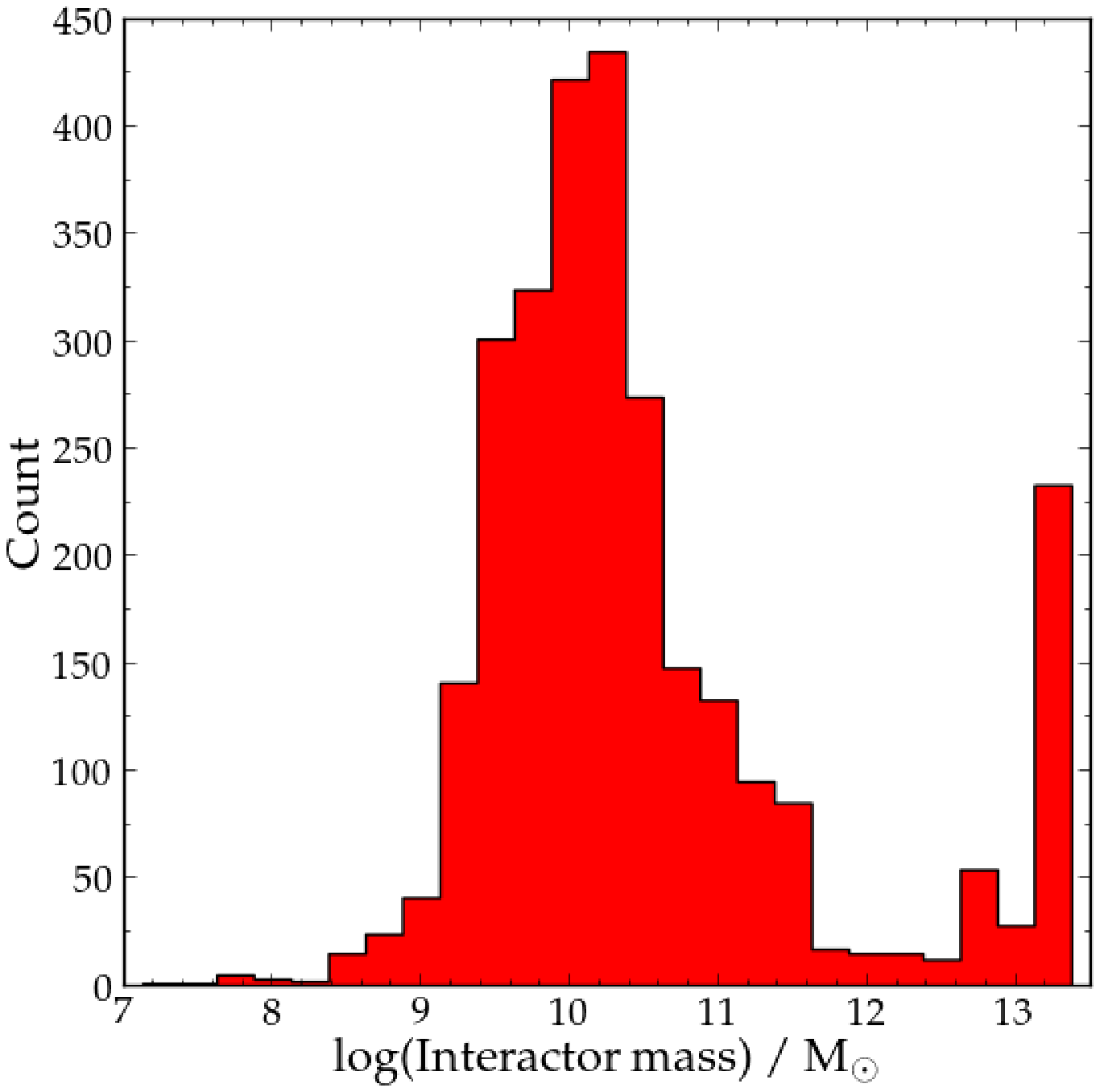}} 
  \subfloat[]{\includegraphics[height=55mm]{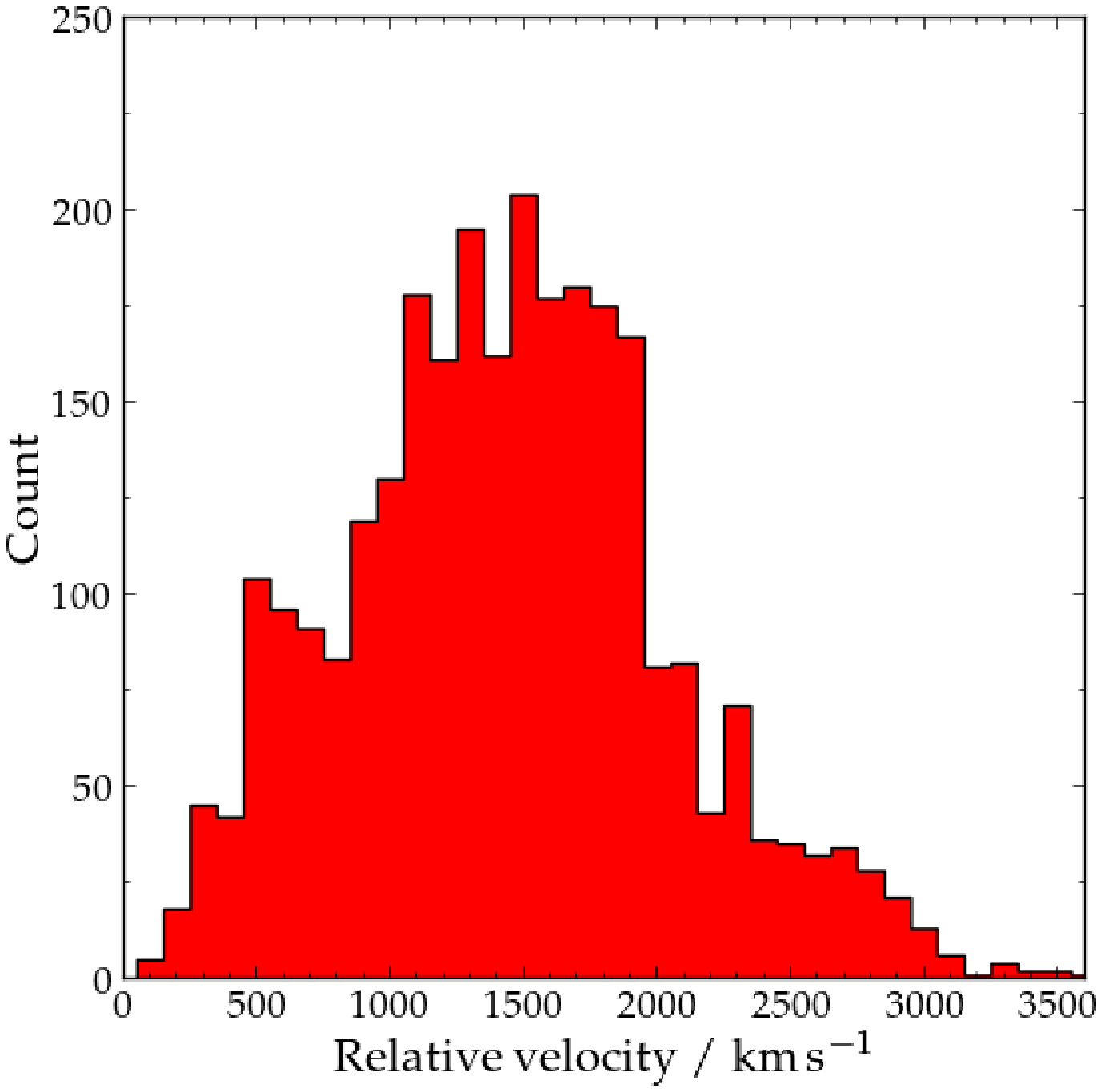}}  
\caption[Streams]{Distribution of the parameters of the subhalos (excluding the main cluster potential) that come within a 100 kpc radius of the particle galaxy in all 27 M2 simulations at any time. In these plots all timesteps are shown - that is, each timestep a subhalo spends within the 100 kpc radius is defined as an encounter. The peak for the mass distribution at the high end is due to the largest subhalo being located near the cluster centre, which the galaxies can encounter several times since they are on plunging orbits.}
\label{fig:interactionparameters}
\end{figure*}

One inherent difficulty, already briefly mentioned, of simulating the gravitational field of both the cluster and its substructure is that it is very difficult to identify which encounters are causing the gas stripping and influencing the resulting streams. Unlike in B05 and DB08 the particle galaxy is interacting with 400 other  subhalos simultaneously, and often (on visual inspection) there is no single clear interactor responsible. The situation is even more difficult and perhaps fundamentally impossible for the stripped gas : streams $>$100 kpc often have different subhalos simultaneously affecting different parts of them by different amounts - it is not always possible to say that one particular encounter is responsible, it's the cumulative effect of many interactions which causes the resulting structures. Therefore, rather than attempting to disentangle which particular subhalo is likely to be most responsible for the gas stripping and cloud formation, which is not the goal of the current work (but see \citealt{rory10} for a detailed discussion), we parameterise the interactions in a simple way by examining the objects that come within 100 kpc of the galaxy. We show the distribution of these in figure \ref{fig:interactionparameters}.

Although the timing of the encounters varies depending on the precise orbit of the galaxy, in broad terms the tidal history of the galaxies tend to be similar. Interaction velocities are typically high, around 1,500 \kms{} though with a strong scatter. Velocities $\sim$1,000 \kms{} (as in DB08) are common, with interactions over the range 800 \kms{} $<$ $v$ $<$ 1,200 \kms{} occuring on average (median) 8 times for each galaxy over the simulated 5 Gyr (here classing the whole duration for which a subhalo spends within the 100 kpc radius as a single encounter, rather than each timestep as in figure \ref{fig:interactionparameters}). Encounters with galaxies as massive as the interloper in DB08 (1.4$\times$10$^{12}$\Msolar{}) are rare, as expected given the mass distribution of the subhalos as shown in figure \ref{fig:halomass}, with the typical interloper mass being around 1.0$\times$10$^{10}$\Msolar{} - lower than the 1.4$\times$10$^{11}$\Msolar{} in B05. DB08-like encounters, with the same velocity range as above but an interactor mass 1.1$\times$10$^{12}$\Msolar{} $<$ M $<$ 1.0$\times$10$^{13}$\Msolar{} are very rare, occurring only twice in the full 27 simulations.

DB08-like encounters are rare primarily due to the high mass of the interloper rather than the relative velocity or close proximity. Encounters similar to those in B05 are also rare, but here the limiting factor is proximity. B05 do not state the interaction velocity, however, we find only three encounters (at any interaction velocity) within 50 kpc with interloper masses  1.0$\times$10$^{11}$\Msolar{} $<$ M $<$ 5.0$\times$10$^{11}$\Msolar{}. Both the interactions described in DB08 and B05 do occasionally occur, but are not common. A larger parameter study would be required to quantify their expected frequency more precisely.

\subsection{The M1 model}
\label{sec:M1results}
\subsubsection{Global properties}
The M1 model is overall less massive than the M2 model, but it has more gas which is also more extended. It is designed to give the best possible chance of gas stripping, given the parameters of NGC 4254 permitted by the observations. As shown in figure \ref{fig:M1run}, on occasion the results can be dramatic, with the galaxy almost torn apart. These instances are unusual, however, with the galaxy typically retaining over half of its gas within its initial radius after 5 Gyr, but clearly the M1 model galaxies are generally more disturbed than the M2 case.

The stripped fraction of both the gas and stars are larger than in the M2 case (see figure \ref{fig:v210props}), and the difference between the two is also stronger. After 5 Gyr, about 60\% of the gas but 80\% of the stars remain within the initial 30 kpc radius of the disc (compared to 92 and 95\% in the M2 case). The greater stellar removal than in the M2 case is not surprising, since the total (dynamical) mass of the galaxy is lower. The \HI{} deficiency of the galaxy is also significantly greater, but still only enough to make the galaxy appear borderline deficient - but again this is very generously assuming the galaxy's initial \HI{} is typical of a spiral of this size. In fact, as discussed, this would represent an extremely gas-rich case. However, interestingly, the median amount of gas remaining within the disc radius (about 5.0$\times$10$^{9}$\Msolar{}) is very close to the observed value in the real NGC 4254.

\subsubsection{Stripped gas and fake dark galaxies} 
The stripped mass in the M1 case is rather larger than for the M2 model, typically around 3.0$\times$10$^{9}$\Msolar{} after 5 Gyr - significantly higher than the 5.0$\times$10$^{8}$\Msolar{} in the real VIRGOHI21\footnote{Not all of the external mass in the simulation would be detectable, and in some cases the detectable mass in the stream can be comparable to the VIRGOHI21 stream. However the variation is very strong and with only 9 simulations it is difficult to describe a `typical' amount of detectable gas.}, but enough that the gas remaining in the disc is comparable to that in the real NGC 4254. In most cases the galaxies end up becoming far more disturbed than the real NGC 4254. While long tails do form, they are usually (though not always) accompanied by other, far more complex structures. This amount of gas stripping still takes several gigayears in the cluster and is rarely attributable to a single interaction, but if anything this model is \textit{too} susceptible to gas stripping to be a plausible progenitor for NGC 4254 - in one case the stripped mass exceeds 3.0$\times$10$^{9}$\Msolar{} within 1 Gyr. The DB08 model may not have revealed this since it only had a single interactor.

While the radial profile of the gas in the M2 model did not evolve significantly even over 5 Gyr, remaining close to the observed profile of NGC 4254, this is not the case for the M1 model (figure \ref{fig:gasprofileM1}). Although the profile evolves significantly over time, becoming approximately exponential over its entire length, it is still very different to the real profile of NGC 4254 even after 5 Gyr - it is much flatter in the outer parts and there is no clear break at the edge of the stellar disc. Although the amount of gas remaining in the disc is comparable to the amount in the real NGC 4254, its distribution is quite different. In contrast the stellar radial profile evolution is little different to that of the M1 case, as shown in figure \ref{fig:starprofileM1}.

\begin{figure}
\centering 
\includegraphics[width=85mm]{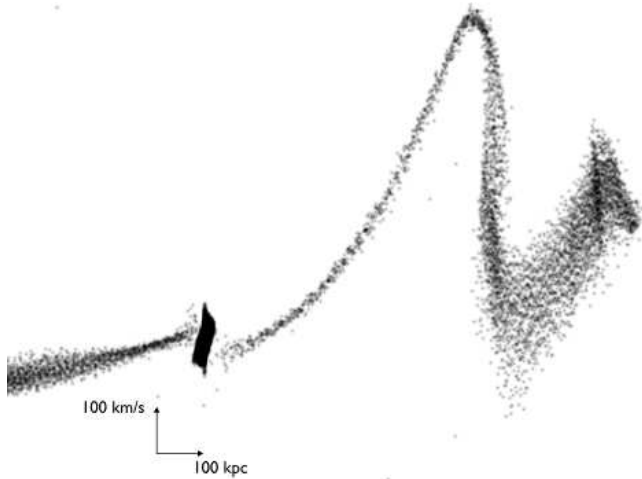}
\caption[Streams]{Complex kinematic features in the stream produced by the galaxy starting at -500, 0.0. 500.0 kpc from the cluster centre at 3.9 Gyr, in the M1 model. The stream on the left has been truncated by the field of view; it remains smooth along its entire $\sim$1 Mpc length.}
\label{fig:anothervhi21analogue}
\end{figure}

Since the gravitational field experienced by the M1 galaxy is the same as for the M2 case, the kinematics of the stripped gas are essentially similar in both cases, though there are major differences in the density and detectability of the stripped material. We show an example of a possible VIRGOHI21-like fake dark galaxy (i.e. a sharp velocity kink in a stream that could be interpreted as a rotating disc) in figure \ref{fig:anothervhi21analogue}. The same structure was also seen in the M2 model, but here there are approximately 4,000 particles in the stream (1.7$\times$10$^{9}$\Msolar{}) whereas previously there were only around 600 (1.6$\times$10$^{8}$\Msolar{}). 

The situation for producing optically dark clouds is more extreme than the M2 model, with AGES-like clouds being slightly rarer but low-width clouds being more common. Again, the velocity width tends to indicate streaming motions along the line of sight and the clouds are generally not self-bound. Clouds with W50 $<$50 \kms{} are found for about 20\% of the total simulation time - in the most extreme simulation they were present for 54\% of the time. Despite the significant increase in the low-width clouds, the high-width clouds are rarer than in the M2 model. No isolated clouds with W50 $>$ 100 \kms{} are ever produced in the M1 simulations. Clouds with W50 $>$ 50 \kms{} occur for a total of six timesteps, or 0.3\% of the total simulated time. Once again, producing fake dark galaxies within long \HI{} streams is much easier than producing isolated dark galaxy candidates. Isolated high-width clouds do form, but at a rate so low it is not really a sensible explanation for the observed clouds in the real Virgo cluster. Conversely, clouds of widths $<$50 \kms{} are again found much more frequently, further emphasising that tidal debris is an entirely plausible explanation for some observed clouds.

\subsection{The M3 model}
\label{sec:M1results}
This model has the same gas and stellar mass with the same distributions as the standard model, but it is more massive with a higher circular velocity. The overall results are very much as one might expect : it is harder to remove the gas, so stream and cloud formation is suppressed. The median amount of gas removed after 5 Gyr is just 2.0$\times$10$^{8}$\Msolar{}, about half that of the M2 case and more than ten times less that of the M1 model. The radial profile evolution of the gas is negligible, but very few detectable streams are produced, let alone streams with sharp kinks. Isolated clouds of all parameters are rarer than in either the M1 or M2 models. Clouds with velocity widths $<$50 \kms{} are found for only 6\% of the total simulation time. Only two such clouds ever exceed a velocity width of 50 \kms{} (a total of 0.1\% of the simulation time) while none at all reach 100 \kms{}. In short, as far as the formation of long detectable streams and fake dark galaxies are concerned, this model is essentially uninteresting. Details are shown in appendix \ref{sec:ap1}.

\section{Summary and Conclusions}
\label{sec:conc}
We simulated the effects of harassment on a variety of giant spiral galaxies with varying \HI{} contents and dynamical masses. We used SPH particles and n-bodies to represent the gas, stars and dark matter of an infalling galaxy, and 400 pre-computed NFW halos as a reasonable model for the cluster potential including substructure. The galaxy was given 27 different initial positions and allowed to fall freely into the cluster potential for 5 Gyr. We created synthetic \HI{} observations to approximate what we would observe with a survey of the same capabilities as AGES. The goal was to see if tidal encounters could produce two types of features known from observations which are sometimes interpreted as dark galaxy candidates : isolated clouds with high velocity widths, and sharp changes in the velocity gradient of long \HI{} streams which can appear as kinematically distinct. Specifically we searched for features similar to the dark clouds described in T12 and streams similar to VIRGOHI21, originally proposed as an optically dark galaxy but later claimed in DB08 to be an effect of harassment.

Our results completely support our earlier findings in T16 but offer several advantages. We again found that apparently isolated clouds do form, and indeed overall are quite common in most of our simulations, being found for approximately 10\% of the time in our standard M2 model. However, the great majority of these clouds do not match the observed AGES clouds we are interested in : they have velocity widths $<$50 \kms{}, whereas the width of the AGES clouds is about 150 \kms{}. Isolated clouds of width $>$100 \kms{} are found only extremely rarely in our simulations - just 0.1\% of the total simulation time. This means that tidal debris is hardly a credible explanation for the six high-width clouds observed in the 20 square degree region of the cluster described in T12. These clouds were only ever produced at all in a few simulations and then only as individual clouds for very short periods. To explain the six observed clouds by this mechanism would require there to have been six unusual encounters, and for us to be observing all of those encounters during their very short periods when the clouds are detectable.

However, we also demonstrated that clouds produced in this way could be optically dark, with their stellar content being negligible (confirming the result of DB08). We were also able to examine the effects on the parent galaxy : in half the cases the galaxy appeared undisturbed, but the other half indicated that we should expect to detect very extended \HI{} emission. This is contrary to the observations since no extended features were detected in the AGES region at all.

Our simulations therefore suggest that (at the given 17 kpc diameter of the features we study) tidal debris is an entirely credible explanation for isolated features of low velocity width ($<$50 \kms) but extremely unlikely for those of higher velocity widths ($>$100 \kms). Tidal encounters also appear to be a sensible explanation for high velocity width features which are \textit{not} isolated, i.e. embedded in long, detectable \HI{} streams. We found examples of sharp changes of velocity gradient - the characteristic of the VIRGOHI21 system that led to its identification as a possible dark galaxy - in almost all cases where long streams were formed, and we found several examples of systems with a strong resemblance to certain aspects of VIRGOHI21. Unlike previous simulations these occurred simply by chance, with no attempt to deliberately simulate the formation of such a system. Moreover, our results reproduce this sharp change of velocity gradient significantly better than in the DB08 paper which proposed this model, as evident by a comparison between figures \ref{fig:ducvsrealvhi21} and \ref{fig:vhi21analogues}. It should be emphasised that our aim was to see if objects that would be mistaken for dark galaxies could really be produced by the mechanism, not to precisely recreate VIRGOHI21 itself.

Having shown that tidal encounters can explain isolated low-width clouds and high-width features within streams such as VIRGOHI21, but not isolated high-width clouds, we note several reasons why this is the case :

\noindent1) We have shown that small clouds are not torn off from the galaxies directly. So they only result (in our models) from the fragmentation of longer streams, though these are commonly produced in interactions.

\noindent2) Small isolated clouds must by definition be the longest-lived features in order to be detectable - but with high velocity widths, they should disintegrate more quickly than the rest of the stream. Thus, given (1), the absence of streams associated with real \HI{} clouds argues against a tidal origin.

\noindent3) At the same total mass, a feature of higher velocity width is intrinsically harder to detect than one of a lower velocity width since the flux is spread over more channels. The low-width features should not only persist for longer than high-width features but, at any moment, they should also have higher S/N levels.

\noindent4) If a section of a stream is detectable and has a high velocity width, adjacent sections also tend to have similar widths. High velocity width features in streams therefore do not necessarily appear as kinematically distinct. While sharp changes in velocity widths over short distances do occur (i.e. VIRGOHI21-like features), because of (3) it is very difficult to render the surrounding parts of the stream undetectable.

Thus, the formation of detectable, isolated, high velocity width features is rare in our simulations not because we have unluckily selected trajectories unfavourable to their formation, but because it is a fundamentally difficult and unlikely event. What our simulations have established is the quantitative values of the parameters of clouds which can be formed by tidal encounters : at diameters $\sim$20 kpc, clouds with velocity widths $\lesssim$50 \kms{} are common, while those with widths $\gtrsim$ 50 \kms{} are much rarer and clouds above 100 \kms{} are essentially negligible. In all cases the velocity widths of the clouds in our models arise from streaming motions along the line of sight - forming gravitationally self-bound blobs of high velocity widths would require them to be so small that they could not remain optically dark. In T12 we stated that it would be very difficult to distinguish tidal debris from dark galaxies, however these and the T16 simulations suggest we may have been pessimistic : in fact, at a given size, the velocity width alone (for isolated features) is a powerful discriminant if not for the actual origin of the features, then certainly as to whether the clouds can be explained as tidal debris or not.

A major caveat is that the results could change if there were other processes acting to render the rest of the stream undetectable, e.g. ionisation that affects the low-density material more strongly\footnote{We do not expect pressure confinement from the ICM to change these conclusions - if anything, it should make it more difficult to form structures with these velocities in the first place. We are running numerical simulations to investigate this in more detail (Taylor \& W\"{u}nsch in preparation), but see \cite{BL} for an alternative view.}. It is difficult to intuitively predict the full effects of the intracluster medium, with turbulence, heat conduction, radiative heating, and cooling all playing a possible role in fragmenting the streams. However, as far as the notion that purely tidal effects are responsible for the observed AGES clouds, we regard this as disproved. Producing a high velocity width or a sharp change in velocity gradient within a stream is relatively easy, but isolating that feature from the rest of the stream is intrinsically very difficult.

We have confirmed earlier findings that objects like VIRGOHI21 may be the result of tidal encounters, but some degree of caution is still warranted with regard to VIRGOHI21 itself. First, the emission from each velocity channel in VIRGOHI21 is brighter than in the rest of the stream, which is not what we might expect from harassment (as in point 3 above). However this could be explained by the stream being non-uniform prior to harassment, and indeed some similar features are seen in our simulations (see figure \ref{fig:vhi21analogues}). Second, the gas content of NGC 4254 makes it unusually gas rich, and the gas content would have been exceptionally high if the gas in the stream also originated from NGC 4254's disc. Third, none of our models showed anything like the one prominent spiral arm in NGC 4254, though this was clearly reproduced in the model of \cite{d08} (and also by ram-pressure stripping as shown in \citealt{sofue}). This may suggest that a slower encounter and/or ram-pressure stripping may be necessary to explain the system, rather than pure high-speed harassment. Fourth, the models which best reproduce the Westerbork observations do not closely match the ALFALFA observations showing that VIRGOHI21 is in the middle of the stream rather than at its end. Fifth, the observed radial profile of the gas was best reproduced in the model which lost the least gas. That is, a massive galaxy which begins with the same profile as the observations (which as we have discussed is not unusual) retains that profile, whereas in the less massive M1 case, which began with an unusual gas profile but experienced more gas removal, the simulation profile was never a close match to the observations. Finally, owing to these differences, we have not examined the synthetic observations so we leave a quantitative analysis of these simulations to a future project\,: reproducing the specific details of the VIRGOHI21 system is beyond the scope of the current work.

Overall, it is by no means clear if the evidence favours a tidal encounter or some other process, since the models to explain the VIRGOHI21 object are currently very limited. For instance, the \HI{} disc of NGC 4254 seems to be experiencing ram pressure stripping, but no simulation has yet explored the effect of this on the stripped material in this system. We do not yet know how this would affect either the harassment scenario or the dark galaxy model - the success of one does not preclude the success of the other. So for this specific object, any conclusion on its most likely origin remains premature. Both our M1 and M2 models show some similarities to the system, suggesting that perhaps the real NGC 4254 orignally had properties somewhere intermediate between the two.

We noted that our results broadly agree with the masses of the observed streams and show how long detectable streams can be found around non-deficient galaxies, but cannot explain why many galaxies have strong deficiencies. In contrast, as we discussed in T16, simulations of ram-pressure stripping can explain these high deficiencies but also predict detectable \HI{} in streams $>$ 100 kpc length. While the highly deficient galaxies are common in Virgo, the streams are very rare. Perhaps ram-pressure by the hot ICM always causes rapid ionization of the \HI{}, rendering it undetectable, whereas tidal encounters occur less frequently but do not significantly heat the gas. We leave a quantitative analysis of this suggestion to a future work.

However, we can certainly answer the three questions posed in the introduction. Firstly, it is possible to produce similar clouds to those described in T13 by purely tidal encounters but it requires very specific circumstances, and this is not at all a plausible explanation for the observed clouds. Secondly, sharp velocity kinks such as that of VIRGOHI21 could indeed have a tidal origin, bearing in mind the caveats we have discussed. Thirdly, the quantitative differences can be significant (an order of magnitude) for a massive spiral galaxy, but they do not change the main conclusions : isolated clouds that mimic dark galaxies are extremely difficult to produce by harassment, but harassment is a plausible explanation for such features embedded in \HI{} streams.

The question as to the true origin of the clouds and VIRGOHI21 remains open. The simulations described here have further weakened the case for the isolated clouds being tidal debris, but they have somewhat strengthened the argument that VIRGOHI21 has a tidal origin. We regard the case for the isolated high velocity width clouds being tidal debris as now being extremely weak, but the origin of VIRGOHI21 remains unclear. Optically dark galaxies remain a perfectly valid explanation for both - especially the isolated clouds, since in T16 we showed these could explain the observations much more closely, and our simulations here added further evidence against the clouds being tidal debris.

The key component missing from our simulations thus far is the intracluster medium, which we will examine in forthcoming papers. Observationally we hope to obtain more AGES-depth observations over a larger area to discover more clouds, and higher resolution observations with the VLA of the currently known clouds. With this combination of better statistics, more detailed examination of individual structures, and improved numerical modelling, we will continue to explore the role of these features in the cluster baryon cycle.

\section*{Acknowledgments}

This work was supported by the Tycho Brahe LG14013 project, the Czech Science Foundation projects P209/12/1795 and RVO 67985815. R.S. acknowledges support from Brain Korea 21 Plus Program (21A20131500002) and the Doyak Grant (2014003730).

This work is based on observations collected at Arecibo Observatory. The Arecibo Observatory is operated by SRI International under a cooperative agreement with the National Science Foundation (AST-1100968), and in alliance with Ana G. M\'{e}ndez-Universidad Metropolitana, and the Universities Space Research Association. 
This research has made use of the NASA/IPAC Extragalactic Database (NED) which is operated by the Jet Propulsion Laboratory, California Institute of Technology, under contract with the National Aeronautics and Space Administration. 

This work has made use of the SDSS. Funding for the SDSS and SDSS-II has been provided by the Alfred P. Sloan Foundation, the Participating Institutions, the National Science Foundation, the U.S. Department of Energy, the National Aeronautics and Space Administration, the Japanese Monbukagakusho, the Max Planck Society, and the Higher Education Funding Council for England. The SDSS Web Site is http://www.sdss.org/.

The SDSS is managed by the Astrophysical Research Consortium for the Participating Institutions. The Participating Institutions are the American Museum of Natural History, Astrophysical Institute Potsdam, University of Basel, University of Cambridge, Case Western Reserve University, University of Chicago, Drexel University, Fermilab, the Institute for Advanced Study, the Japan Participation Group, Johns Hopkins University, the Joint Institute for Nuclear Astrophysics, the Kavli Institute for Particle Astrophysics and Cosmology, the Korean Scientist Group, the Chinese Academy of Sciences (LAMOST), Los Alamos National Laboratory, the Max-Planck-Institute for Astronomy (MPIA), the Max-Planck-Institute for Astrophysics (MPA), New Mexico State University, Ohio State University, University of Pittsburgh, University of Portsmouth, Princeton University, the United States Naval Observatory, and the University of Washington.

{}

\clearpage

\appendix
\section{The M1 and M3 model}
\label{sec:ap1}

\begin{figure*}
\centering
\includegraphics[width=175mm]{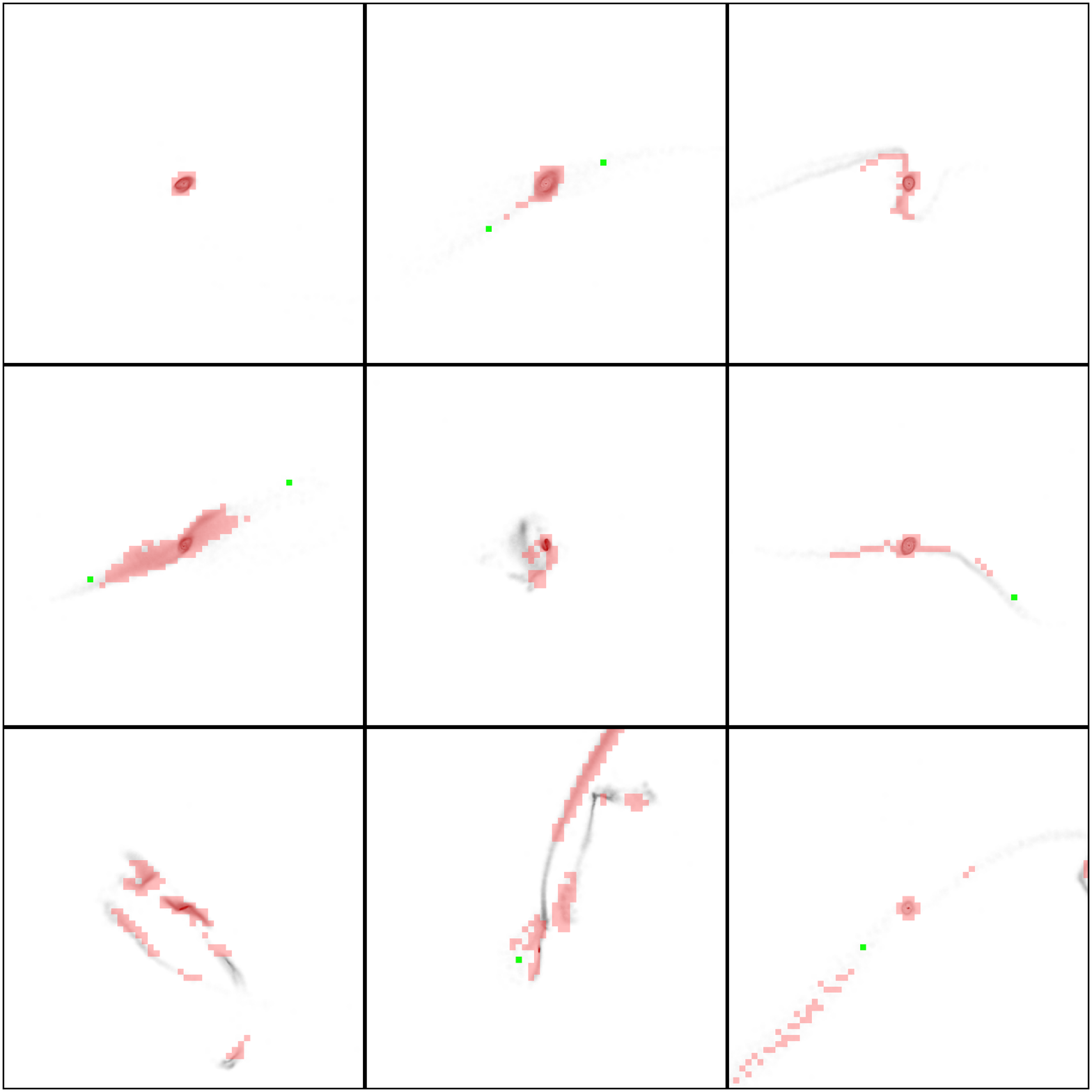}
\caption[vmap]{Snapshots of the M1 simulations, each after 4.2 Gyr in the cluster, shown with a 1 Mpc field of view. Particle data is shown in greyscale (black is the most intense), red uses gridded data to show anything with a S/N $>$ 4.0 to an AGES-class survey, while the green squares show \HI{} clouds detectable to AGES at least 100 kpc and 500 \kms{} in projected space/velocity from the nearest other \HI{} detection. The top row show initial positions which showed negligible disturbance in the M2 case, the middle row show cases where the M2 model showed minor disturbance, and the bottom row cases where the M2 model demonstrated major disruption. Movies can be seen at the following URL : \href{http://tinyurl.com/hhttkml}{http://tinyurl.com/hhttkml}.}
\label{fig:M1run}
\end{figure*}

\begin{figure}
\centering
\includegraphics[width=84mm]{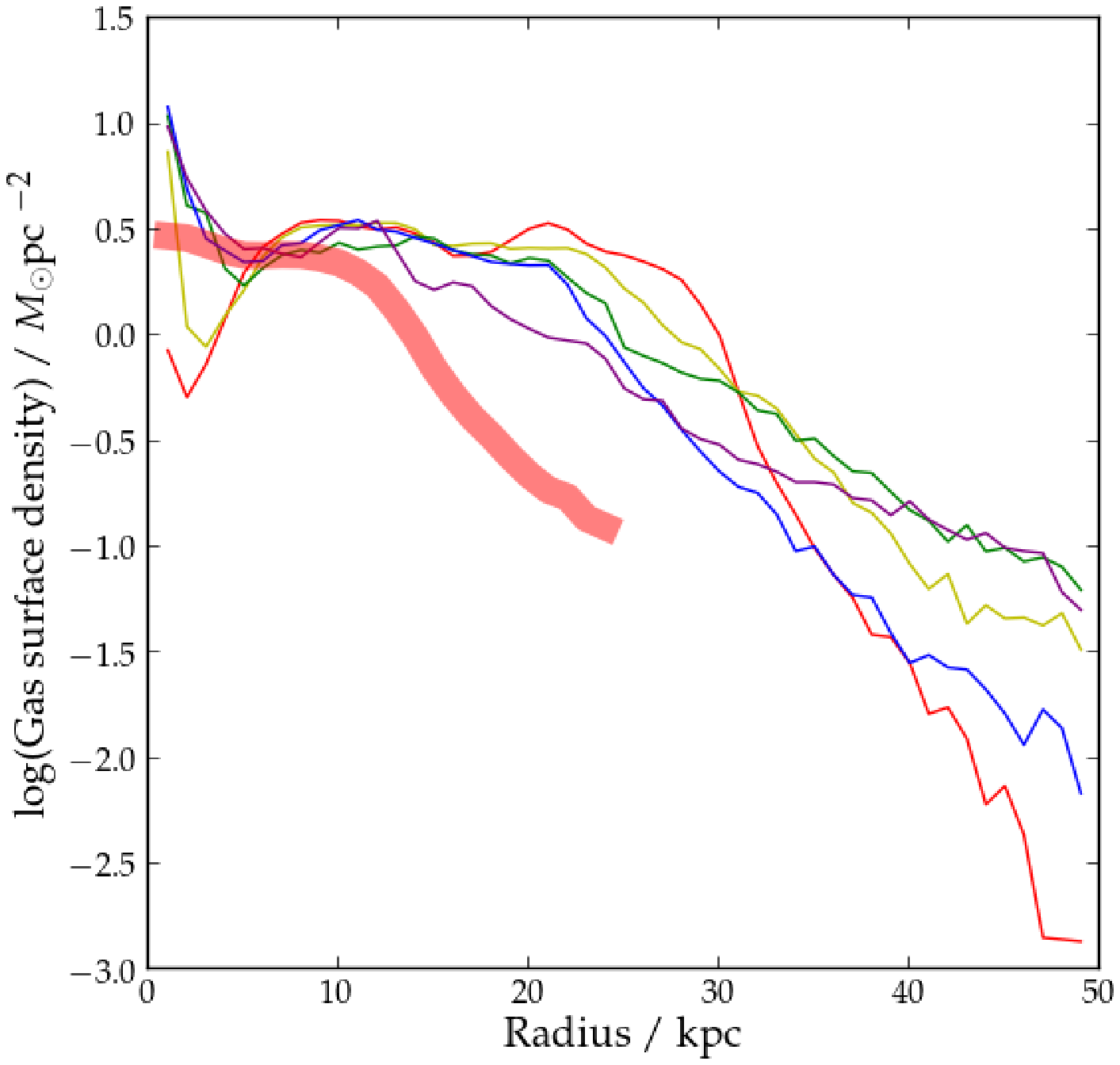}
\caption[vmap]{Evolution of the surface density profile of the gas component in our M1 simulation. The thick red line shows the observed profile of the real NGC 4254, while the thin lines indicate the median simulated profile in the cluster at different timesteps\,: red at 1 Gyr, yellow at 2 Gyr, green at 3 Gyr, blue at 4 Gyr, and purple at 5 Gyr.}
\label{fig:gasprofileM1}
\end{figure}

\begin{figure}
\centering
\includegraphics[width=84mm]{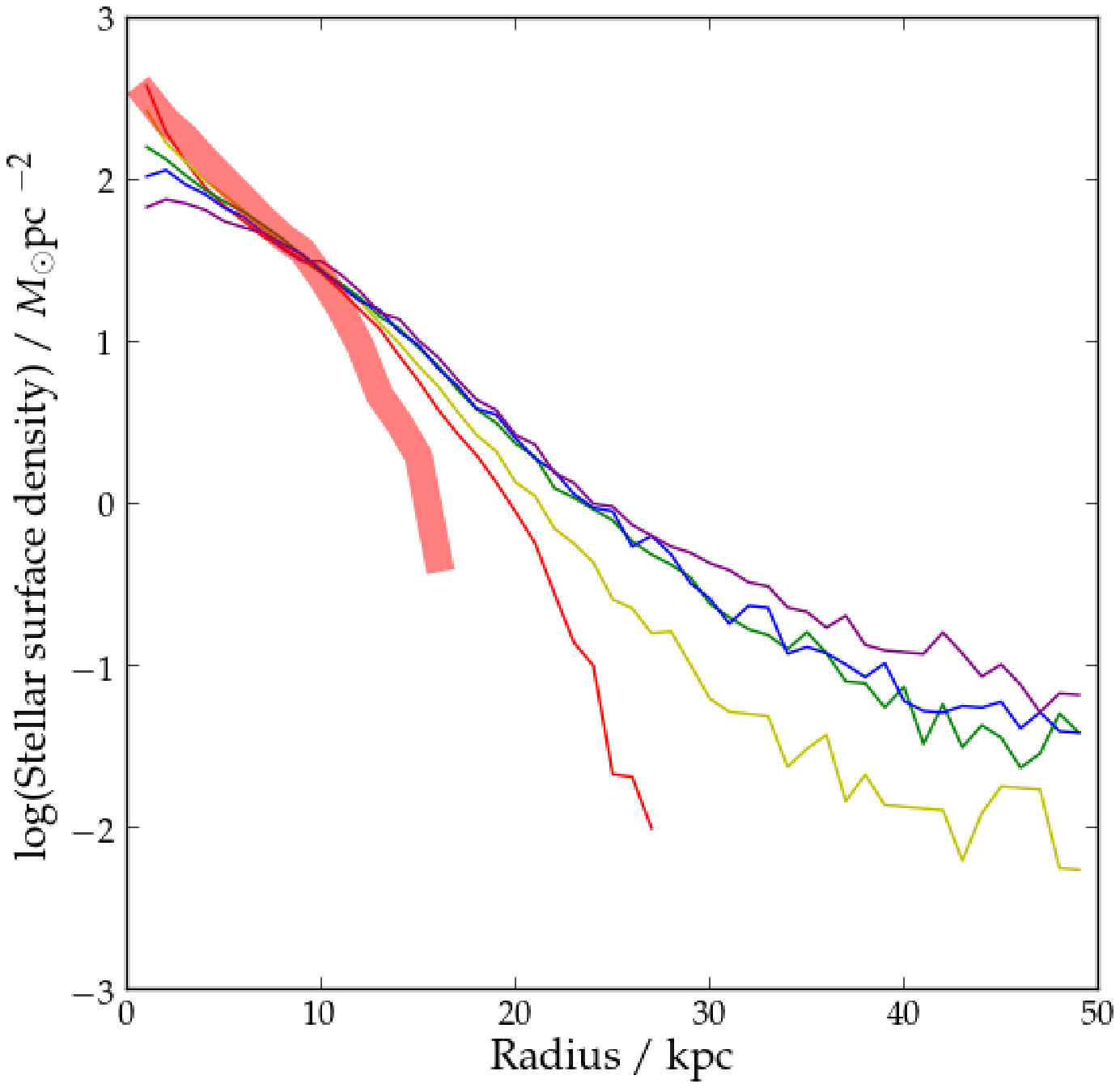}
\caption[vmap]{Evolution of the surface density profile of the stellar component in our M1 simulation in isolation, using the same colour scheme as figure \ref{fig:gasprofileM1}.}
\label{fig:starprofileM1}
\end{figure}

\begin{figure}
\centering 
  \subfloat[]{\includegraphics[height=40mm]{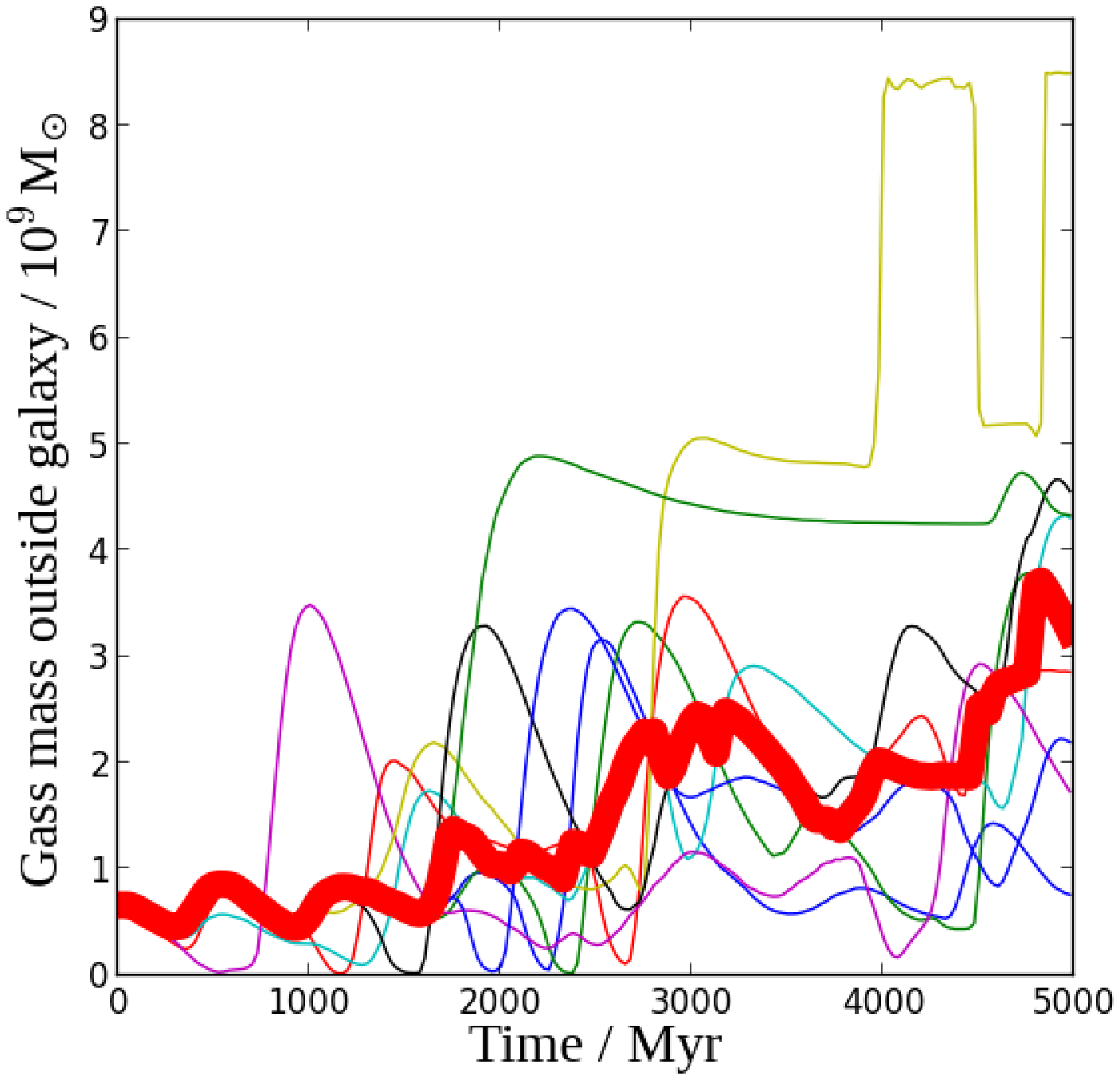}}
  \subfloat[]{\includegraphics[height=40mm]{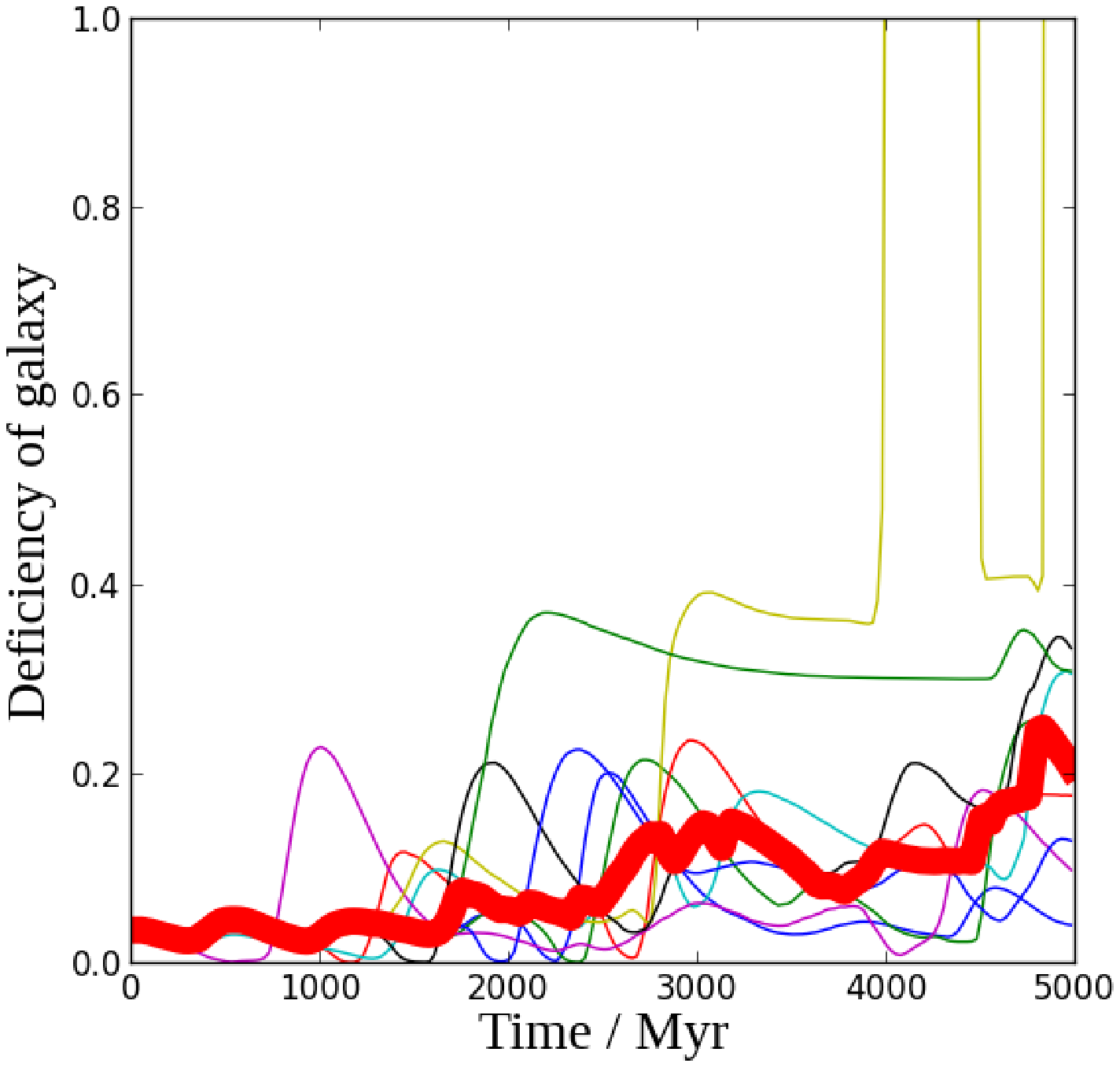}}\\ 
  \subfloat[]{\includegraphics[height=40mm]{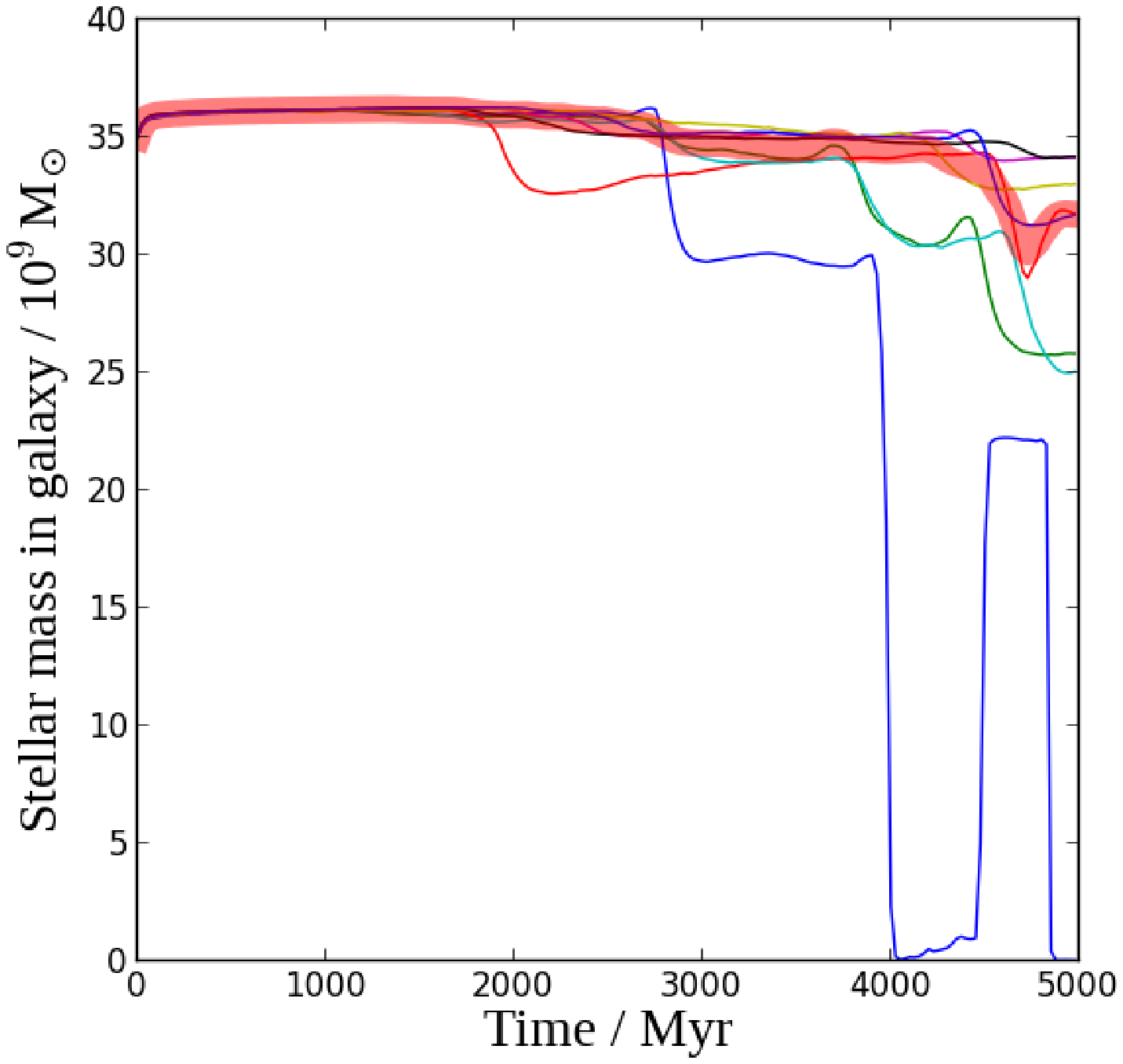}}
  \subfloat[]{\includegraphics[height=40mm]{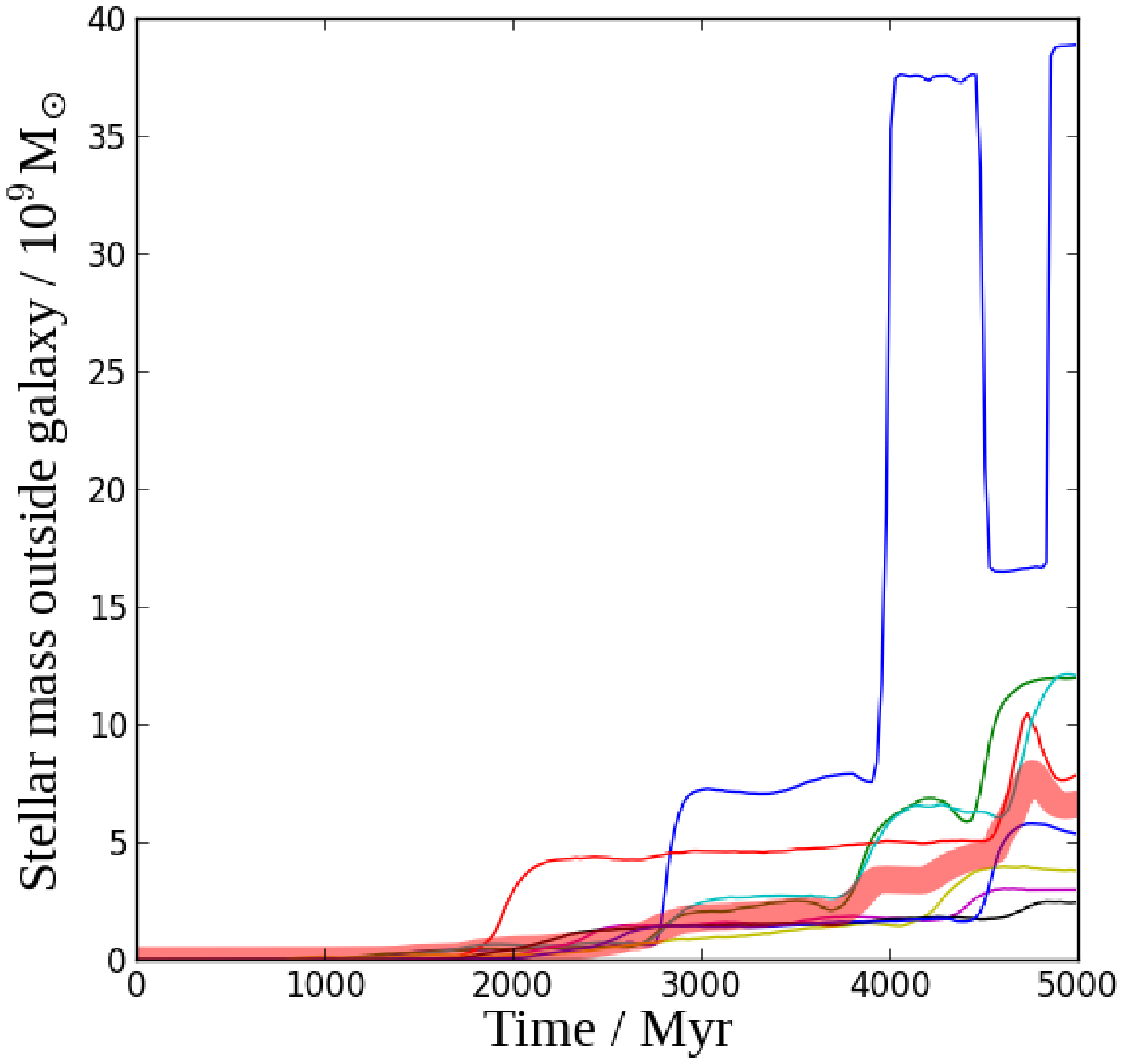}}  
\caption[Streams]{Evolution of the major properties of the stars and gas in the M1 model. In all cases we the define, `the galaxy' to be a sphere of radius 30 kpc centred on the median particle position. Panel (a) shows the gas mass outside the galaxy; (b) shows the measured \HI{} deficiency of the galaxy assuming its initial deficiency is zero; (c) shows the total stellar mass within the galaxy; (d) shows the total stellar mass outside the galaxy. The thick red line indicates the median value while the thin coloured lines show individual simulations. Note that the median particle position was sometimes well outside the true centre of the galaxy (yellow line in the upper panels, blue in the lower panels).}
\label{fig:v210props}
\end{figure}

\begin{figure*}
\centering 
  \subfloat[]{\includegraphics[height=55mm]{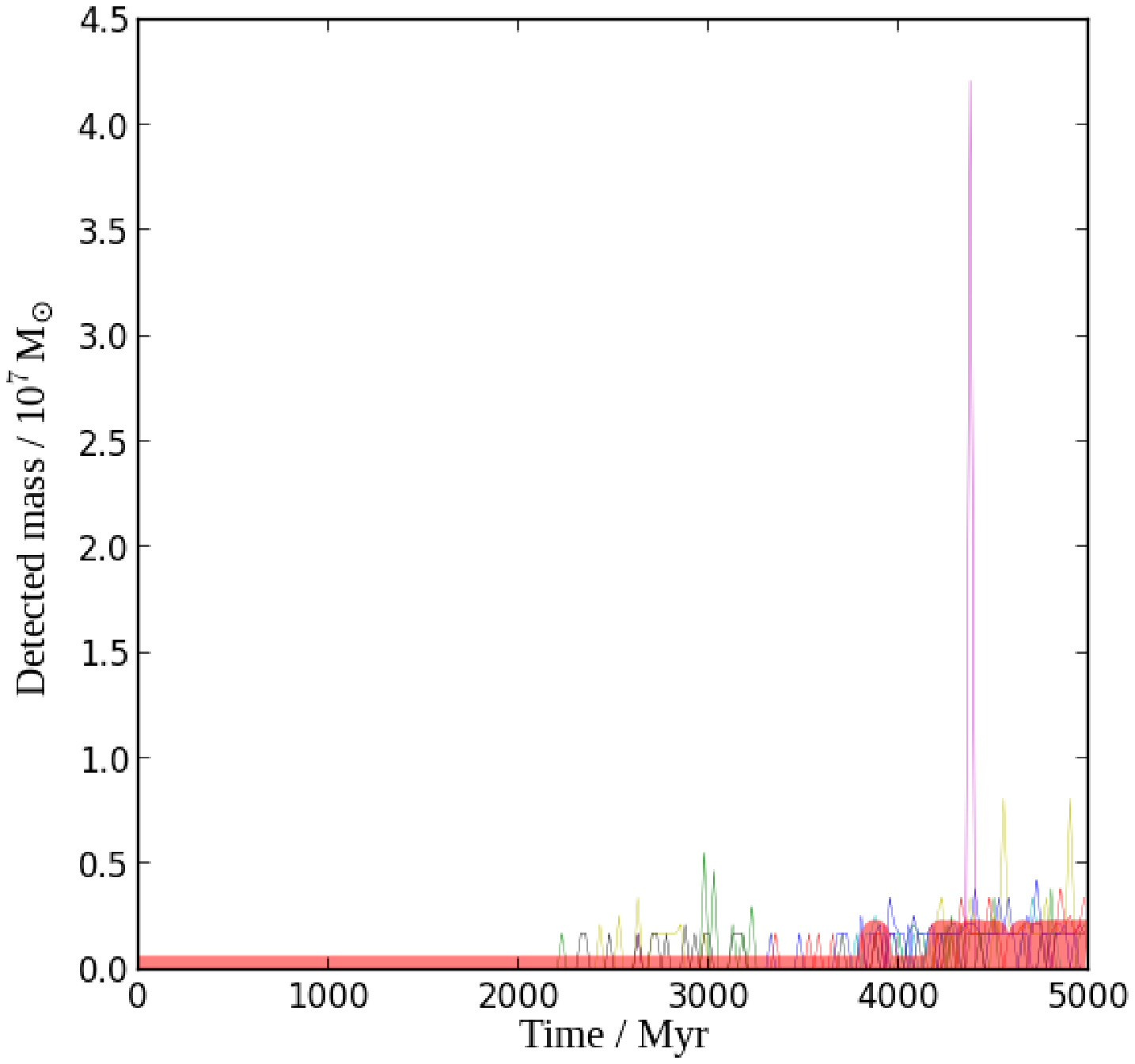}}
  \subfloat[]{\includegraphics[height=55mm]{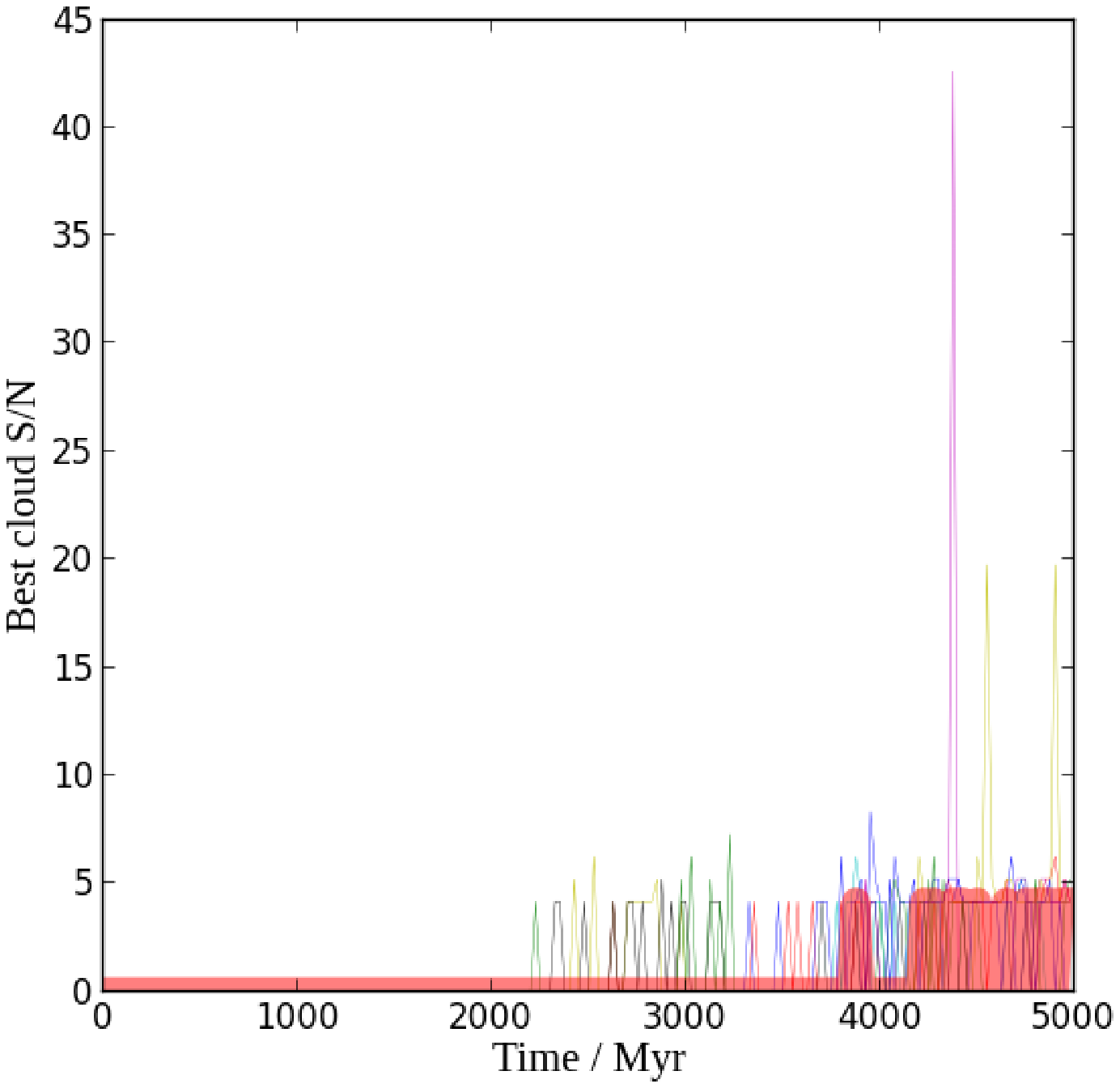}} 
  \subfloat[]{\includegraphics[height=55mm]{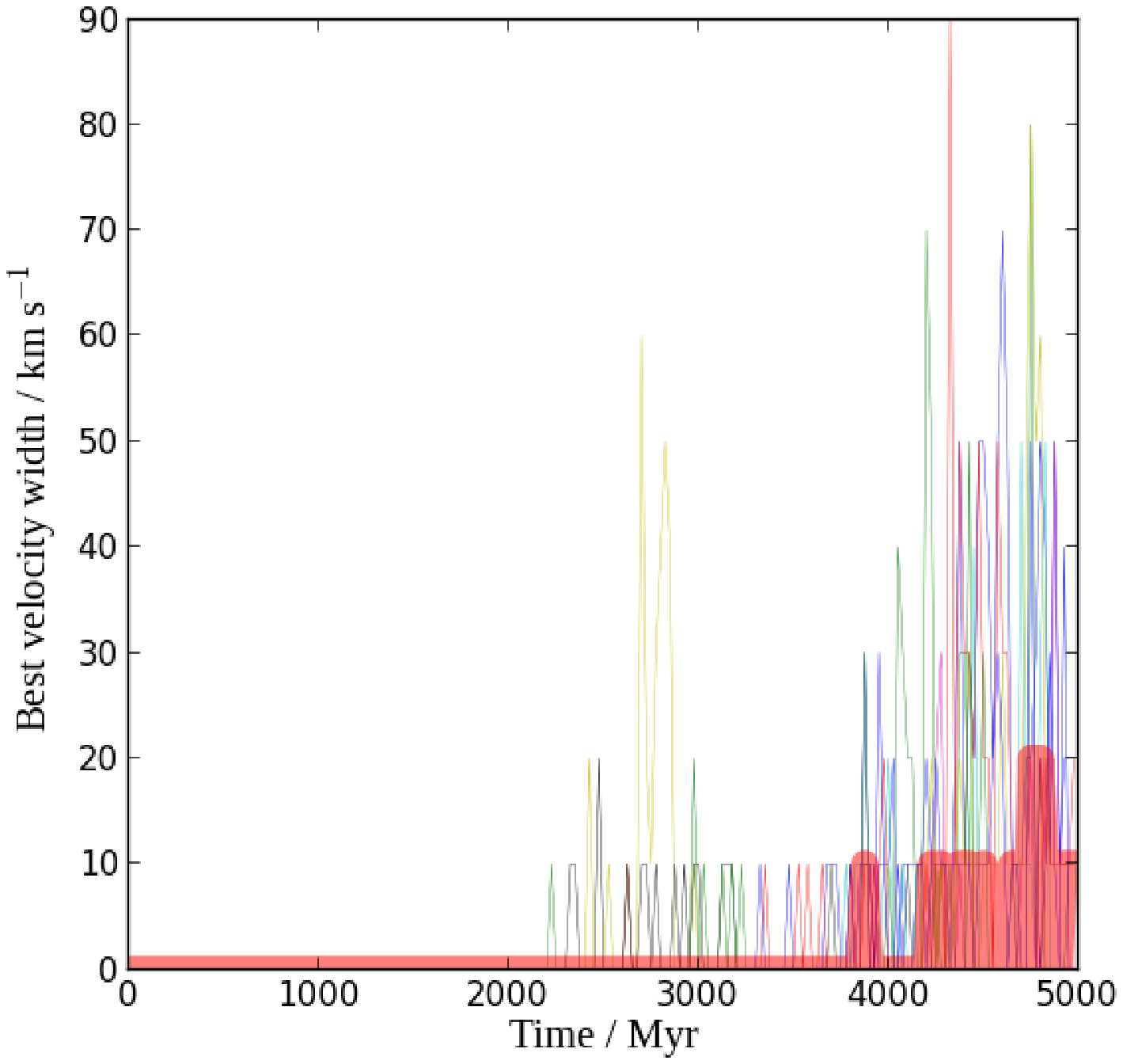}}
\caption[Streams]{Evolution of the unresolved, isolated \HI{} clouds as seen with an AGES-class survey for the M1 model. As in T16 we plot only the properties of the cloud with the highest velocity width for each simulation (each plotted as a different coloured line), since the high velocity widths appear to be the limiting factor in reproducing clouds similar to those described in T12 and T13. Panel (a) shows the detected mass in the cloud; (b) shows its signal to noise ratio; (c) shows its velocity width.}
\label{fig:v210clouds}
\end{figure*}

\begin{figure*}
\centering
\includegraphics[width=175mm]{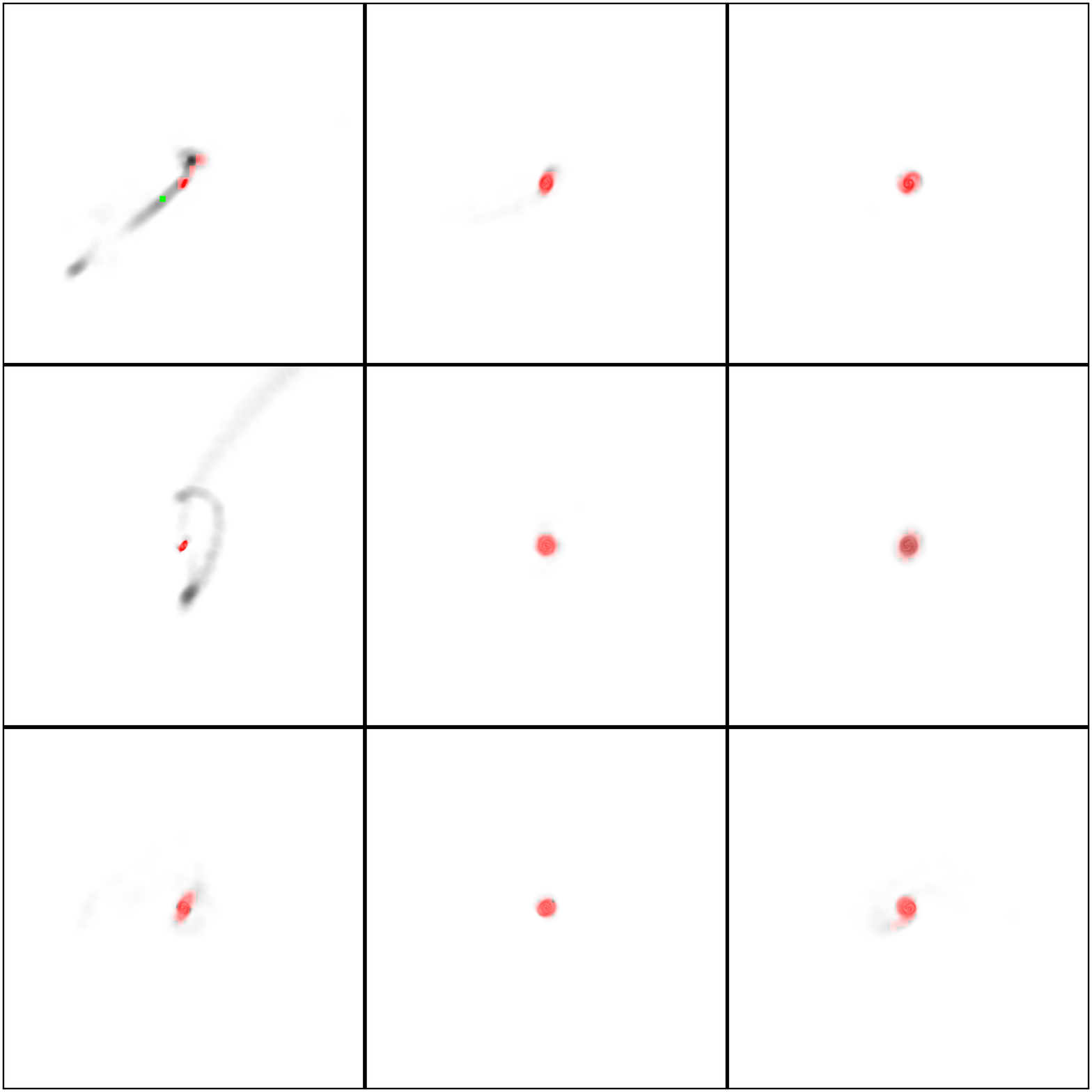}
\caption[vmap]{Snapshots of the M2 simulations (part 1), each after 4.2 Gyr in the cluster, shown with a 1 Mpc field of view. Particle data is shown in greyscale (black is the most intense), red uses gridded data to show anything with a S/N $>$ 4.0 to an AGES-class survey, while the green squares show \HI{} clouds detectable to AGES at least 100 kpc and 500 \kms{} in projected space/velocity from the nearest other \HI{} detection. Movies can be seen at the following URL : \href{http://tinyurl.com/ja2auqq}{http://tinyurl.com/ja2auqq}.}
\label{fig:M2run1}
\end{figure*}

\begin{figure*}
\centering
\includegraphics[width=175mm]{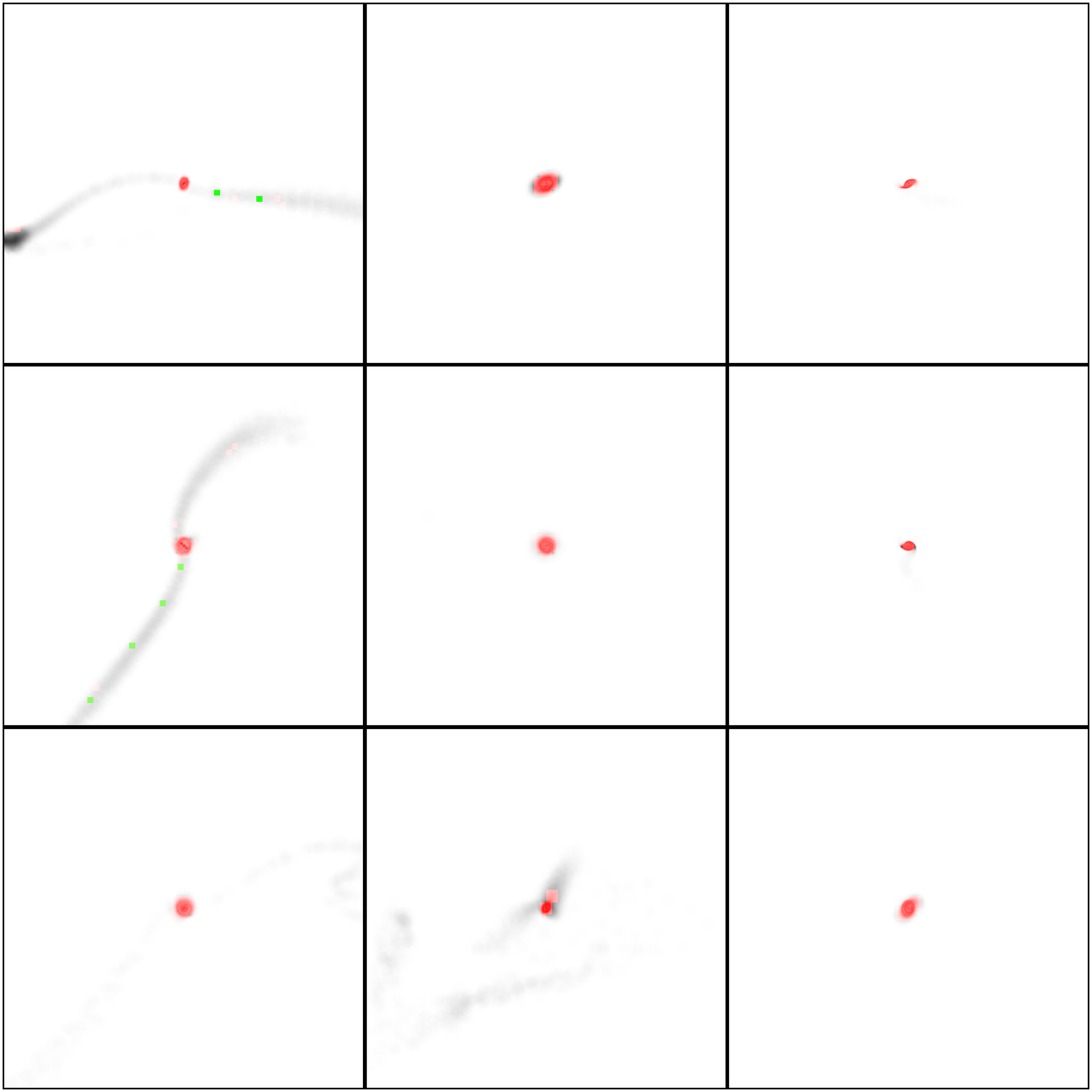}
\caption[vmap]{Snapshots of the M2 simulations \textbf{(part 2) - for description see caption of figure \ref{fig:M2run1}.}}
\label{fig:M2run2}
\end{figure*}

\begin{figure*}
\centering
\includegraphics[width=175mm]{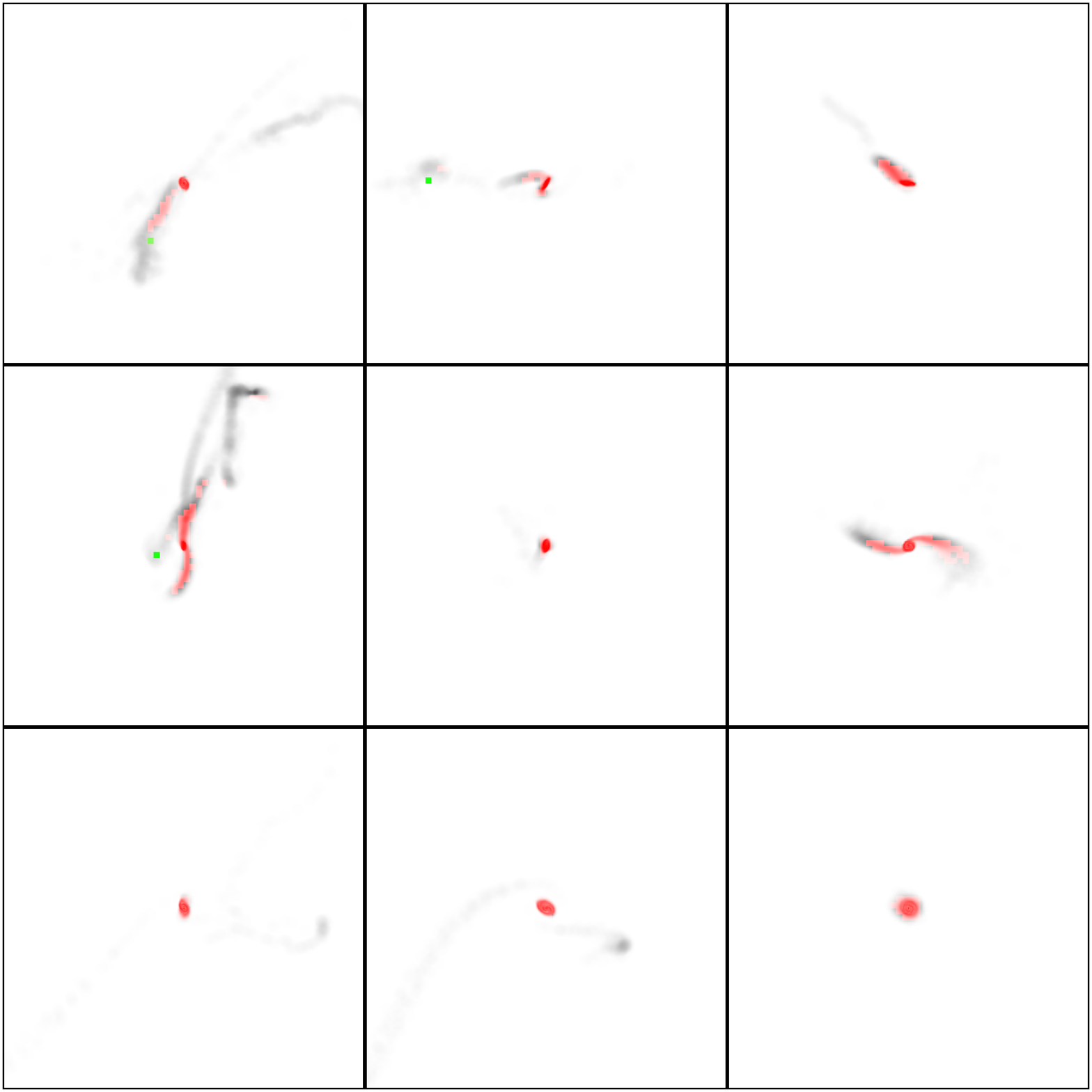}
\caption[vmap]{Snapshots of the M2 simulations \textbf{(part 3) - for description see caption of figure \ref{fig:M2run1}.}}
\label{fig:M2run3}
\end{figure*}

\begin{figure*}
\centering
\includegraphics[width=175mm]{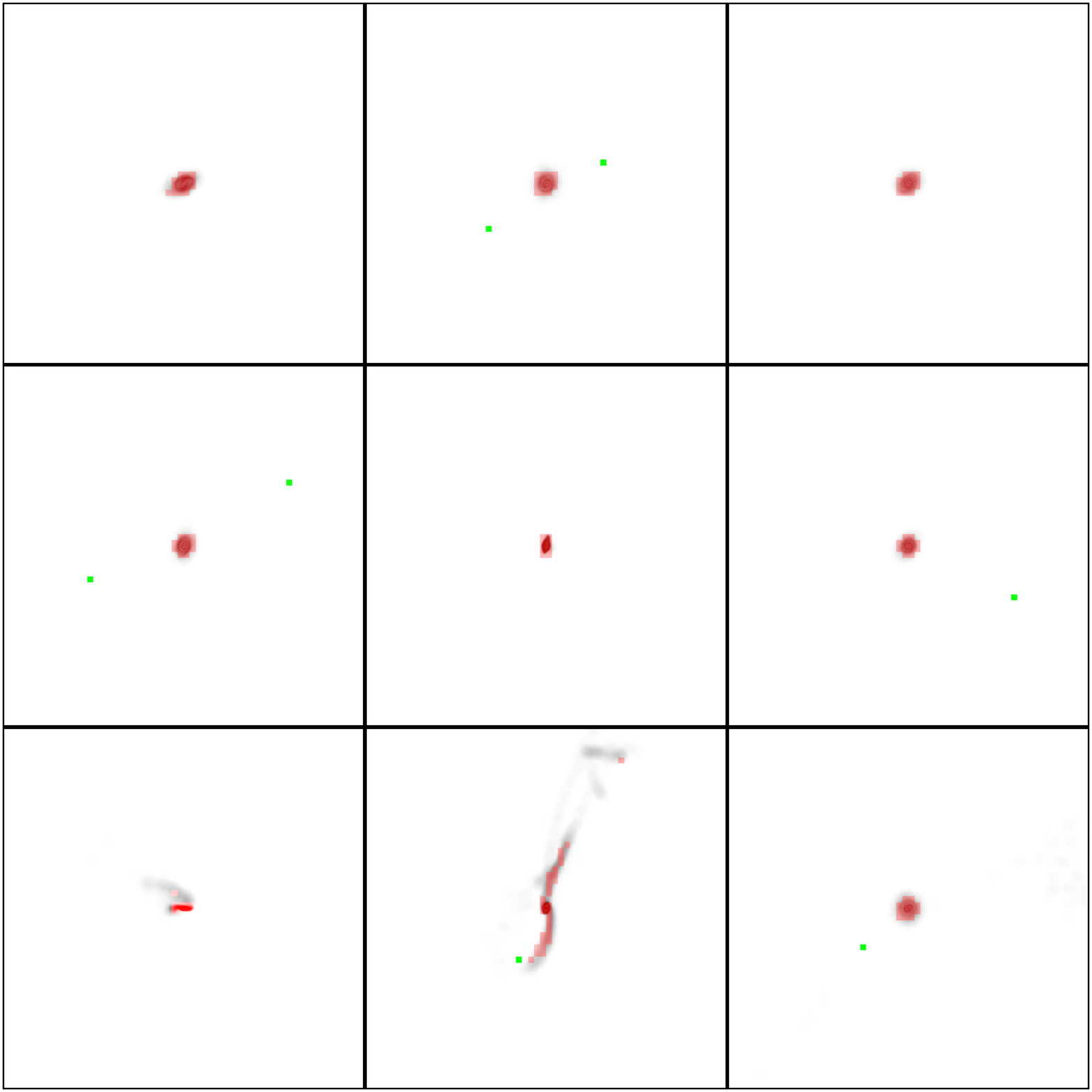}
\caption[vmap]{Snapshots of the M3 simulations, each after 4.2 Gyr in the cluster, shown with a 1 Mpc field of view. Particle data is shown in greyscale (black is the most intense), red uses gridded data to show anything with a S/N $>$ 4.0 to an AGES-class survey, while the green squares show \HI{} clouds detectable to AGES at least 100 kpc and 500 \kms{} in projected space/velocity from the nearest other \HI{} detection. movies can be seen at the following URL : \href{http://tinyurl.com/j8ken43}{http://tinyurl.com/j8ken43}.}
\label{fig:M3run}
\end{figure*}

\begin{figure}
\centering
\includegraphics[width=84mm]{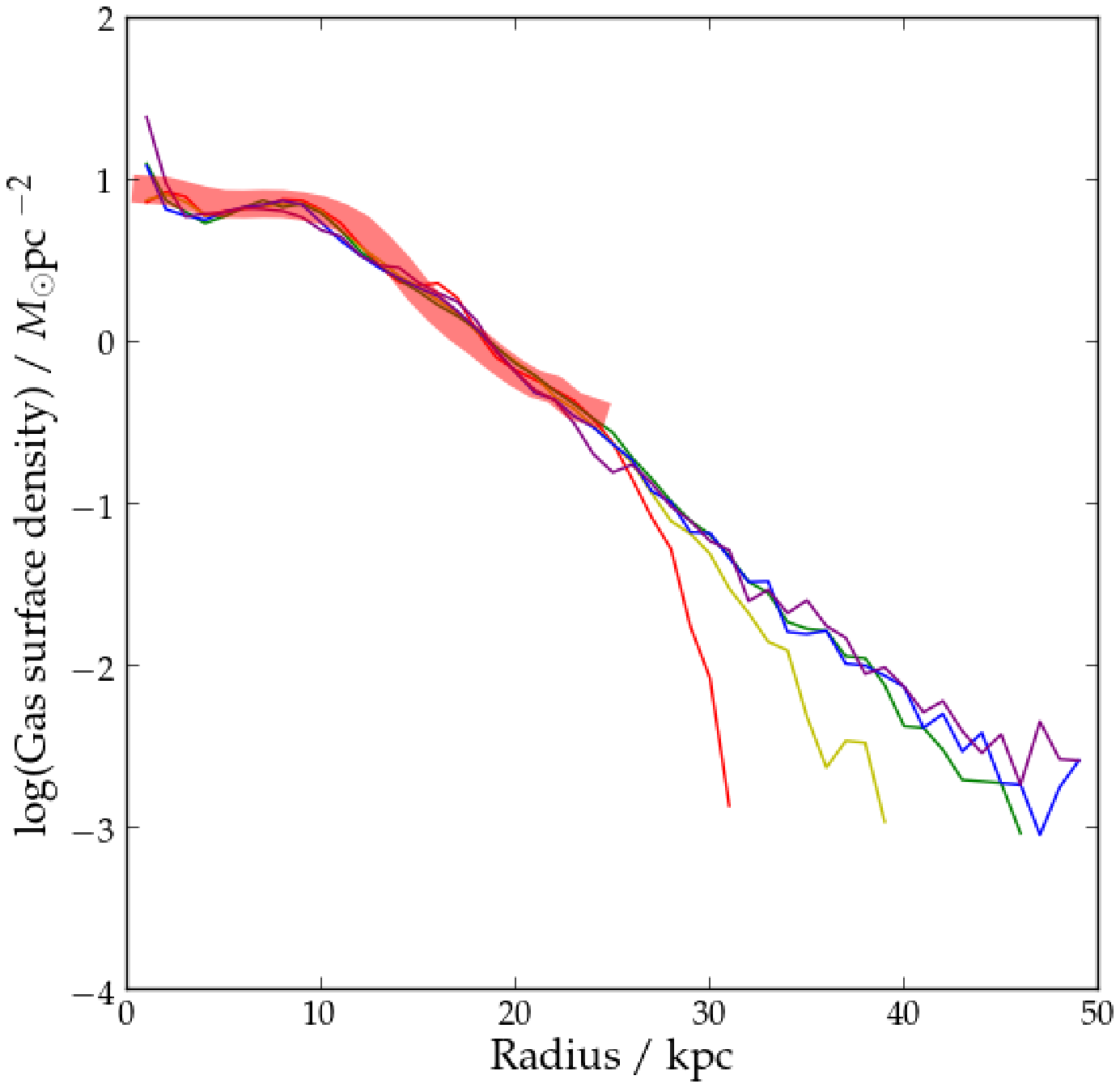}
\caption[vmap]{Evolution of the surface density profile of the gas component in our M3 simulation, using the same colour scheme as in figure \label{fig:gasprofileM3}.}
\label{fig:gasprofileM3}
\end{figure}

\begin{figure}
\centering
\includegraphics[width=84mm]{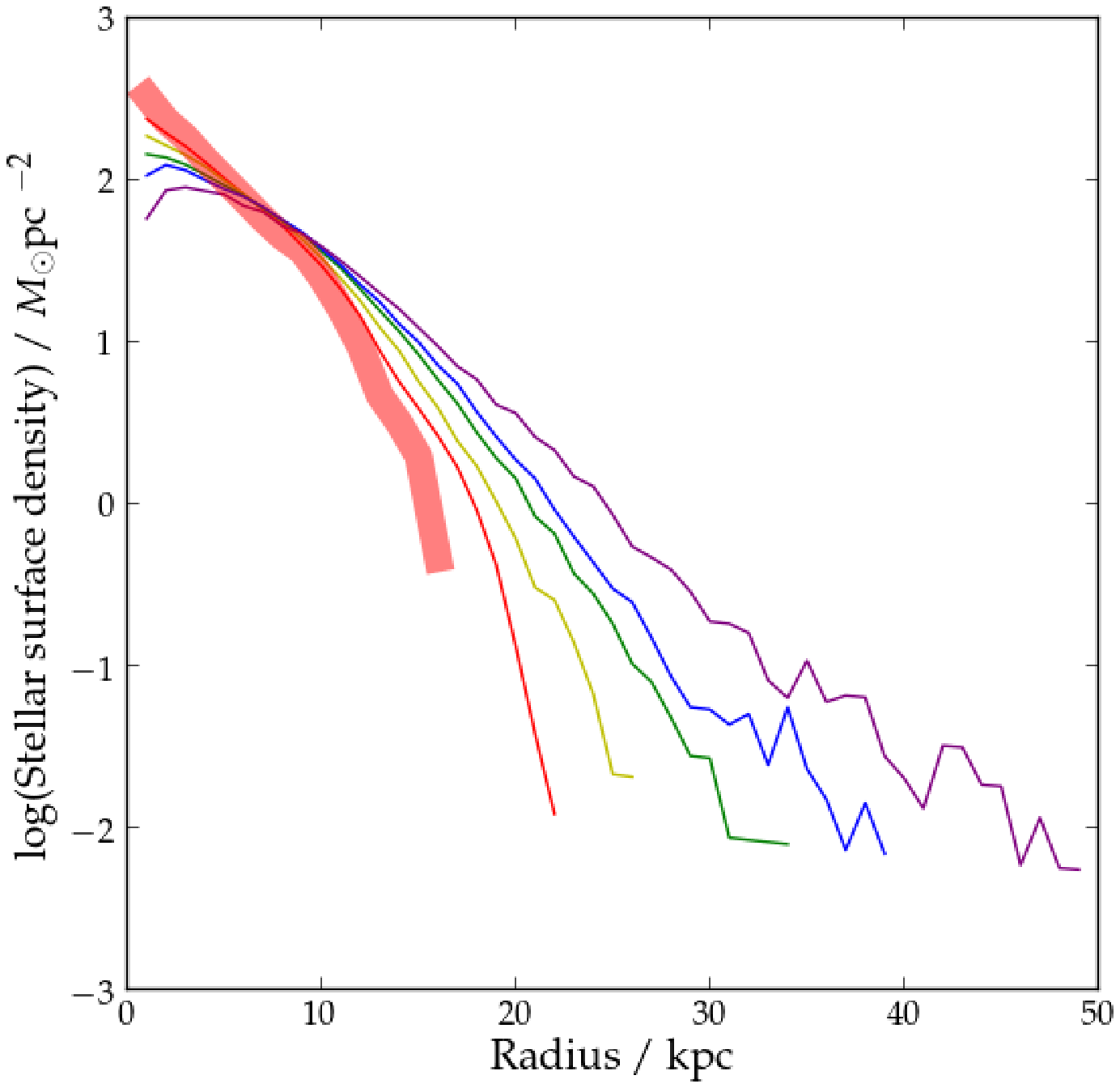}
\caption[vmap]{Evolution of the surface density profile of the stellar component in our M3 simulation in isolation, using the same colour scheme as figure \ref{fig:gasprofileM1}.}
\label{fig:starprofileM3}
\end{figure}

\begin{figure}
\centering 
  \subfloat[]{\includegraphics[height=40mm]{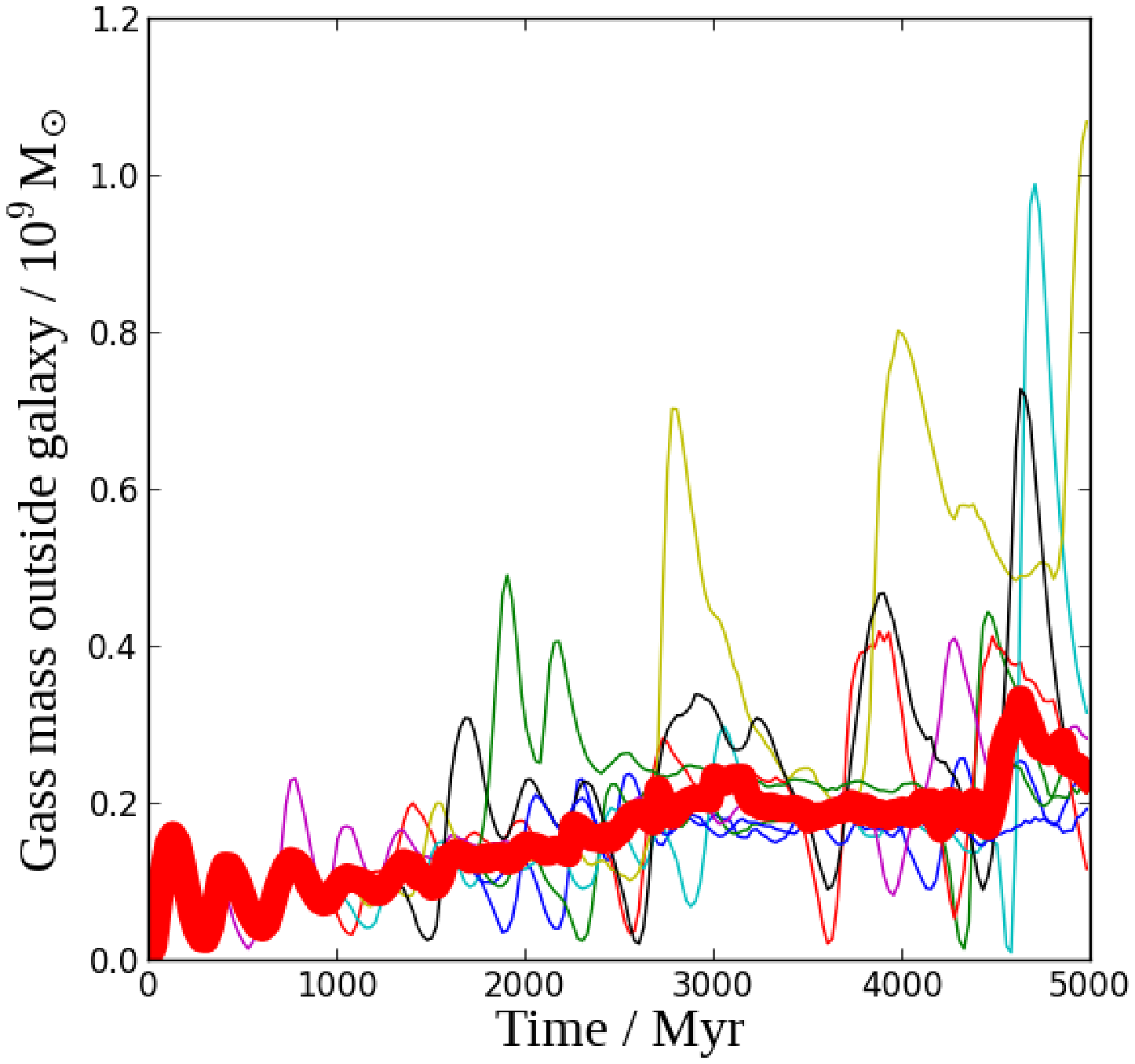}}
  \subfloat[]{\includegraphics[height=40mm]{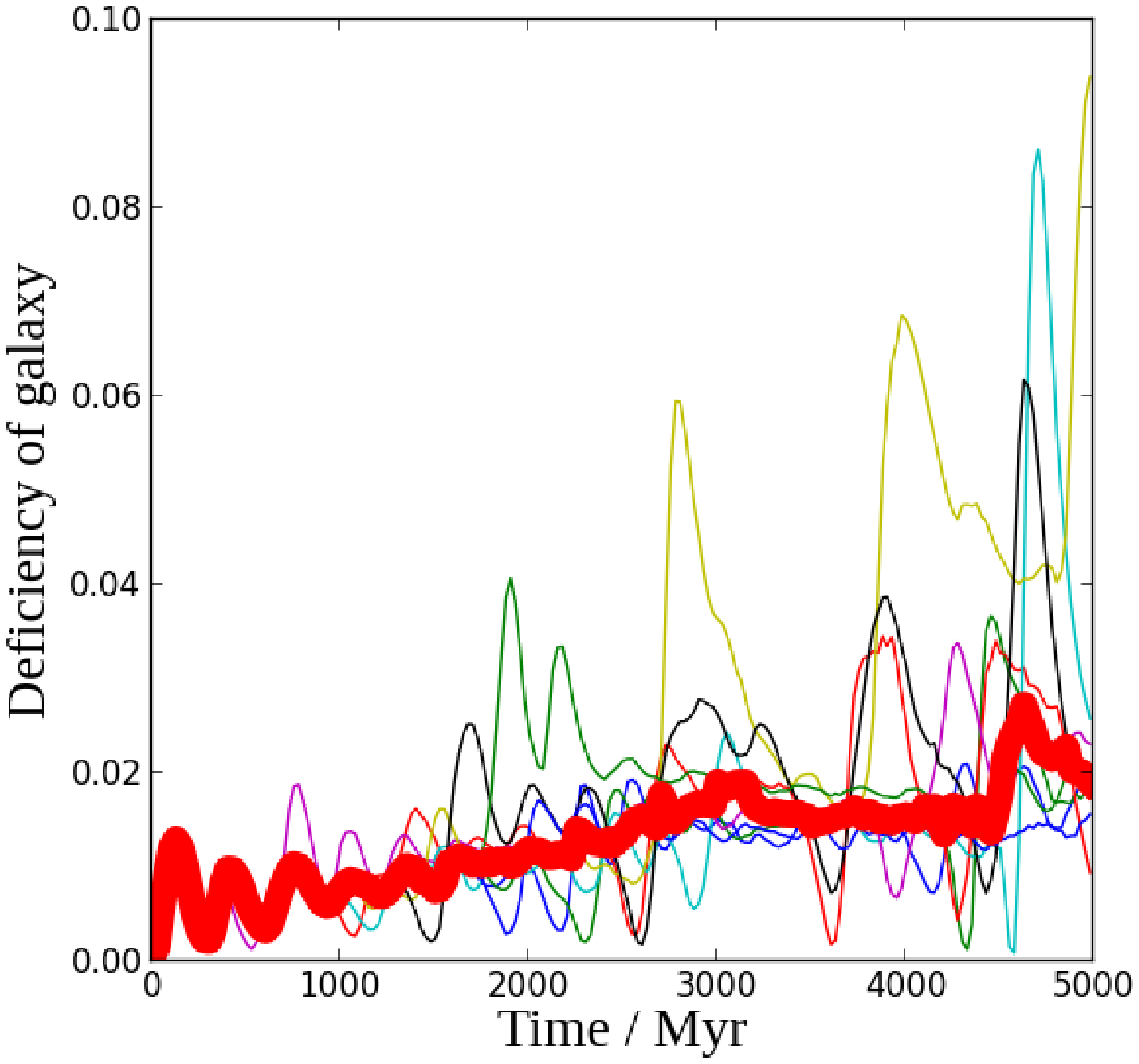}}\\ 
  \subfloat[]{\includegraphics[height=40mm]{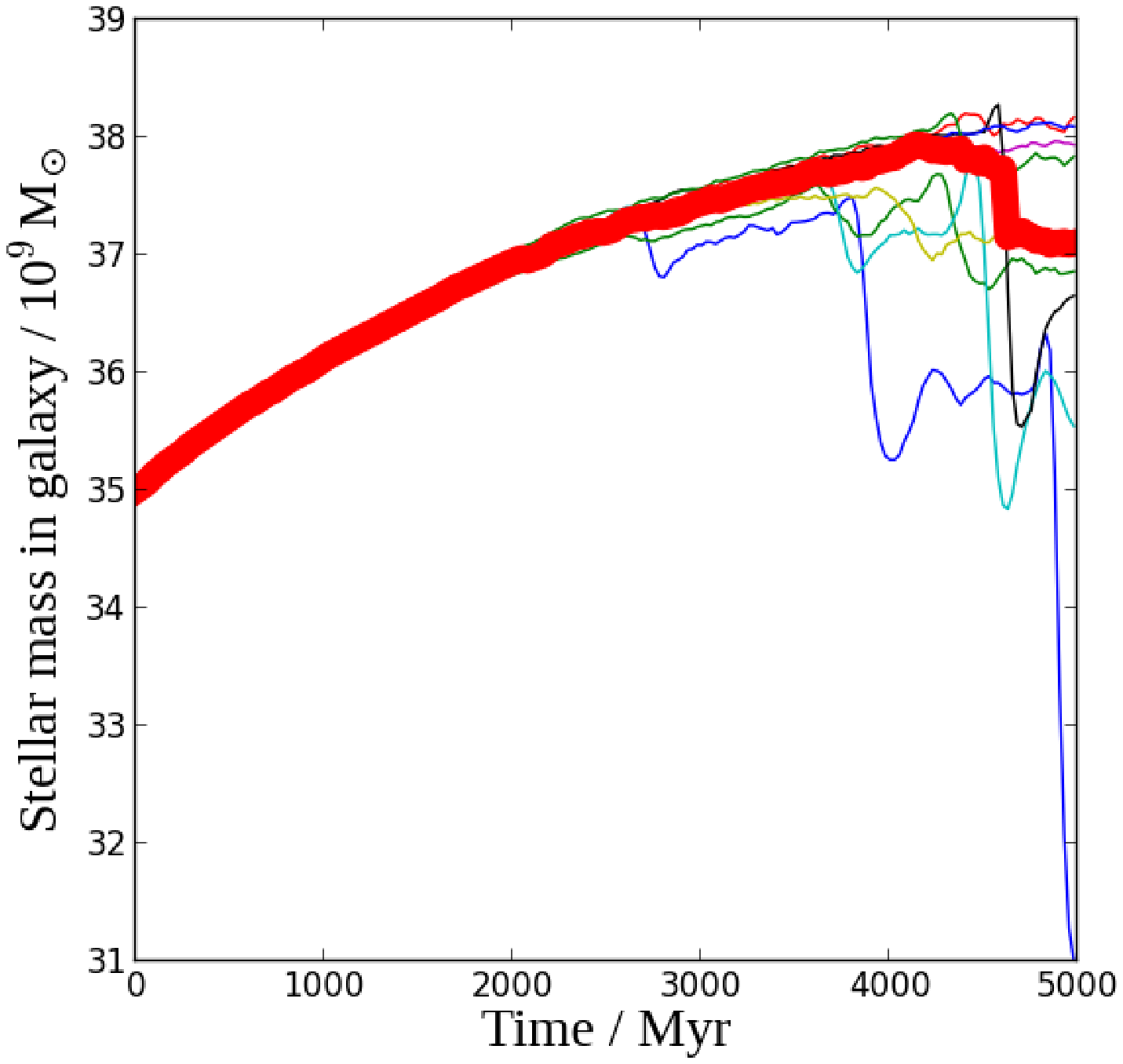}} % DO THESE TWO !!!
  \subfloat[]{\includegraphics[height=40mm]{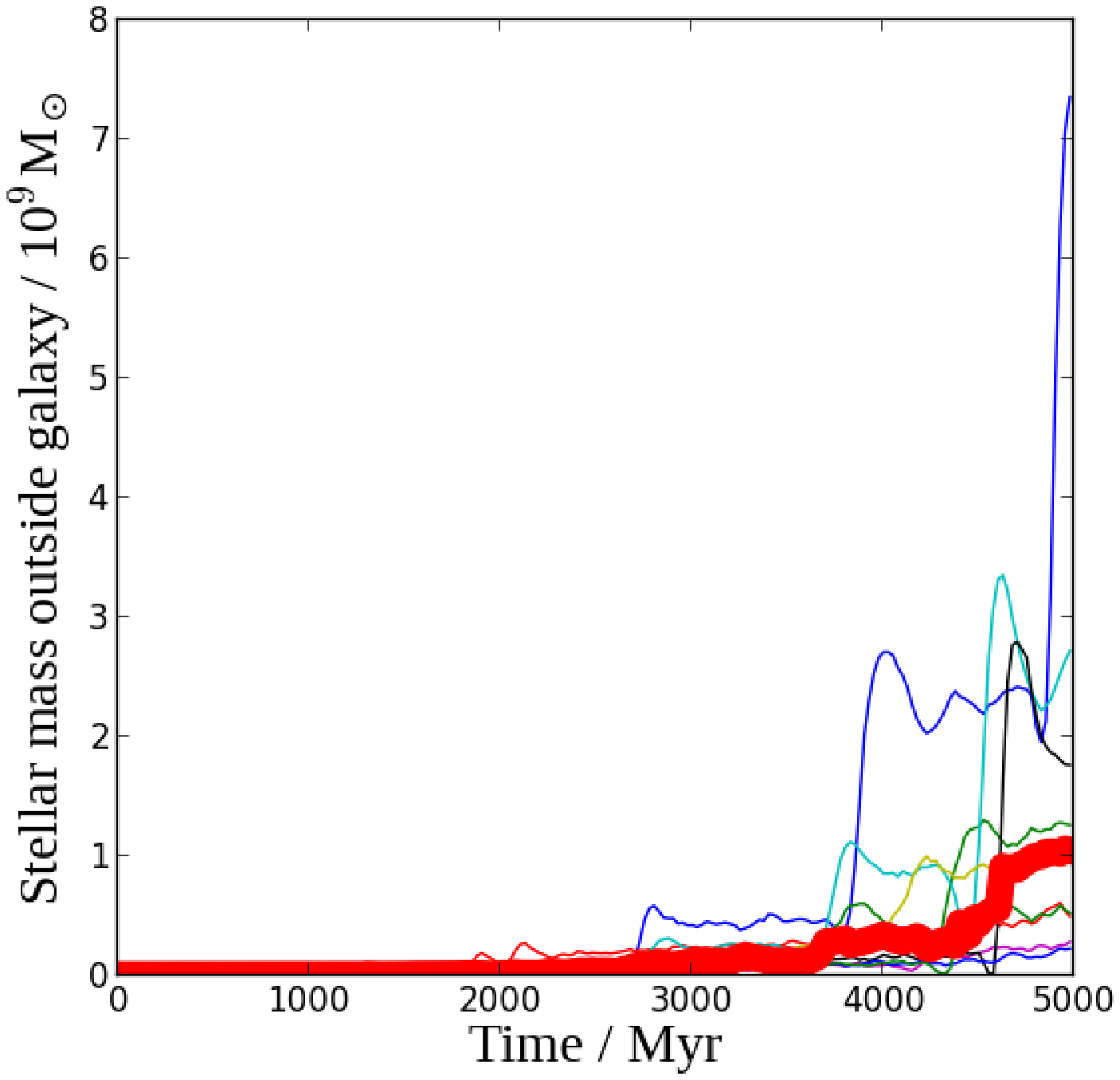}}  
\caption[Streams]{Evolution of the major properties of the stars and gas in the M3 model. In all cases we the define, `the galaxy' to be a sphere of radius 30 kpc centred on the median particle position. Panel (a) shows the gas mass outside the galaxy; (b) shows the measured \HI{} deficiency of the galaxy assuming its initial deficiency is zero; (c) shows the total stellar mass within the galaxy; (d) shows the total stellar mass outside the galaxy. The thick red line indicates the median value while the thin coloured lines show individual simulations. Note that in the case of the yellow line the median particle position was sometimes well outside the true centre of the galaxy.}
\label{fig:v390props}
\end{figure}

\begin{figure*}
\centering 
  \subfloat[]{\includegraphics[height=55mm]{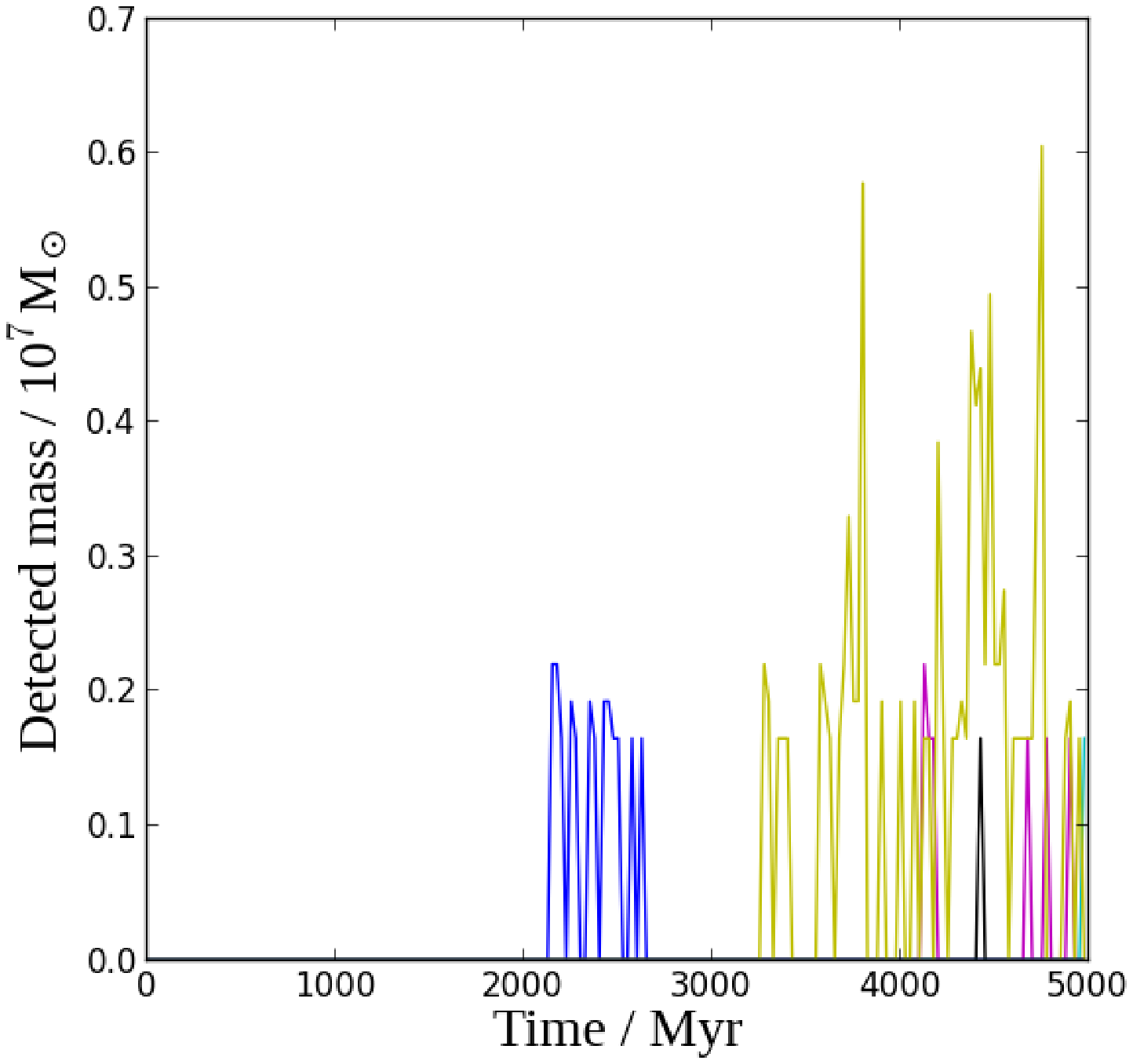}}
  \subfloat[]{\includegraphics[height=55mm]{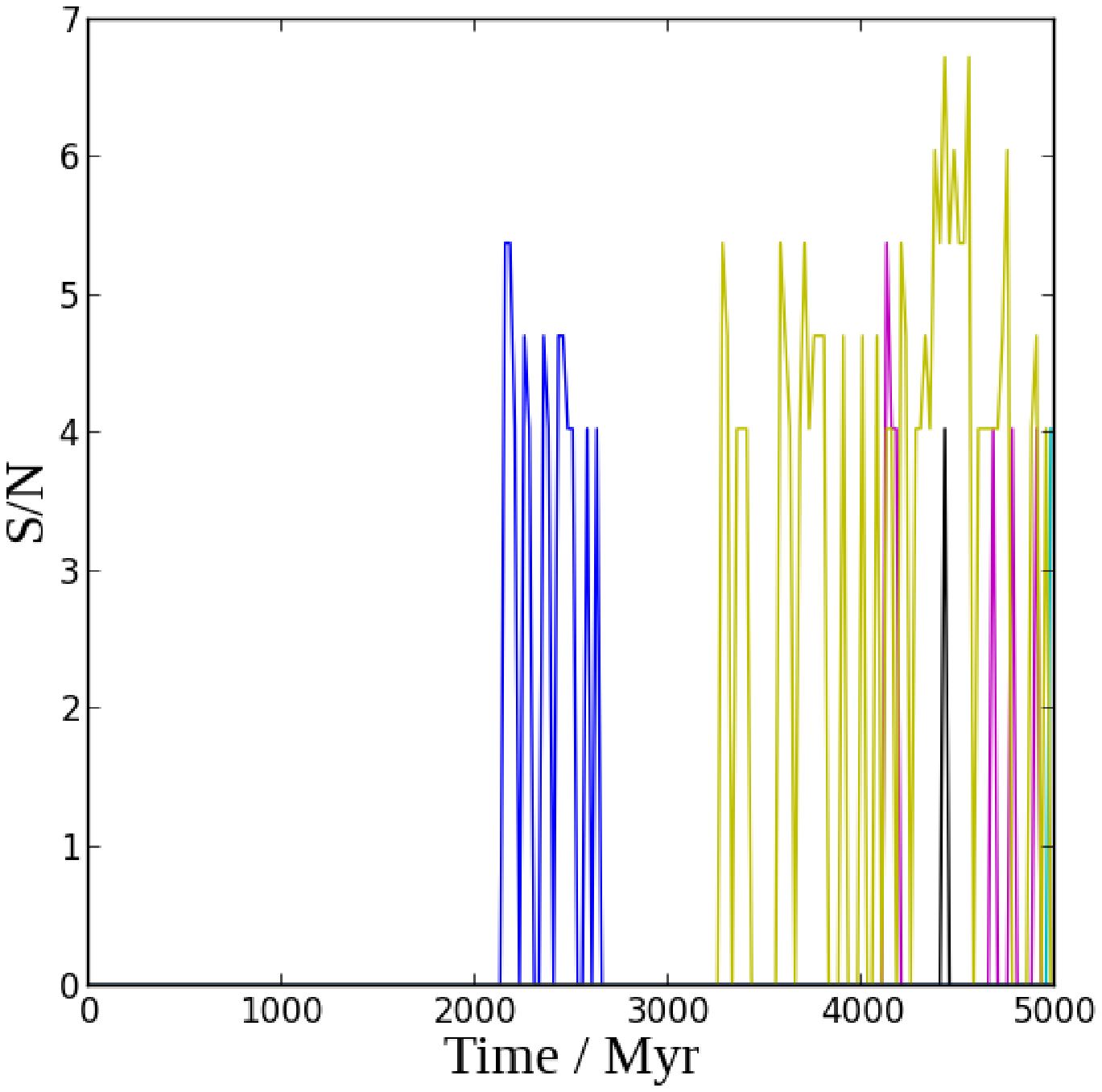}} 
  \subfloat[]{\includegraphics[height=55mm]{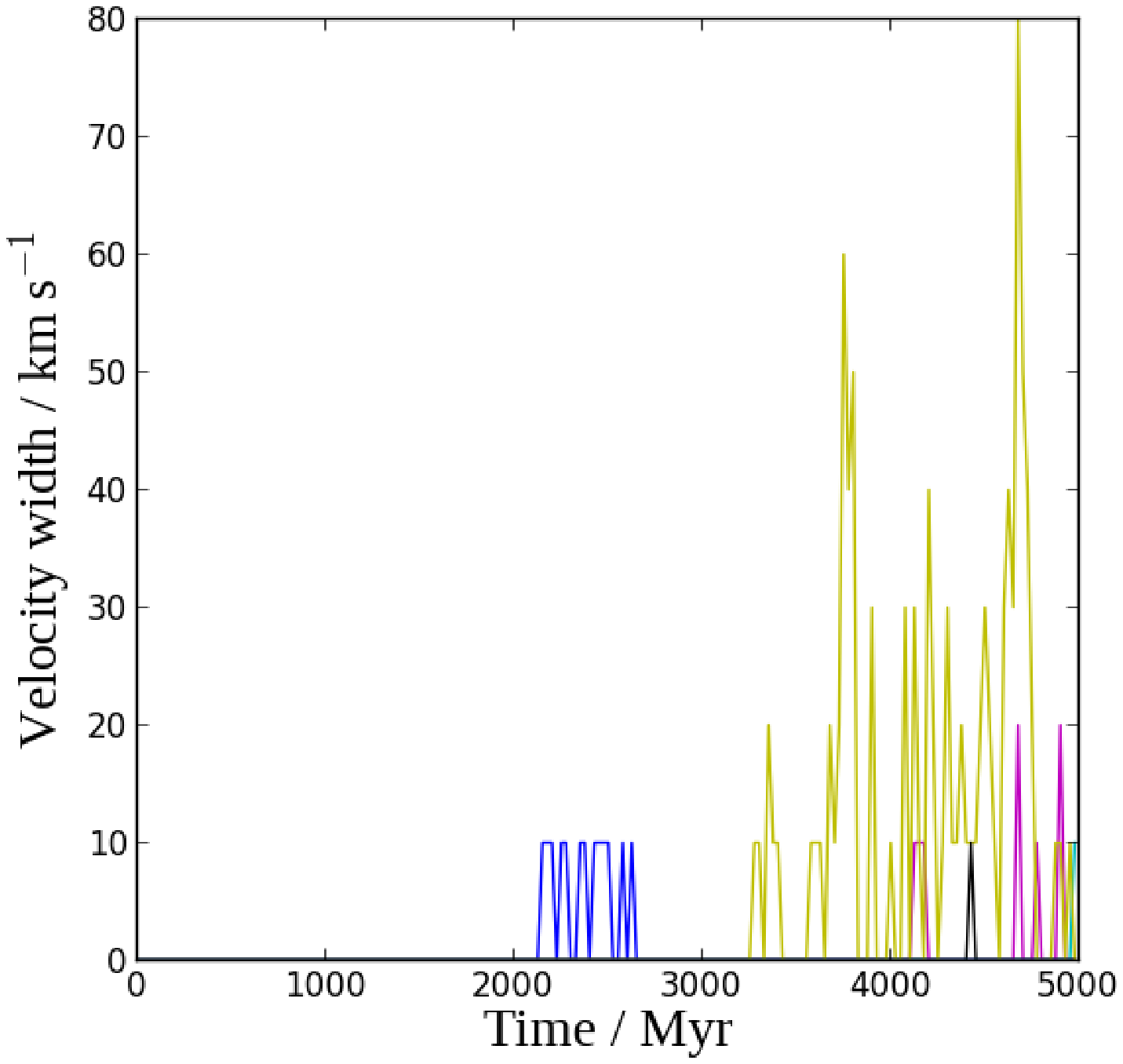}}
\caption[Streams]{Evolution of the unresolved, isolated \HI{} clouds as seen with an AGES-class survey for the M3 model. As in T16 we plot only the properties of the cloud with the highest velocity width for each simulation (each plotted as a different coloured line), since the high velocity widths appear to be the limiting factor in reproducing clouds similar to those described in T12 and T13. Panel (a) shows the detected mass in the cloud; (b) shows its signal to noise ratio; (c) shows its velocity width.}
\label{fig:v390clouds}
\end{figure*}

\end{document}